%% file: low-radio.tex
\documentclass[prd,aps,10pt,nofootinbib,twocolumn,superscriptaddress,preprintnumbers,balancelastpage,longbibliography,floatfix]{revtex4-2}

\usepackage{amsmath,amssymb}	
\usepackage{mathtools}
\usepackage{fontawesome}
\usepackage[dvipsnames]{xcolor}
\usepackage{hyperref}
\usepackage{xspace}
\usepackage{graphicx}
\usepackage{siunitx}
\usepackage{bbm}
\usepackage{float}
\usepackage{tabularx}
\usepackage{multirow}
\usepackage{dcolumn}
\usepackage{makecell}
\usepackage{afterpage}
\usepackage{longtable}
\usepackage{tikz}

\tikzset{every picture/.style={line width=0.75pt}} %set default line width to 0.75pt        

\input{definitions}

\hypersetup{colorlinks=true,
linkcolor=linkcolor,
citecolor=linkcolor,
urlcolor=linkcolor,
linktocpage=true,
pdfproducer=medialab,
}

\definecolor{nicegreen}{rgb}{0.1,0.5,0.1}

\DeclareSIUnit \h {\ensuremath{\mathit{h}}}
\DeclareSIUnit\eV{e\kern-.05em V}
\DeclareSIUnit\parsec{pc}
\DeclareSIUnit\year{yr}

\newcommand\aap{Astronomy and Astrophysics}
\newcommand\aaps{Astronomy and Astrophysics, Supplement}
\newcommand\aj{The Astronomical Journal}
\newcommand\mnras{Monthly Notices of the Royal Astronomical Society}

\makeatletter 
    
\renewcommand\onecolumngrid{
\do@columngrid{one}{\@ne}
\def\set@footnotewidth{\onecolumngrid}
\def\footnoterule{\kern-6pt\hrule width 1.5in\kern6pt}
}

\renewcommand\twocolumngrid{
        \def\footnoterule{
        \dimen@\skip\footins\divide\dimen@\thr@@
        \kern-\dimen@\hrule width.5in\kern\dimen@}
        \do@columngrid{mlt}{\tw@}
}

\makeatother  

\newcolumntype{C}[1]{>{\centering\let\newline\\\arraybackslash\hspace{0pt}}m{#1}}

\begin{document}

%\preprint{DESY 20-143, MIT-CTP/5441}
\preprint{MIT-CTP/5441}

\title{A Stimulating Explanation of the Extragalactic Radio Background}

\author{Andrea~Caputo}
\email{andrea.caputo@uv.es}
\thanks{ORCID: \href{https://orcid.org/0000-0003-1122-6606}{0000-0003-1122-6606}}
\affiliation{School  of  Physics  and  Astronomy,  Tel-Aviv  University,  Tel-Aviv  69978,  Israel}
\affiliation{Department  of  Particle  Physics  and  Astrophysics,Weizmann  Institute  of  Science,  Rehovot  7610001,  Israel}%

\author{Hongwan~Liu}
\email{hongwanl@princeton.edu}
\thanks{ORCID: \href{https://orcid.org/0000-0003-2486-0681}{0000-0003-2486-0681}}
\affiliation{Center for Cosmology and Particle Physics, Department of Physics, New York University, New York, NY 10003, USA}
\affiliation{Department of Physics, Princeton University, Princeton, NJ 08544, USA}

\author{\mbox{Siddharth~Mishra-Sharma}}
\email{smsharma@mit.edu}
\thanks{ORCID: \href{https://orcid.org/0000-0001-9088-7845}{0000-0001-9088-7845}}
\affiliation{The NSF AI Institute for Artificial Intelligence and Fundamental Interactions}
\affiliation{Center for Theoretical Physics, Massachusetts Institute of Technology, Cambridge, MA 02139, USA}
\affiliation{Department of Physics, Harvard University, Cambridge, MA 02138, USA}

\author{Maxim~Pospelov}
\affiliation{School of Physics and Astronomy, University of Minnesota, Minneapolis, MN 55455, USA}
\affiliation{William I. Fine Theoretical Physics Institute, School of Physics and Astronomy, University of Minnesota, Minneapolis, MN 55455, USA}

\author{Joshua~T.~Ruderman}
\email{ruderman@nyu.edu}
\thanks{ORCID: \href{https://orcid.org/0000-0001-6051-9216}{0000-0001-6051-9216}}
\affiliation{Center for Cosmology and Particle Physics, Department of Physics, New York University, New York, NY 10003, USA}
%\affiliation{Deutsches Elektronen-Synchrotron (DESY), D-22607 Hamburg, Germany}

\date{\protect\today}

\begin{abstract}

Despite an intense theoretical and experimental effort over the past decade, observations of the extragalactic radio background at multiple frequencies below 10 GHz are not understood in terms of known radio sources, and may represent a sign of new physics. In this \emph{Letter} we identify a new class of dark sector models with feebly interacting particles, where dark photons oscillate into ordinary photons that contribute to the radio background. Our scenario can explain both the magnitude and the spectral index of the radio background, while being consistent with other cosmological and astrophysical constraints. These models predict new relativistic degrees of freedom and spectral distortions of the cosmic microwave background, which could be detected in the next generation of experiments. 

\end{abstract}

\maketitle

\noindent

\noindent
{\bf Introduction.---}The cosmic microwave background (CMB) between \SIrange{60}{600}{\giga\hertz} is one of the most well-studied electromagnetic signals in science. In this frequency range, the CMB monopole dominates over astrophysical backgrounds, and is consistent with a blackbody distribution, with distortions limited to less than 1 part in $10^{4}$~\cite{Fixsen:1996nj}. Spatial fluctuations in the brightness of the CMB are on the level of 1 part in $10^5$, and the power spectrum of these fluctuations forms a key pillar of modern cosmology~\cite{Planck:2018vyg}. These facts constitute strong evidence for the primordial origins of the CMB. 

Much less is known about the extragalactic radio background (ERB) at frequencies $\nu \lesssim \SI{10}{\giga\hertz}$. The ARCADE~2 collaboration combined their own observations~\cite{Fixsen:2009xn} with measurements from other radio telescopes~\cite{Roger:1999jy,1999A&AS..140..145M,1981A&A...100..209H,1986A&AS...63..205R}, spanning a frequency range of \SI{22}{\mega\hertz}--\SI{90}{\giga\hertz}, and found that the ERB can be modeled as
\begin{equation}
	T(\nu)=T_0 + T_\text{R} \left(\frac{\nu}{\SI{310}{\mega\hertz}}\right)^\beta \,,
	\label{eq:T_power_law}
\end{equation}
where $T(\nu)$ denotes the brightness temperature. $T$ is related to the photon spectrum via 
$T(\omega) = (\pi^2 / \omega) \dd n_\gamma / \dd \omega$, where $\dd n_\gamma/\dd \omega$ is the number density of photons per unit energy (taking $\hbar = c = k_\text{B} = 1$), and $\omega \equiv 2 \pi \nu$. The power-law fit in Ref.~\cite{Fixsen:2009xn} found $T_0 = \SI{2.725 \pm 0.001}{\kelvin}$, $T_\text{R} = \SI{24.1 \pm 2.1}{\kelvin}$ and $\beta = -2.599(36)$. For $\nu \gtrsim \SI{100}{\giga\hertz}$, this is consistent with the CMB temperature of $T_0^\text{FIRAS} = \SI{2.7255(85)}{\kelvin}$ measured by FIRAS~\cite{Fixsen:1996nj,Fixsen:2009ug}. At lower frequencies, however, the power law is an excellent fit to $T_\text{exc} \equiv T - T_0$. A more recent re-analysis including all-sky maps from the LWA1 Low Frequency Sky Survey (LLFSS) reached a similar conclusion~\cite{Dowell:2018mdb}.

Explaining $T_\text{exc}$ with known sources of radio emission has thus far been surprisingly difficult~\cite{Singal:2017jlh}. Almost all attempts to do so have relied on synchrotron emission, since synchrotron sources typically produce $-3 \lesssim \beta \lesssim -2.5$~\cite{Cas1999ApJ,1968ARA&A...6..321S}. Extragalactic radio synchrotron sources such as active galactic nuclei and star-forming galaxies can produce $\beta \approx -2.7$~\cite{Platania:1997zn, Carilli:1991zz,Protheroe:1996si,Gervasi:2008rr,Nitu:2020vzn}, and the source counts distribution is thought to be relatively well-understood. However, numerous studies have estimated the emission from extragalactic radio sources to be 3-10 times smaller than $T_\text{exc}$~\cite{Gervasi:2008rr,Vernstrom_2011,Condon_2012,Fornengo:2014mna,2015MNRAS.447.2243V,Hardcastle:2020dfj} between \SI{100}{\mega\hertz}--\SI{10}{\giga\hertz}. 
Electrons that are reaccelerated during cluster mergers have been proposed as a significant contribution to $T_\text{exc}$~\cite{Fang:2015dga}, albeit under optimistic assumptions; this mechanism also produces a softer power law than is required. 

An alternative explanation of $T_\text{exc}$ is contamination from unmodeled Galactic sources of radio emission, which needs to be subtracted from radio data to obtain the extragalactic component~\cite{Subrahmanyan:2013eqa,Orlando:2013ysa,Krause:2021xav}. However, proposed additional sources of Galactic radio emission are strongly constrained~\cite{Singal_2010,Singal:2015tta,Krause:2021xav}. More sophisticated modeling using cosmic ray propagation models disfavors the possibility of significant Galactic contamination~\cite{Fornengo:2014mna}. 

A more exotic class of solutions involves radio emission from hypothetical early structures~\cite{Feng:2018rje}, including black holes~\cite{Biermann:2014lna,Ewall-Wice:2018bzf,Ewall-Wice:2019may} and star-forming galaxies~\cite{Mirocha:2018cih} at high redshifts. These solutions typically require large, persistent magnetic fields~\cite{Ewall-Wice:2019may}, whose origin and impact on inverse Compton cooling are debated~\cite{Sharma:2018agu}.

The ERB has also been studied in conjunction with 21-cm experiments, including EDGES~\cite{Bowman:2018yin}, LEDA~\cite{Bernardi:2016pva} and LOFAR~\cite{Mertens:2020llj}. These measurements are in tension with a $T_\text{exc}$ that is fully produced at high redshifts. In particular, the LEDA and LOFAR results constrain any cosmological radio background with $T_\text{exc} \propto \omega^{-2.6}$ at $\sim 10\%$ ($13.2 < z < 27.4$)~\cite{Feng:2018rje} and $\sim 46\%$ ($z = 9.1$)~\cite{Fialkov:2019vnb} of the present-day radio excess respectively.

Another aspect of the ERB that makes astrophysical explanations difficult is the spatial smoothness of the emission~\cite{Holder_2013}, deduced from measurements of the anisotropy of the radio sky~\cite{1988AJ.....96.1187F,1997ApJ...483...38P,Subrahmanyan:2000df,Holder_2013,Choudhuri:2020dgd,Offringa:2021rwp}. These measurements find that fluctuations in $T_\text{exc}$, $\Delta T / T_\text{exc}$, are $\sim 10^{-2}$ across a range of angular scales and frequencies; this is smoother than expected if radio emission is correlated with the present-day dark matter distribution~\cite{Holder_2013}. 

New physics explanations proposed thus far have primarily focused on synchrotron emission from DM annihilation and decay~\cite{Fornengo:2011cn, Hooper:2012jc, Cline:2012hb, Fang:2014joa}. However, these models also run into similar issues: they may result in nonsmooth emission; underproduce $T_\text{exc}$, unless the magnetic fields responsible for synchrotron production have unusual or unlikely properties~\cite{Cline:2012hb,Fang:2014joa}; require a new, large population of faint sources~\cite{2015MNRAS.447.2243V}, or require a large portion of the isotropic, extragalactic gamma-ray background to come from DM annihilation~\cite{Hooper:2012jc}, which is disfavored~\cite{Ajello:2015mfa,Lisanti:2016jub}.

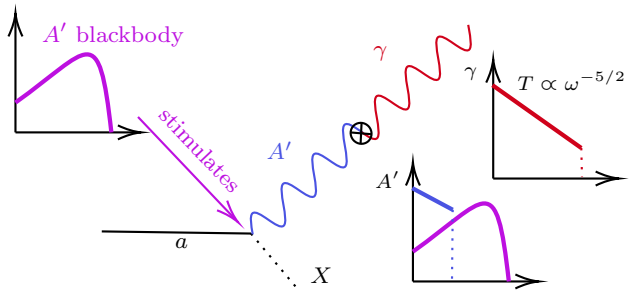
\begin{figure}
\resizebox*{0.48\textwidth}{!}{
\centering

\tikzset{every picture/.style={line width=0.75pt}} %set default line width to 0.75pt        

\begin{tikzpicture}[x=0.75pt,y=0.75pt,yscale=-1,xscale=1]
%uncomment if require: \path (0,184); %set diagram left start at 0, and has height of 184

%Straight Lines [id:da4183519610943698] 
\draw    (129,120) -- (199,121) ;
%Straight Lines [id:da17035911627855094] 
\draw  [dash pattern={on 0.84pt off 2.51pt}]  (199,121) -- (221.5,147.75) ;
%Shape: Wave [id:dp2975008266063859] 
\draw  [color={rgb, 255:red, 74; green, 83; blue, 226 }  ,draw opacity=1 ] (199.82,121.36) .. controls (199.07,116.98) and (198.37,112.8) .. (199.67,111.54) .. controls (200.98,110.29) and (203.93,112.31) .. (207.03,114.43) .. controls (210.12,116.55) and (213.07,118.56) .. (214.38,117.31) .. controls (215.68,116.05) and (214.98,111.88) .. (214.23,107.49) .. controls (213.48,103.11) and (212.78,98.93) .. (214.08,97.68) .. controls (215.39,96.42) and (218.34,98.44) .. (221.44,100.56) .. controls (224.53,102.68) and (227.49,104.7) .. (228.79,103.44) .. controls (230.09,102.19) and (229.39,98.01) .. (228.64,93.63) .. controls (227.9,89.24) and (227.19,85.06) .. (228.49,83.81) .. controls (229.8,82.55) and (232.75,84.57) .. (235.85,86.69) .. controls (238.94,88.81) and (241.9,90.83) .. (243.2,89.57) .. controls (244.5,88.32) and (243.8,84.14) .. (243.05,79.76) .. controls (242.31,75.37) and (241.6,71.2) .. (242.91,69.94) .. controls (244.21,68.69) and (247.16,70.7) .. (250.26,72.82) .. controls (250.69,73.12) and (251.12,73.42) .. (251.55,73.7) ;
%Shape: Wave [id:dp04088717791861263] 
\draw  [color={rgb, 255:red, 208; green, 2; blue, 27 }  ,draw opacity=1 ] (250.82,73.36) .. controls (253.92,75.48) and (256.87,77.5) .. (258.17,76.24) .. controls (259.48,74.99) and (258.77,70.81) .. (258.03,66.43) .. controls (257.28,62.04) and (256.57,57.87) .. (257.88,56.61) .. controls (259.18,55.36) and (262.13,57.37) .. (265.23,59.49) .. controls (268.33,61.62) and (271.28,63.63) .. (272.58,62.38) .. controls (273.89,61.12) and (273.18,56.94) .. (272.44,52.56) .. controls (271.69,48.18) and (270.98,44) .. (272.29,42.74) .. controls (273.59,41.49) and (276.55,43.5) .. (279.64,45.63) .. controls (282.74,47.75) and (285.69,49.76) .. (286.99,48.51) .. controls (288.3,47.25) and (287.59,43.07) .. (286.85,38.69) .. controls (286.1,34.31) and (285.4,30.13) .. (286.7,28.87) .. controls (288,27.62) and (290.96,29.64) .. (294.05,31.76) .. controls (297.15,33.88) and (300.1,35.89) .. (301.41,34.64) .. controls (302.71,33.38) and (302,29.21) .. (301.26,24.82) .. controls (301.15,24.21) and (301.05,23.6) .. (300.95,23) ;
%Flowchart: Summing Junction [id:dp41680594745796307] 
\draw   (247.05,77.11) .. controls (245.23,75.03) and (245.37,71.92) .. (247.37,70.17) .. controls (249.38,68.42) and (252.48,68.69) .. (254.3,70.77) .. controls (256.12,72.85) and (255.97,75.95) .. (253.97,77.71) .. controls (251.97,79.46) and (248.87,79.19) .. (247.05,77.11) -- cycle ; \draw   (245.78,73.52) -- (255.57,74.36) ; \draw   (250.9,69.03) -- (250.44,78.85) ;
%Straight Lines [id:da971552845509386] 
\draw    (88.5,73.5) -- (88.5,19.5) ;
\draw [shift={(88.5,17.5)}, rotate = 90] [color={rgb, 255:red, 0; green, 0; blue, 0 }  ][line width=0.75]    (10.93,-3.29) .. controls (6.95,-1.4) and (3.31,-0.3) .. (0,0) .. controls (3.31,0.3) and (6.95,1.4) .. (10.93,3.29)   ;
%Straight Lines [id:da046886369861139476] 
\draw    (88.5,73.5) -- (145.5,73.5) ;
\draw [shift={(147.5,73.5)}, rotate = 180] [color={rgb, 255:red, 0; green, 0; blue, 0 }  ][line width=0.75]    (10.93,-3.29) .. controls (6.95,-1.4) and (3.31,-0.3) .. (0,0) .. controls (3.31,0.3) and (6.95,1.4) .. (10.93,3.29)   ;
%Curve Lines [id:da18411153181208006] 
\draw [color={rgb, 255:red, 189; green, 16; blue, 224 }  ,draw opacity=1 ][line width=1.5]    (88.5,59.5) .. controls (123.5,34.5) and (127.5,19.5) .. (133.5,73.5) ;
%Straight Lines [id:da05681387626649981] 
\draw [color={rgb, 255:red, 189; green, 16; blue, 224 }  ,draw opacity=1 ]   (146,66.25) -- (191.62,114.3) ;
\draw [shift={(193,115.75)}, rotate = 226.48] [color={rgb, 255:red, 189; green, 16; blue, 224 }  ,draw opacity=1 ][line width=0.75]    (10.93,-3.29) .. controls (6.95,-1.4) and (3.31,-0.3) .. (0,0) .. controls (3.31,0.3) and (6.95,1.4) .. (10.93,3.29)   ;
%Straight Lines [id:da7517460100162514] 
\draw    (275,143.5) -- (275,89.5) ;
\draw [shift={(275,87.5)}, rotate = 90] [color={rgb, 255:red, 0; green, 0; blue, 0 }  ][line width=0.75]    (10.93,-3.29) .. controls (6.95,-1.4) and (3.31,-0.3) .. (0,0) .. controls (3.31,0.3) and (6.95,1.4) .. (10.93,3.29)   ;
%Straight Lines [id:da9045488750955897] 
\draw    (275,143.5) -- (332,143.5) ;
\draw [shift={(334,143.5)}, rotate = 180] [color={rgb, 255:red, 0; green, 0; blue, 0 }  ][line width=0.75]    (10.93,-3.29) .. controls (6.95,-1.4) and (3.31,-0.3) .. (0,0) .. controls (3.31,0.3) and (6.95,1.4) .. (10.93,3.29)   ;
%Curve Lines [id:da6211099293503983] 
\draw [color={rgb, 255:red, 189; green, 16; blue, 224 }  ,draw opacity=1 ][line width=1.5]    (275,129.5) .. controls (310,104.5) and (314,89.5) .. (320,143.5) ;
%Straight Lines [id:da7505380851996788] 
\draw [color={rgb, 255:red, 74; green, 83; blue, 226 }  ,draw opacity=1 ][fill={rgb, 255:red, 74; green, 99; blue, 226 }  ,fill opacity=1 ][line width=1.5]    (274.67,99.67) -- (293.67,109.83) ;
%Straight Lines [id:da723539440340389] 
\draw [color={rgb, 255:red, 74; green, 83; blue, 226 }  ,draw opacity=1 ] [dash pattern={on 0.84pt off 2.51pt}]  (293.67,109.83) -- (293.33,143.17) ;
%Straight Lines [id:da5808483054053262] 
\draw    (312.67,95.17) -- (312.67,41.17) ;
\draw [shift={(312.67,39.17)}, rotate = 90] [color={rgb, 255:red, 0; green, 0; blue, 0 }  ][line width=0.75]    (10.93,-3.29) .. controls (6.95,-1.4) and (3.31,-0.3) .. (0,0) .. controls (3.31,0.3) and (6.95,1.4) .. (10.93,3.29)   ;
%Straight Lines [id:da31550266927756665] 
\draw    (312.67,95.17) -- (369.67,95.17) ;
\draw [shift={(371.67,95.17)}, rotate = 180] [color={rgb, 255:red, 0; green, 0; blue, 0 }  ][line width=0.75]    (10.93,-3.29) .. controls (6.95,-1.4) and (3.31,-0.3) .. (0,0) .. controls (3.31,0.3) and (6.95,1.4) .. (10.93,3.29)   ;
%Straight Lines [id:da47613162748698734] 
\draw [color={rgb, 255:red, 208; green, 2; blue, 27 }  ,draw opacity=1 ][fill={rgb, 255:red, 74; green, 99; blue, 226 }  ,fill opacity=1 ][line width=1.5]    (312.33,51.33) -- (354,80.83) ;
%Straight Lines [id:da48718108191917997] 
\draw [color={rgb, 255:red, 208; green, 2; blue, 27 }  ,draw opacity=1 ] [dash pattern={on 0.84pt off 2.51pt}]  (354,80.83) -- (354.33,94.5) ;

% Text Node
\draw (162,121.9) node [anchor=north west][inner sep=0.75pt]  [font=\footnotesize]  {$\mathnormal{a}$};
% Text Node
\draw (205,76.4) node [anchor=north west][inner sep=0.75pt]  [font=\footnotesize]  {$\textcolor[rgb]{0.29,0.33,0.89}{A'}$};
% Text Node
\draw (255,31.4) node [anchor=north west][inner sep=0.75pt]  [font=\footnotesize]  {$\textcolor[rgb]{0.82,0.01,0.11}{\gamma }$};
% Text Node
\draw (98.5,20.5) node [anchor=north west][inner sep=0.75pt]  [font=\footnotesize,color={rgb, 255:red, 189; green, 16; blue, 224 }  ,opacity=1 ] [align=left] {\begin{minipage}[lt]{51.41pt}\setlength\topsep{0pt}
\begin{center}
$\displaystyle A'${ blackbody}
\end{center}

\end{minipage}};
% Text Node
\draw (160.2,58.1) node [anchor=north west][inner sep=0.75pt]  [font=\footnotesize,color={rgb, 255:red, 189; green, 16; blue, 224 }  ,opacity=1 ,rotate=-46.57] [align=left] {\begin{minipage}[lt]{38.81pt}\setlength\topsep{0pt}
\begin{center}
{stimulates}
\end{center}

\end{minipage}};
% Text Node
\draw (255.83,90.4) node [anchor=north west][inner sep=0.75pt]  [font=\scriptsize,color={rgb, 255:red, 0; green, 0; blue, 0 }  ,opacity=1 ]  {$A'$};
% Text Node
\draw (297.5,41.4) node [anchor=north west][inner sep=0.75pt]  [font=\scriptsize,color={rgb, 255:red, 0; green, 0; blue, 0 }  ,opacity=1 ]  {$\gamma $};
% Text Node
\draw (322.5,43.07) node [anchor=north west][inner sep=0.75pt]  [font=\scriptsize]  {$T\varpropto \omega ^{-5/2}$};
% Text Node
\draw (225.5,135.4) node [anchor=north west][inner sep=0.75pt]  [font=\footnotesize]  {$X$};

\end{tikzpicture}
}
\caption{A schematic representation of the key aspects of the proposed class of models.}
\label{fig:model}
\end{figure}

In this \emph{Letter}, we show that a simple class of experimentally viable new-physics models can explain the amplitude, power-law dependence and smoothness of $T_\text{exc}$. These models rely on three basic ingredients: \textit{1)} a particle decaying into dark photons $A'$; \textit{2)} the presence of a thermal bath of $A'$ which stimulates this decay; and \textit{3)} $A'$ resonantly oscillating into radio photons.  This class of models leads to $T_\text{exc} \propto \omega^{-5/2}$, close to the observed power-law dependence. Relatively low-$z$ resonant oscillations as well as $\dd n_{A'}/\dd \omega \propto \omega^{-1/2}$ prior to oscillations are both crucial elements of such models; we have not been able to identify viable alternatives. One possibility is the decay of DM particles to final states charged under $A'$, and $A'$ appearing as ``dark bremsstrahlung''. While having a spectrum similarly enhanced in the infrared, the DM has to be light and millicharged, with an effective electromagnetic charge exceeding current experimental bounds~\cite{Vogel:2013raa}. Another alternative is DM decay into a dark photon $A''$, which then resonantly converts twice via $A'' \to A' \to \gamma$,  with some dark states charged under a dark photon $A''$ facilitating the $A'' \to A'$ transition. However, two resonant conversions makes it difficult to produce a sufficiently large $T_\text{exc}$ while ensuring that $T_\text{exc} \propto \omega^{-5/2}$ over nearly three decades of frequency.

The remainder of this \emph{Letter} is organized as follows. We begin by introducing the class of models and deriving the expected $T_\text{exc}$ from them. Next, we discuss our fit of a particular example model to the radio data, and relevant experimental constraints. We then move on to discuss the anisotropy of the ERB produced by the model. We conclude with future prospects for confirmation of this model. We adopt \emph{Planck} 2018 cosmology throughout~\cite{Planck:2018vyg}.

\noindent
{\bf Model.---}A particle physics model that has the following three features can generate an ERB with $T_\text{exc} \propto \omega^{-5/2}$:
\begin{enumerate}
	\item  a cold component of DM $a$ which is stable on cosmological timescales, but undergoes a two-body decay with lifetime $\tau_\text{vac}$ in vacuum (for simplicity, we take $a$ to be all of DM);
	  \item one of the daughter particles of the decay is a dark photon, $A'$, with an existing blackbody distribution, characterized by a temperature $T'(z) = T_0'(1+z)$, where $T_0'$ is its temperature today; and
	\item the $A'$ has  mass $\SI{e-15}{\eV}\lesssim m_{A'} \lesssim \SI{e-9}{\eV}$ and is emitted relativistically with energy $\omega_{A'}$. $A'$ mixes with the Standard Model photon $\gamma$ with kinetic mixing parameter $\epsilon$.
 
\end{enumerate}
These three features are summarized in Fig.~\ref{fig:model}. The existence of the thermal population of $A'$ enhances the decay rate of $a$ due to Bose enhancement~\cite{Caputo:2018vmy,Bolliet:2020ofj}, leading to a redshift-dependent effective decay lifetime $\tau(z)$, where
\begin{alignat}{1}
	\tau(z) = \tau_\text{vac} \left[1 + n f^\text{BB}_{A'}(z) \right]^{-1} \,, 
	\label{eq:tau_stimulated}
\end{alignat}
with $f^\text{BB}_{A'} = (e^{\omega_{A'} / T'} - 1)^{-1}$ being the blackbody occupation number of $A'$ with energy $\omega_{A'}$. 

There are several parameters that depend on the specifics of the model. The value of $n$ depends on the occupation number of the other daughter particle. In addition, $a$ can decay into $\alpha = 1$ or $2$ dark photons. Finally, $\omega_{A'}$ depends on the kinematics of the decay of $a$, with $\omega_{A'} = m_a/2$ if it decays into a pair of relativistic final states. In App.~\ref{Sec:FiducialModel}, we describe a concrete particle physics model that realizes all three features. To guarantee $T_\text{exc} \propto \omega^{-5/2}$, processes such as inverse decays of $A'$ into $a$ cannot significantly distort either the power-law index of the $A'$ spectrum produced by the decay, or the blackbody distribution of $A'$ across all relevant frequencies. In accordance with this fiducial model, we fix $n = 2$, $\alpha = 1$, and $\omega_{A'} = m_a/2$ for all results that are specific to it, although we emphasize that this is only one example, out of potentially many, that we have studied in detail.

The small kinetic mixing between $A'$ and $\gamma$ enables resonant conversion between the two particles whenever $m_{A'}^2 = m_\gamma^2$~\cite{Mirizzi:2009iz,Caputo:2020rnx}, where $m_\gamma^2$ is the effective photon plasma mass squared. This quantity is proportional to the free electron number density, $n_\text{e}$. The converted photons ultimately form the present-day $T_\text{exc}$. We calculate the sky-averaged conversion probability per redshift $\dd \langle P_{A' \to \gamma} \rangle / \dd z$ taking into account inhomogeneities using the formalism developed in Refs.~\cite{Caputo:2020bdy,Caputo:2020rnx} (see also Refs.~\cite{Bondarenko:2020moh,Garcia:2020qrp,Witte:2020rvb}).

\noindent
{\bf Radio background production.---}The particle $a$ decays throughout cosmic history, producing a number density of $A'$ per unit energy given by (see App.~\ref{Sec:SpectrumDerivation} for a complete derivation, which follows from Ref.~\cite{Pospelov:2018kdh})
\begin{alignat}{1}
	\frac{\dd n_{A'}}{\dd \omega}(\omega, z) = \frac{\alpha  \rho_a(z)}{m_a \omega \tau(z_\star) H(z_\star)} \Theta(z_\star - z) \,,
	\label{eq:Ap_spectrum}
\end{alignat}
where $1 + z_\star \equiv \omega_{A'}(1+z)/ \omega$ is the redshift at which a daughter $A'$ with frequency $\omega$ at redshift $z$ was produced, $H$ is the Hubble parameter, $\rho_a(z)$ is the DM energy density, and $\Theta$ is the Heaviside step function. The total spectrum of $A'$ is then a sum of this decay spectrum and the blackbody spectrum of the thermal $A'$ distribution, which is subdominant in the energy range of interest.

In comoving coordinates, the spectrum of photons produced is obtained by integrating the spectrum of dark photons times $\dd \langle P_{A' \to \gamma} \rangle/ \dd z$, \textit{i.e.},
\begin{alignat}{1}
	\frac{1}{(1+z)^3}\frac{\dd n_\gamma}{\dd x} = \int_z^\infty \dd z' \, \frac{\dd \langle P_{A' \to \gamma} \rangle}{\dd z'} \frac{1}{(1+z')^3} \frac{\dd n_{A'}}{\dd x}(x, z') \,,
	\label{eq:dn_gamma_d_omega_basic_expr}
\end{alignat}
where $x \equiv \omega/[T_0(1+z)]$. For the range of $m_{A'}$ that is of interest, resonant conversions occur only after recombination. In the range $3 \times 10^{-4} \lesssim x \lesssim 0.2$, which are relevant for $T_\text{exc}$, these photons only evolve via redshifting, with the baryonic fluid being essentially transparent to them~\cite{Bolliet:2020ofj}. 

Substituting Eq.~\eqref{eq:Ap_spectrum} into Eq.~\eqref{eq:dn_gamma_d_omega_basic_expr}, we find
\begin{alignat}{1}
	\frac{\dd n_\gamma}{\dd x}(x,z) = \frac{\rho_a(z)}{m_a} \frac{\alpha}{x} \underbrace{\frac{1}{\tau(z_\star)}}_{\propto \, x^{-1}}\underbrace{\frac{1}{H(z_\star)}}_{\propto \, x^{3/2}} \underbrace{\int_z^{z_\star} \dd z' \frac{\dd \langle P_{A' \to \gamma} \rangle}{\dd z'}}_{\propto \, x^{-1}} \,.
	\label{eq:dn_gamma_d_omega_final}
\end{alignat}
This gives $\dd n_\gamma / \dd x \propto x^{-3/2}$, or $T_\text{exc} \propto \omega^{-5/2}$, the desired frequency dependence. Note that $z_\star$ must occur during matter domination in order for $H(z_\star) \propto x^{-3/2}$, while $\dd \langle P_{A' \to \gamma} \rangle / \dd z \propto x^{-1}$ is derived in Refs.~\cite{Mirizzi:2009iz,Caputo:2020rnx}. $\tau(z_\star) \propto x$ follows from Eq.~\eqref{eq:tau_stimulated} when $T' \gg \omega_{A'}$. 

\begin{figure*}[t!]
	\centering
	\includegraphics[width=0.5\textwidth]{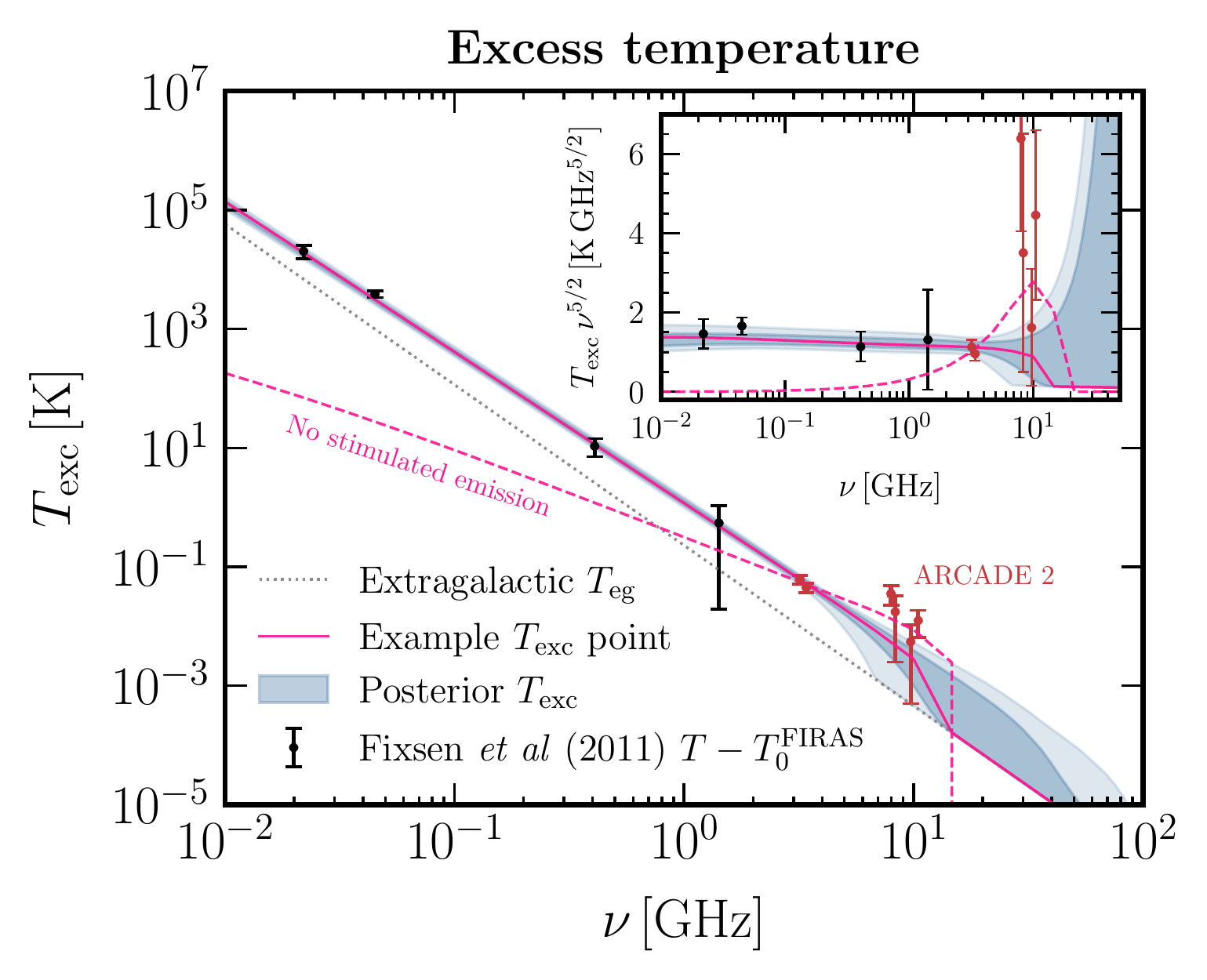} \qquad
	\includegraphics[width=0.44\textwidth]{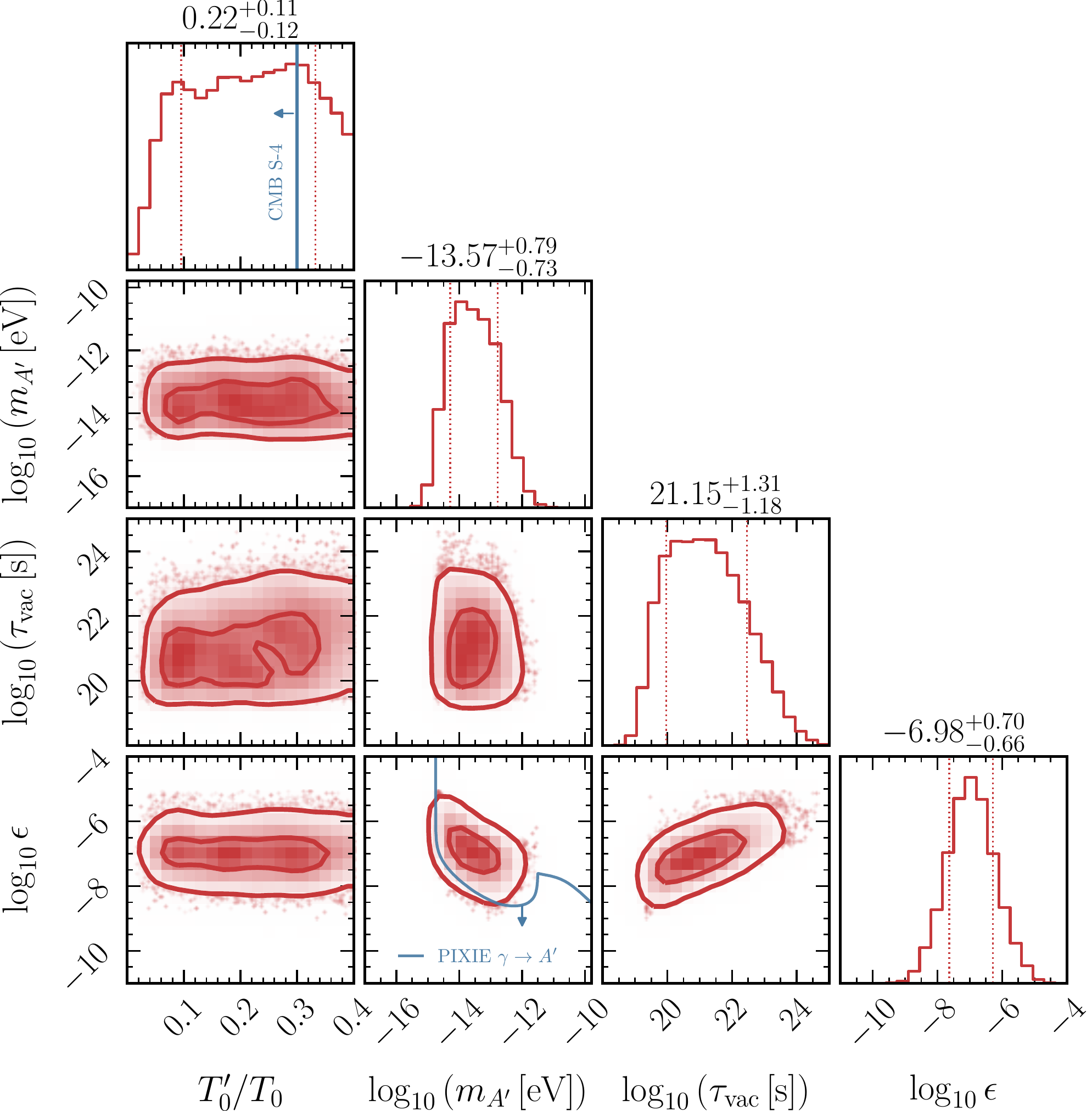}
	\caption{\emph{(Left)} Point-wise posterior for $T_\text{exc}$ within our proposed model, showing the middle-68\% and 95\% regions (dark and light blue regions, respectively). We include a subdominant but irreducible contribution from unresolved extragalactic sources $T_\text{eg}$~\cite{Gervasi:2008rr} (dashed grey) for completeness. The spectrum for a single point in parameter space is shown in pink. Radio data, plotted as $T - T_0^\text{FIRAS}$, include measurements from ARCADE~2 are shown in red~\cite{Fixsen:2009xn}, with results from other telescopes compiled in the same reference shown in black. A fit to the ARCADE~2 data points assuming no stimulated emission and $T_\text{eg} = 0$ is shown by the pink dashed line. The inset shows the same quantities scaled by a factor of $\nu^{5/2}$ in order to more clearly expose the posterior in relation to the measured data. \emph{(Right)} The inferred marginal and joint posterior distributions over a subset of parameters---$T_0'/T_0$, $m_{A'}, \tau_a$, and $\epsilon$---in our fiducial model. Median and middle-68\% containment values are indicated in the subtitles, while for $\log_{10}(m_a / \SI{}{\eV})$ we obtain $-3.66^{+0.31}_{-0.29}$ (not shown). The potential reaches of PIXIE~\cite{Kogut:2011xw,Caputo:2020bdy}, and CMB-S4~\cite{Abazajian:2019eic} in our parameter space are shown in blue, with projected unconstrained regions indicated by the arrow.}
	\label{fig:T_exc_fiducial}
\end{figure*}

Eq.~\eqref{eq:dn_gamma_d_omega_final} is the main result of this paper. For our fiducial model, we find the following approximate parametric dependence of $T_\text{exc}$ at $z = 0$: 

\begin{multline}
\label{Eq:Tex2}
	T_\text{exc}(\omega) \approx \SI{10}{\kelvin} \left(\frac{\nu}{\SI{310}{\mega\hertz}}\right)^{-5/2} \left(\frac{T_0'}{0.2 T_0}\right) \left(\frac{\SI{e21}{\second}}{\tau_\text{vac}}\right) \\ 
	\times \left(\frac{\SI{e-4}{\eV}}{\omega_{A'}}\right)^{3/2} \left(\frac{\SI{2e-4}{\eV}}{m_a}\right)  \left(\frac{P_\text{GHz}}{10^{-5}}\right) \,,
\end{multline}
where $P_\text{GHz}$ is $\langle P_{A' \to \gamma} \rangle$ for $\nu = \SI{1}{\giga\hertz}$ today, which the FIRAS measurement of the CMB energy spectrum limits to be $P_\text{GHz} \lesssim 10^{-2}$~\cite{Mirizzi:2009iz,Caputo:2020bdy}.

\noindent
{\bf Fit and constraints.---}
We can now perform a fit of $T_\text{exc}$ predicted by our model to the measured data from Ref.~\cite{Fixsen:2009xn}. There are six parameters in our fit: five parameters $\{ m_a, m_{A'}, \tau_\text{vac}, T_0', \epsilon \}$ from our new physics model, and $T_0$.  Fitting the radio data places some requirements on the model parameters. For $T_\text{exc}$ to be present across all data points, we require $m_a/2 \gtrsim 2 \pi \times \SI{10}{\giga\hertz}$. On the other hand, photons at \SI{22}{\mega\hertz} must originate from decays of $a$ during matter domination to satisfy $T_\text{exc} \propto \omega^{-5/2}$, leading to the requirement that $m_a / 2 \lesssim 2\pi (1+z_\text{eq}) \times \SI{22}{\mega\hertz}$, where $z_\text{eq}$ is the redshift of matter-radiation equality. Together, this means $\SI{8e-5}{\eV} \lesssim m_a \lesssim \SI{6e-4}{\eV}$. Resonant conversion must predominantly occur after $z_\star$ for \SI{10}{\giga\hertz} photons so that the full power law extends to at least that frequency; the requirements on $m_a$ show that $z_\star \lesssim 6$, which favors $\SI{e-14}{\eV} \lesssim m_{A'} \lesssim \SI{3e-13}{\eV}$. Finally, $T'$ must be sufficiently large for $f_{A'}^\text{BB} \gg 1$ and $\tau(z_\star) \propto x$ for the \SI{10}{\giga\hertz} data points; this imposes $T_0'/T_0 \gtrsim 0.06$. 

Several important experimental constraints exist on the class of models under consideration. First, $T_0' \lesssim 0.4 T_0$, in order to avoid introducing excessive effective relativistic degrees of freedom~\cite{Planck:2018vyg}.  
The kinetic mixing between $A'$ and $\gamma$ also leads to spectral distortions due to $\gamma \to A'$ resonant conversions; 
the FIRAS measurement of the CMB spectrum~\cite{Fixsen:1996nj} leads to constraints on the order of $\epsilon < 10^{-7}$--$10^{-5}$ in the $m_{A'}$ range of interest~\cite{Caputo:2020bdy,Caputo:2020rnx}.

There are also limits on the DM decay lifetime obtained from the CMB power spectrum, large scale structure and the Milky Way satellite population~\cite{Poulin:2016nat,Nygaard:2020sow,DES:2020mpv,Mau:2022sbf,Simon:2022ftd,Alvi:2022aam}; however, these results assume a constant decay width. Conservatively requiring the total energy density of $a$ that has decayed away with stimulated decay from the $A'$ thermal population by the present-day to be less than 2.2\%, we find $\tau_\text{vac} > \SI{e19}{\second} \times \max \left[2.1, 6.3 \left(\SI{e-4}{\eV} / \omega_{A'}\right)\left(T_0' / 0.2 T_0\right)\right]$ (see App.~\ref{Sec:ExptConstraints} for details). We also note that the decaying particle $a$ may be a subcomponent $f_{\rm DM}$ of DM, which evades the lifetime bound altogether if $f_{\rm DM} \lesssim 2.2\%$~\cite{Simon:2022ftd}.

21-cm power spectrum measurements from both LEDA and LOFAR are both in tension with $T_\text{exc}$ being fully produced at high redshifts ($z \gtrsim 9.1$), placing strong constraints on models that produce $T_\text{exc}$ primordially. In our model, however, $A' \to \gamma$ resonant conversions occur predominantly at $z \lesssim 6$. 

We explore the posterior on the model parameters using nested sampling~\cite{10.1214/06-BA127,Feroz:2008xx,2004AIPC..735..395S} implemented in \texttt{dynesty}~\cite{Speagle_2020}. Our priors, described in detail in App.~\ref{Sec:DataAnalysis}, are constructed such that they have zero probability density in parameter regions incompatible with \textit{1)} the FIRAS spectral distortion limits of Refs.~\cite{Caputo:2020bdy,Caputo:2020rnx} and \textit{2)} the DM lifetime limit discussed above. Priors on $T_0'$ and $T_0$ are also chosen to account for the $N_\text{eff}$ and FIRAS constraints on these parameters, respectively. The posterior on the excess temperature is shown on the left in Fig.~\ref{fig:T_exc_fiducial}. In computing $T_\text{exc}$ from our model, we include an irreducible contribution from unresolved extragalactic radio sources of $T_\text{eg} = \SI{0.23}{\kelvin}(\nu / \SI{}{GHz})^{-2.7}$~\cite{Gervasi:2008rr,Holder_2013}, which is at least 3 times smaller than the measured brightness temperature. This parametrization of $T_\text{eg}$ is in excellent agreement with other independent estimates~\cite{2010MNRAS.404..532M}. Although this power-law expression strictly applies in the range \SI{100}{\mega\hertz}--\SI{10}{\giga\hertz}, we estimate the contribution outside of this frequency range by extrapolation. The marginal and joint posterior distributions are shown on the right in Fig.~\ref{fig:T_exc_fiducial}. 
The marginal posterior medians correspond to $m_a \simeq \SI{2e-4}{\eV}$, $m_{A'} \simeq \SI{2.5e-14}{\eV}$, $\tau \simeq \SI{1.3e21}{\second}$, $T_0' \simeq 0.22 \, T_0 $ and $\epsilon \simeq 10^{-7}$. We find an excellent fit to the data over a wide range of allowed model parameters. 
In App.~\ref{Sec:ExtendedResults}, we show the extended corner plot for posterior distributions of all parameters of interest, along with other systematic variations; these do not qualitatively change our main result.

\noindent 
{\bf Smoothness.---}Upper limits on the anisotropy of the ERB have been reported for $4 \times 10^3 \lesssim \ell \lesssim 6 \times 10^4$ in the \SIrange{4}{8}{\giga\hertz} range by VLA~\cite{1988AJ.....96.1187F,1997ApJ...483...38P} and ATCA~\cite{Subrahmanyan:2000df}, while actual measurements of the power spectrum have been made by LOFAR~\cite{Offringa:2021rwp} and TGSS~\cite{Choudhuri:2020dgd} at $\sim$\SI{140}{\mega\hertz} for $10^2 \lesssim \ell \lesssim 10^4$. Such measurements can be challenging: observations at $\sim$\SI{140}{\mega\hertz} disagree with each other by a factor of 3, and face issues such as galactic foreground contamination and calibration errors~\cite{Offringa:2021rwp}. To assess the smoothness of $T_\text{exc}$ in our model, we take the \SIrange{4}{8}{\giga\hertz} upper limits and results from LOFAR---which reports a lower power---as approximate upper limits, noting that astrophysical sources can contribute more power~\cite{Offringa:2021rwp}, and that these observations are expected to improve.

There are two possible contributions to the ERB anisotropy produced by our model that are essentially independent: \textit{1)} decay anisotropy, due to DM density correlations from the point at which $a$ decays, and \textit{2)} conversion anisotropy, due to correlations in free electron density fluctuations $\delta_\text{e}$ during $A' \to \gamma$ conversions, since $m_\gamma^2 \propto n_\text{e}$. The decay anisotropy was found to exceed the radio anisotropy power spectrum in the $\SIrange{4}{8}{\giga\hertz}$ range, unless the decay that produces these photons happens at $z_\star \gtrsim 5$~\cite{Holder_2013,Cline:2012hb}. This is easily satisfied in the range of parameters providing a good fit. 

We compute the conversion anisotropy power spectrum by first writing down the two-point correlation function of the conversion probability of $A' \to \gamma$ in two different directions in the sky. As such, it depends on the two-point function of $\delta_\text{e}$, which we model as either a Gaussian or a log-normal random field. The anisotropy power spectrum can then be obtained by performing a spherical harmonics decomposition, and using the Limber approximation~\cite{Limber, Kaiser:1991qi, Kaiser:1996tp} following the method outlined in Ref.~\cite{Fornengo:2013rga}. Further details of our calculation can be found in App.~\ref{Sec:Anisotropy}.  

\begin{figure}[t!]
	\centering
	\includegraphics[width=0.48\textwidth]{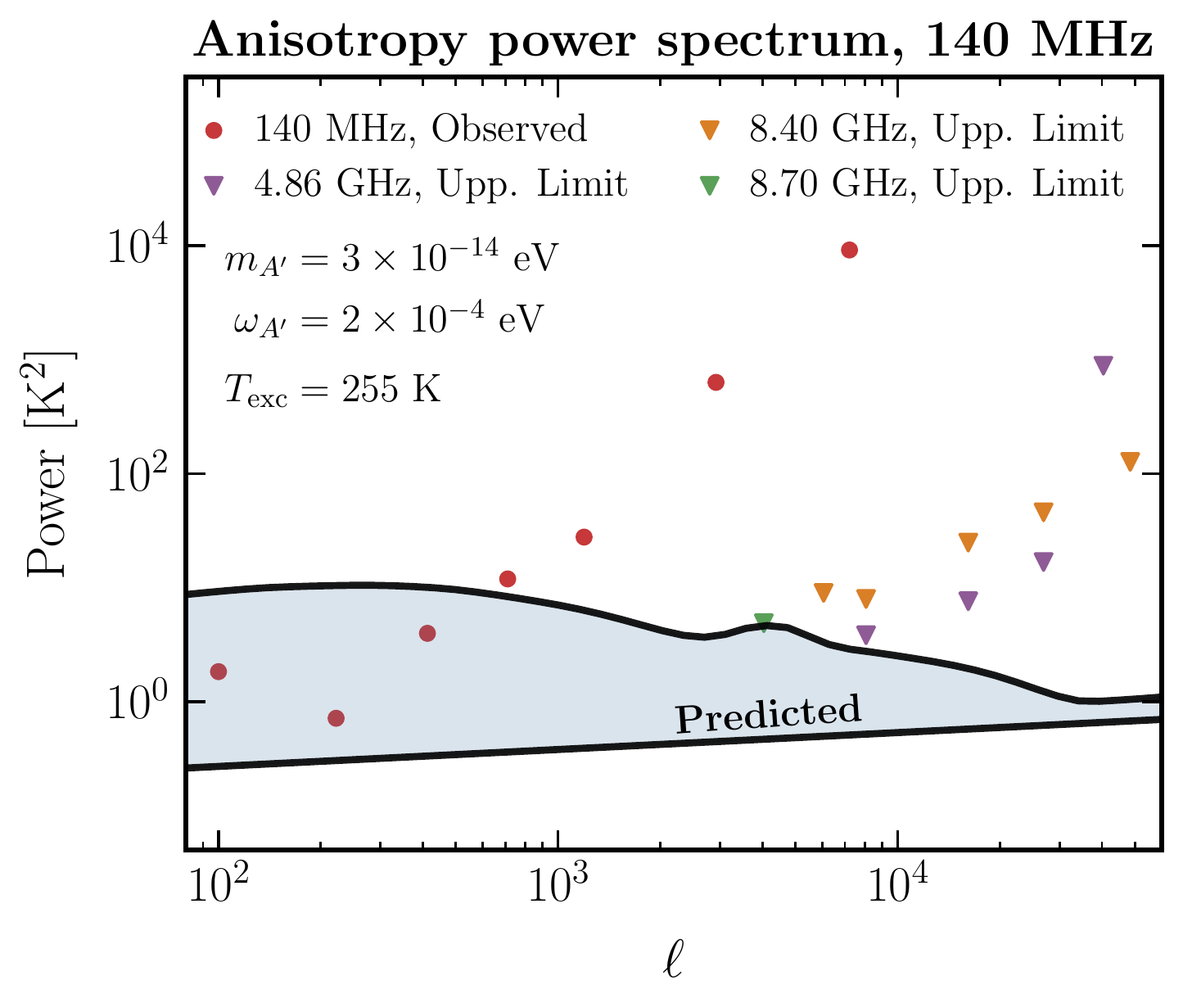}
	\caption{Predicted anisotropy power spectrum with normally (solid black line, below) and log-normally (solid black line, above) distributed baryon fluctuations. The model parameters are $m_{A'} = \SI{3e-14}{\eV}$, and $\omega_{A'} = \SI{2e-4}{\eV}$. Upper limits from VLA at \SI{4.86}{\giga\hertz}~\cite{1988AJ.....96.1187F} (purple triangles), \SI{8.4}{\giga\hertz}~\cite{1997ApJ...483...38P} (orange triangles) and ATCA at \SI{8.7}{\giga\hertz}~\cite{Subrahmanyan:2000df} (green triangle) are shown, and have been rescaled to $T_\text{exc} = \SI{255}{\kelvin}$, the expected value at \SI{140}{\mega\hertz}. Representative data points of the power spectrum measured by LOFAR are shown in red~\cite{Offringa:2021rwp}.}
	\label{fig:anisotropy_power_spec_letter}
\end{figure}

Fig.~\ref{fig:anisotropy_power_spec_letter} shows the predicted anisotropy power spectrum in units of \SI{}{\kelvin\squared}, defined using the same conventions as in the CMB power spectrum~\cite{Planck:2018vyg}, for normally and log-normally distributed $\delta_\text{e}$, with the result between these two choices shaded in blue. This can be taken as an estimate for the uncertainty associated with these distributions. This power should be compared to the LOFAR data, as well as the upper limits from VLA and ATCA at their respective frequencies $\nu$, which have been rescaled by $(\SI{140}{\mega\hertz} / \nu)^{2\beta}$, with $\beta = -2.6$; this assumes that $\Delta T / T_\text{exc}$ is independent of frequency. Our calculated anisotropy power spectrum for a fiducial choice of $m_{A'} = \SI{3e-14}{\eV}$ and $\omega_{A'}=\SI{2e-4}{\eV}$ lies below the \SIrange{4}{8}{\giga\hertz} measurements for both choices of the two-point PDF of $\delta_\text{e}$, while only the log-normal two-point PDF exceeds the low-$\ell$ measurements by LOFAR. However, significant scatter exists between adjacent frequency bands in the data for $\ell \lesssim 10^3$~\cite{Offringa:2021rwp}. Producing a sufficiently smooth radio background is thus highly plausible in our model.

\noindent
{\bf Conclusion.---}In this \textit{Letter}, we have provided a new physics explanation for the ERB, which is brighter than expected from known astrophysical sources. Our model consists of a DM candidate which decays into dark photons in the presence of a thermal bath of the latter; the dark photons then resonantly convert to ordinary photons, producing the ERB. The anisotropy of the signal is expected to be relatively smooth, consistent with measurements of the anisotropy of the radio background~\cite{Holder_2013,Offringa:2021rwp}. Future experiments may corroborate the predictions of our model. PIXIE~\cite{Kogut:2011xw} may be sensitive to spectral distortions expected from our model, and is expected to almost fully cover the 95\% region of our posterior distribution in the $\epsilon$-$m_{A'}$ plane. The thermal population of $A'$ may also lead to a value of $N_\text{eff}$ that is detectable in future CMB experiments such as CMB-S4~\cite{Abazajian:2019eic}. The potential reach of both experiments are shown in Fig.~\ref{fig:T_exc_fiducial}.

\noindent
{\bf Acknowledgments.---}We thank J.~Dowell for useful correspondence, and D.~Aloni, A.~Hook, J.~Mu\~{n}oz, N.~Outmezguine, J.~Pradler, M.~Regis, M.~Schmaltz, N.~Sehgal and A.~Urbano for helpful discussions. 
AC is supported by the Foreign Postdoctoral Fellowship Program of the Israel Academy of Sciences and Humanities and also acknowledges support from the Israel Science Foundation (Grant 1302/19), the US-Israeli BSF (Grant 2018236), the German-Israeli GIF (Grant I-2524-303.7) and the European Research Council (ERC) under the EU Horizon 2020 Programme (ERC-CoG-2015-Proposal n. 682676 LDMThExp). 
HL and JTR are supported by NSF grant PHY-1915409.
HL is also supported by the DOE under Award Number DE-SC0007968 and the Simons Foundation. 
JTR is also supported by the NSF CAREER grant PHY-1554858.  
MP is supported in part by U.S. Department of Energy Grant No. DE-SC0011842. 
This work is supported by the National Science Foundation under Cooperative Agreement PHY-2019786 (The NSF AI Institute for Artificial Intelligence and Fundamental Interactions, \url{http://iaifi.org/}).
This work is partly supported by the U.S. Department of Energy, Office of Science, Office of High Energy Physics of U.S. Department of Energy under grant Contract Number DE-SC0012567.
This work was performed in part at the Aspen Center for Physics, which is supported by NSF grant PHY-1607611.
This work made use of the NYU IT High Performance Computing resources, services, and staff expertise. This work was also performed using the Princeton Research Computing resources at Princeton University which is a consortium of groups including the Princeton Institute for Computational Science and Engineering and the Princeton University Office of Information Technology's Research Computing department. 
This research has made use of NASA's Astrophysics Data System. This research made use of the {ArviZ}~\cite{arviz_2019}, \texttt{astropy}~\cite{Price-Whelan:2018hus,Robitaille:2013mpa}, CLASS~\cite{Blas:2011rf}, \texttt{dynesty}~\cite{Speagle_2020}, \texttt{HyRec}~\cite{AliHaimoud:2010dx}, \texttt{IPython}~\cite{PER-GRA:2007}, Jupyter~\cite{Kluyver2016JupyterN}, \texttt{matplotlib}~\cite{Hunter:2007}, \texttt{NumPy}~\cite{harris2020array}, \texttt{pyfftlog}~\cite{TALMAN197835,Hamilton:1999uv,dieter_werthmuller_2020_3830476}, \texttt{seaborn}~\cite{seaborn}, \texttt{pandas}~\cite{pandas:2010}, \texttt{SciPy}~\cite{2020SciPy-NMeth}, and \texttt{tqdm}~\cite{da2019tqdm} software packages.

\bibliography{low-radio}

\clearpage
\input{supp_input2.tex}

\end{document}

%% file: definitions.tex
\definecolor{linkcolor}{rgb}{0.7752941176470588, 0.22078431372549023, 0.2262745098039215}

\newcommand{\nbicon}{{\color{linkcolor}\faFileCodeO}\xspace}
\newcommand{\nblink}[1]{\href{https://github.com/smsharma/edges-endpoints-21cm/notebooks/#1.ipynb}{\nbicon}}

\newcommand{\dd}{\mathrm{d}}

\definecolor{deepgreen}{rgb}{0.2,0.8,0.2}

%% file: supp_input2.tex
\onecolumngrid
\appendix
\makeatletter

\begin{center}
\vspace{0.05in}
{\bf \large Appendix}\\ 
\vspace{0.05in}
\end{center}

\newcommand\ptwiddle[1]{\mathord{\mathop{#1}\limits^{\scriptscriptstyle(\sim)}}}

This appendix is organized as follows. In App.~\ref{Sec:SpectrumDerivation} we derive in detail the dark photon spectrum as a function of frequency and redshift in the class of models considered in the \textit{Letter}. App.~\ref{Sec:FiducialModel} gives an extended description of a particular model which presents all the ingredients necessary to explain the extragalactic radio background (ERB). App.~\ref{Sec:ExptConstraints} presents some of the important experimental limits on our model, including a derivation of the cosmological limits on the dark matter lifetime in the presence of stimulated decay, as well as limits on the dark sector temperature due to the presence of extra relativistic degrees of freedom. App.~\ref{Sec:Anisotropy} presents the calculation of the anisotropy of the ERB due to photon-dark photon resonant conversions. In App.~\ref{Sec:DataAnalysis} we describe our data analysis procedure, before presenting in App.~\ref{Sec:ExtendedResults} our extended results, including a complete corner plot of our parameter space, and other systematics checks such as the use of different datasets and fits using the ARCADE 2 data points only. Finally, in App.~\ref{Sec:Synchrotron} we give a brief summary of the spectrum of radiation expected from synchrotron sources. 

\section{Dark photon spectrum}\label{Sec:SpectrumDerivation}

In this section, we provide a detailed derivation of the dark photon spectrum as a function of frequency and redshift, in the presence of stimulated emission. 

Consider a two-body decay of a heavy particle $a$ into two daughter particles, at least one of which is a dark photon $A'$, emitted relativistically with energy $\omega_{A'}$. In the presence of a blackbody distribution of $A'$ with temperature $T'$, $a$ undergoes stimulated decay, with a decay width $\Gamma(z)$ that is related to the usual vacuum decay rate $\Gamma_\text{vac}$ via
\begin{alignat*}{1}
    \Gamma(z) = (1 + n f_{A'}^\text{BB}) \Gamma_\text{vac} \,,
\end{alignat*}
where $f_{A'}^\text{BB} \equiv [\exp(\omega_{A'}/T') - 1]^{-1}$ is the blackbody occupation number at energy $\omega_{A'}$. $n$ is an integer that depends on the details of the model; $n = 2$ in the fiducial particle physics model that we describe below. Note that for $\omega_{A'} \ll T'$, we obtain $\Gamma(z) \simeq (n T' / \omega_{A'}) \Gamma_\text{vac}$, an approximation that we will use frequently to obtain parametric estimates. 

Given $\Gamma(z')$, the number density of $A'$ produced per unit time at redshift $z'$ is
\begin{alignat*}{1}
    \frac{\dd n_{A'}}{\dd t} = \alpha \Gamma(z') \frac{\rho_a(z')}{m_a} \,, 
\end{alignat*}
where $\alpha = 2$ if both daughter particles are $A'$ and $\alpha = 1$ otherwise, $\rho_a(z)$ is the energy density of $a$ at redshift $z$, and $m_a$ is the mass of $a$. With this, we can find the contribution to the dark photon number density at redshift $z$ per unit frequency $\omega$, due to a decay occuring at redshift $z'$: 
\begin{alignat*}{1}
    \frac{\dd n_{A'}(z)}{\dd \omega \, dz'} = \frac{\dd n_{A'}(z')}{\dd t} \frac{\dd t}{\dd z'} \frac{(1+z)^3}{(1+z')^3} \delta_\text{D} \left(\omega \frac{1+z'}{1+z} - \omega_{A'}\right) \,,
\end{alignat*}
where the redshift factors rescale the number density of particles produced at redshift $z'$ down to redshift $z$, and the Dirac delta function enforces the fact that the decay produces dark photons with energy $\omega_{A'}$. Integrating over $z'$ gives 
\begin{alignat}{1}
    \frac{\dd n_{A'}}{\dd \omega} &= \int_z^\infty \dd z' \, \frac{\rho_a(z')}{m_a} \frac{\alpha \Gamma(z')}{H(z)(1+z)} \frac{(1+z)^3}{(1+z')^3} \delta_\text{D} \left(\omega \frac{1+z'}{1+z} - \omega_{A'} \right) = \frac{\alpha \rho_a(z)}{m_a} \frac{\Gamma(z_\star)}{\omega H(z_\star)} \Theta \left(z_\star - z \right) \,,
    \label{eq:dnAp_domega}
\end{alignat}
where
\begin{alignat}{1}
    1+z_\star \equiv \frac{\omega_{A'}}{\omega}(1+z)
\end{alignat}
is the redshift at which the decay of $a$ produced the dark photon at frequency $\omega$, with $\omega \leq \omega_{A'}$. We have also assumed that $\rho_a \propto (1+z)^3$. This expression agrees with Ref.~\cite{Pospelov:2018kdh}, if we simply set $\Gamma(z_\star) \to \Gamma_\text{vac}$. In terms of comoving number density and $x \equiv \omega / T_0$ where $T_0$ is the CMB temperature today, we find
\begin{alignat}{1}
    \frac{1}{(1+z)^3} \frac{\dd n_{A'}}{\dd x} = \frac{\alpha \rho_{a,0}}{m_a} \frac{\Gamma(z_\star)}{x H(z_\star)} \Theta(z_\star - z) \,,
\end{alignat}
which is almost constant in redshift, aside from the step function. $\rho_{a,0}$ is the dark matter energy density today. Eq.~\eqref{eq:dnAp_domega} is the expression we use in the \textit{Letter}. 

\section{Particle physics model}\label{Sec:FiducialModel}

\subsection{Model building considerations}
\label{sec:model_building_considerations}

As we explained in the \textit{Letter}, the particle physics model we need to explain the radio background requires three features that we repeat here:
\begin{enumerate}
    \item  A cold component of DM $a$ which is stable on cosmological timescales, but undergoes a two-body decay with lifetime $\tau_\text{vac}$ in vacuum. For simplicity, we take $a$ to be all of DM;  
    \item One of the daughter particles is a dark photon $A'$, which has an existing blackbody distribution of $A'$ characterized by a temperature $T'(z) = T_0'(1+z)$, where $T_0'$ is its temperature today; and   
	\item  $A'$ has a mass $\SI{e-15}{\eV}\lesssim m_{A'} \lesssim \SI{e-9}{\eV}$, and is emitted relativistically with energy $\omega_{A'}$. $A'$ mixes with the Standard Model (SM) photon $\gamma$ with kinetic mixing parameter $\epsilon$.
\end{enumerate}
These conditions appear simple to meet; for example, the model proposed in Ref.~\cite{Pospelov:2018kdh} where the DM is a dark axion decaying into two identical dark photons which oscillates into SM photons, could be supplemented by an additional thermal population of dark photons. However, there are several challenges to building a successful model. In order to produce the right power law in the ERB, the $A'$ power-law spectrum must maintain $\dd n_{A'}/\dd \omega \propto \omega^{-1/2}$ between $3 \times 10^{-4} \lesssim x \lesssim 0.2$, where $x \equiv \omega / T_\text{CMB}$, from the redshift of production $z_\star(x)$ until resonant conversions $A' \to \gamma$ are mostly complete. Furthermore, the $A'$ blackbody spectrum also cannot be significantly distorted; specifically, $A'$ particles from the bath at each value of $x$ must be described by a blackbody spectrum at $z_\star(x)$. In the model of Ref.~\cite{Pospelov:2018kdh}, both of these requirements can be violated by inverse decays $A' A' \to a$ or other $A'$ scattering processes. Moreover, it is difficult to maintain thermal equilibrium in the blackbody spectrum through scattering with another particle in the dark sector bath, since such scattering processes likely distort the power-law spectrum significantly. For typical parameters required to produce the full $T_\text{exc}$, scattering between a low-energy $A'$ in the power-law spectrum and a blackbody $A'$ is too rapid to guarantee that the $A'$ blackbody spectrum remains undistorted at all relevant times. While it is possible that even with significant distortion to the blackbody spectrum, a reasonable fit to the ERB can still result, checking this possibility would require us to integrate the Boltzmann equation over a wide range of $A'$ frequencies. In this paper, we avoid this computational challenge by building a slightly different fiducial model without significant distortion. 

Before discussing the details of our fiducial model, we will first discuss how to check the size of any $A'$ spectral distortion due to a single process. We begin by writing down the Boltzmann equation with the relevant process contributing to the collision term. Neglecting for simplicity the existence of entropy dumps in the Standard Model so that $x$ for any $A'$ particle is a constant, the Boltzmann equation governing the occupation number $f_{A'}$ can be written
\begin{alignat}{1}
    \frac{\dd f_{A'}(x,t)}{\dd t} = \frac{C[f_{A'}]}{\omega} \,,
    \label{eq:boltzmann_equation}
\end{alignat}
where $C[f_{A'}]$ is the collision integral, which we will define for particular processes below. Here, we adopt the convention that $f_{A'}$ is related to the number density via
\begin{alignat*}{1}
    n_{A'} = 3 \int \frac{\dd ^3 \vec{p}}{(2\pi)^3} f_{A'} \,,
\end{alignat*}
where the factor of 3 accounts for the degeneracy in the spin state of $A'$. To determine if any process causes a significant distortion to the spectrum of blackbody $A'$ or low-energy $A'$ from the decay of $a$, we divide Eq.~\eqref{eq:boltzmann_equation} by $f_{A'}$ and integrate with respect to time or, equivalently, redshift, to obtain the change in $\log f_{A'}$, which gives a measure of the fraction of $A'$ at that value of $x$ that undergoes a scattering process. Our criterion for a small distortion to the $A'$ spectra is therefore
\begin{alignat}{1}
    |\Delta \log f_{A'}(x)| = \int_{z_{\min}}^{z_{\max}} \frac{\dd z}{H(z) (1+z)} \frac{|C[f_{A'}]|}{\omega f_{A'}(x,z)} \ll 1 \,,
    \label{eq:distortion_condition}
\end{alignat}
where $z_{\min}$ and $z_{\max}$ are the lowest and highest redshifts respectively for which the collision term is important.

Since we are only interested in producing the radio excess over a finite frequency range, we only need to check that distortions are small over a limited range of $x$. The radio frequency data points with an excess temperature over the CMB temperature span the frequency range \SI{22}{\mega\hertz}--\SI{10}{\giga\hertz}, which we can cover by considering $3 \times 10^{-4} < x < 0.2$. 

The power-law component at fixed $x$ is produced by decays of $a$ at a redshift $1+z_\star \equiv \omega_{A'} / (x T_0)$, where $\omega_{A'}$ is the energy of the emitted $A'$ at the point of decay (for $a$ decaying into two massless particles, this is simply $\omega_{A'} = m_a/2$), and $T_0$ is the blackbody CMB temperature today. After it is produced, it must stay in this power law without undergoing significant distortions until the present day.\footnote{Technically, the dark photons can become significantly distorted after all resonant oscillations into photons have been completed, but for simplicity we set this more stringent requirement.} For the blackbody component to effectively produce the stimulated emission that we need, on the other hand, the blackbody spectrum must not be significantly distorted prior to redshift $z_\star(x)$; distortions after this redshift are unimportant. This means that for the power-law spectrum, for each value of $x$, we need to consider $z_\text{min} = 0$ and $z_\text{max} = z_\star(x)$, while for the blackbody spectrum, $z_\text{min} = z_\star(x)$.

\subsection{Fiducial model}

We will now describe our fiducial particle physics model, which has all the three properties required to produce the radio background laid out in the previous section. We will show that both the low-energy and blackbody distribution of the dark photons do not undergo any significant distortion. 

Our fiducial model is a modified version of the model proposed in Ref.~\cite{Pospelov:2018kdh}. The important particles in this model are: \textit{i)} an axion-like dark matter $a$ with mass $m_a$ and decay constant $f_a$, and \textit{ii)} two dark photons, which we label $A_\psi$ and $A_\text{osc}$. The three requirements for producing the radio excess as discussed in the main body of the paper are satisfied as follows:  
\begin{enumerate}
    \item the DM decays through the process $a \to A_\psi A_\text{osc}$; 
    
        \item a blackbody distribution of $A_\text{osc}$ described by a temperature $T'$ is also present, leading to stimulated decay of $a$, and
    
    \item the dark photon $A_\text{osc}$ possesses a small kinetic mixing term with the SM photon, and has a mass $\SI{e-15}{\eV} \lesssim m_\text{osc} \lesssim \SI{e-9}{\eV}$, so that it undergoes resonant conversions into the SM photon after recombination.

\end{enumerate}
The terms in the dark sector Lagrangian that are relevant to us are therefore
\begin{alignat}{1}
    \mathcal{L} \supset \frac{1}{2}(\partial_\mu a)^2 - \frac{1}{2} m_a^2 a^2 - \frac{a}{4 f_a} F_{\psi,\mu \nu} \tilde{F}_\text{osc}^{\mu \nu} - \frac{\epsilon}{2} F^{\mu\nu}_\text{osc} F_{\mu\nu} + \mathcal{L}_{K} \,,
\end{alignat}
where $\mathcal{L}_{K}$ contains the other kinetic terms of the dark photons, including a mass of $m_\text{osc}$ for $A_\text{osc}$. Note that a discrete symmetry under which $A_\psi \to - A_\psi$ and $a \to -a$ explicitly forbids decays of $a$ to a pair of dark photons of the same species, as well as mixing between $A_\psi$ and the SM photon, naturally leading to the Lagrangian shown above. By allowing $a$ to decay into two different dark photons, we can now introduce a fermion $\psi$ that scatters rapidly with $A_\psi$, to ensure that $A_\psi$ always equilibrates into a thermal distribution with temperature $T'$, without destroying the low-energy spectrum of $A_\mathrm{osc}$.\footnote{For simplicity, we assume that the entire dark sector is described by a common temperature $T'$.} With this modification, we overcome the main difficulty faced by the model in Ref.~\cite{Pospelov:2018kdh}: there are now no low-energy $A_\psi$ to interact with the blackbody distribution of $A_\mathrm{osc}$, which was the main source of distortion for the blackbody spectrum.\footnote{While this completes the basic description of our model, there are some requirements on $\psi$ to keep this model tractable and avoid tracking the spectra of all of these particles as a function of time. First, Compton $\psi A_\psi \to \psi A_\psi$ and double-Compton $\psi A_\psi A_\psi \leftrightarrow \psi A_\psi$ scattering must be sufficiently rapid to ensure that $A_\psi$ is always described by a simple, blackbody distribution. This is easily satisfied, as long as $\psi$ has a large coupling to $A_\psi$, and is sufficiently abundant. Second, the process $a A_\mathrm{osc} \leftrightarrow \psi \overline{\psi}$ must not lead to significant distortion of $A_\mathrm{osc}$. This is hard to determine without tracking the full evolution of the spectrum of $A_\mathrm{osc}$, but it is possible to avoid this entirely by choosing a sufficiently massive $\psi$ to kinematically forbid $a A_\mathrm{osc} \to \psi \overline{\psi}$, and making $\psi$ asymmetric. We find that $m_\psi \sim \SI{30}{\eV}$ with a coupling to $A_\psi$ of $\alpha_D = 1$, with an asymmetric number density today of $n_{\psi,0} = \SI{2.5e-3}{\per\centi\meter\cubed}$ can provide sufficiently efficient scattering with $A_\psi$. This value of $m_\psi$ is large enough to kinematically forbid $a A_\mathrm{osc} \to \psi \overline{\psi}$. Introducing a similar fermion $\chi$ with $m_\chi \sim m_\psi$ that couples to $A_\mathrm{osc}$ instead guarantees that no distortion can occur when $T' > m_\psi, m_\chi$, while allowing distortions to build up once $T' < m_\psi, m_\chi$ and $\psi, \chi$ has frozen out. This also prevents $\psi \overline{\psi} \to a A_\mathrm{osc}$ from being significant at $T' < m_\psi, m_\chi$, since $\overline{\psi}$ annihilates away at $T' \sim m_\psi$.}
 
We are now ready to check that distortions to both the low-energy $A_\text{osc}$ power-law spectrum and the $A_\text{osc}$ blackbody  are small. The most important process, which enters at order $1/f_a^2$, is inverse decays, $A_\psi A_\text{osc} \to a$, which leads to the largest distortion. The collision integral is given by
\begin{alignat*}{1}
    C_{A_\text{osc} A_\psi \to a} [f_\text{osc}] = - \frac{f_\text{osc}}{6} \int \frac{\dd^3 \vec{k}_{A_\psi}}{(2\pi)^3 2 \omega_{A_\psi}} \int \frac{\dd^3 \vec{p}_a}{(2\pi)^3 2 E_a} (2\pi)^4 \delta_\text{D}^4 (k_\text{osc} + k_{A_\psi} - p_a ) |\mathcal{M}|^2_{A_\text{osc} A_\psi \to a} f_{A_\psi} (\vec{k}_{A_\psi}) \,.
\end{alignat*}
We use $k$, $\vec{k}$ and $\omega$ for incoming four-momentum, three-momentum, and energy; likewise, we have $p$, $\vec{p}$ and $E$ for outgoing states. The subscript `osc' represents quantities associated with $A_\text{osc}$. There is no Bose enhancement in the outgoing state, since we can treat the entire population of $a$ as having zero momentum. This also allows us to neglect the contribution from the backward reaction to the collision integral. The squared matrix element for this process (summed over initial and final states) is
\begin{alignat*}{1}
    |\mathcal{M}|^2_{A_\text{osc} A_\psi \to a} = \frac{m_a^4}{2 f_a^2} \,,
\end{alignat*}
while the occupation number of $A_\psi$ is given by the blackbody occupation number, $f_{A_\psi} = [\exp(\omega_{A_\psi}/T') - 1]^{-1}$. These simple expressions allow us to perform the collision integral relatively easily, to obtain
\begin{alignat}{1}
    C_{A_\text{osc} A_\psi \to a} [f_\text{osc}] = - \frac{f_\text{osc} m_a^4 T'}{96 \pi \omega_\text{osc} f_a^2} \log \left[1 - \exp \left(- x \frac{(1+z_\star)^2}{(1+z)^2}\right)\right] \,,
    \label{eq:inverse_decay_collision_full}
\end{alignat}
where $\omega_\text{osc} = x T_\text{CMB}$. With this expression, we can now assess the total distortion to the blackbody and power-law distributions of $A_\text{osc}$, and check that for typical model parameters, these distortions are small, based on the criterion set out in Eq.~\eqref{eq:distortion_condition}. For the blackbody distribution, we want to perform the integral in Eq.~\eqref{eq:distortion_condition} starting from $z_{\min} = z_\star$, allowing us to approximate the collision term as 
\begin{alignat}{1}
    C_{A_\text{osc} A_\psi \to a} [f_\text{osc}] = - \frac{f_\text{osc} m_a^4 T'}{96 \pi \omega_\text{osc} f_a^2} \log \left( x \frac{(1+z_\star)^2}{(1+z)^2} \right) \,.
    \label{eq:inverse_decay_collision_approx}
\end{alignat}
Integrating this gives
\begin{alignat}{1}
    |\Delta \log f_\text{osc}| &\sim \frac{2^{5/2} m_a^{3/2} \gamma x^{1/2} T_0^{3/2}}{96 \pi H_0 \sqrt{\Omega_m} f_a^2} \sim 1.8 \times 10^{-4} \bigg( \frac{\gamma}{0.2} \bigg) \left(\frac{\SI{2e-4}{\eV}}{m_a}\right)^{3/2} \left(\frac{\SI{e21}{\second}}{\tau_\text{vac}}\right) \bigg(\frac{x}{0.2}\bigg)^{1/2} \,,
\end{alignat}
where $\gamma \equiv T_0' / T_0$, a redshift invariant quantity. This indicates that the total distortion to the blackbody distribution is small. 

For the low-energy spectrum, we integrate Eq.~\eqref{eq:distortion_condition} from $z_{\min} = 0$ and $z_{\max} = z_\star$. The integral can be approximately done in two parts: \textit{i)} $0 < 1+z < (1+z_\star)/\sqrt{x}$, where we can expand the expression in Eq.~\eqref{eq:inverse_decay_collision_full} using $\log(1-\alpha) \simeq -\alpha$, and \textit{ii)} $(1+z_\star)/\sqrt{x} < 1+z < 1+z_\star$, where we can use the approximation in Eq.~\eqref{eq:inverse_decay_collision_approx}. The second part of the integral dominates, but the total contribution to the distortion is given numerically by 
\begin{alignat}{1}
    |\Delta \log f_\text{osc}| \sim 0.21 \bigg( \frac{\gamma}{0.2} \bigg) \left(\frac{\SI{2e-4}{\eV}}{m_a}\right)^{3/2} \left(\frac{\SI{e21}{\second}}{\tau_\text{vac}}\right) \left(\frac{3 \times 10^{-4}}{x}\right)^{3/4} \,.
\end{alignat}
This distortion appears to be somewhat large for the fiducial values shown here, but the relative uncertainty on the data point at $x = 3.8 \times 10^{-4}$ or \SI{22}{\mega\hertz} is $\sim$25\%, and so is enough to absorb the distortion obtained here. Moreover, the posterior distribution from our fits include larger values of $m_a$ and $\tau_\text{vac}$, which decreases the size of the distortion. We therefore conclude that inverse decays do not significantly distort either component of the $A_\text{osc}$ spectrum.\footnote{Although elastic scattering processes should be subleading at order $1/f_a^4$, they can, in fact, be very rapid for two reasons. First, processes like $a A_\text{osc} \to a A_\text{osc}$ receive a large Bose enhancement from the existing population of $A_\text{osc}$, which significantly increases the scattering rate. Secondly, the matrix element of $a A_\text{osc} \to a A_\text{osc}$ contains a divergence. Such divergences are commonplace even in Standard Model processes, such as $e^- Z \to e^- Z$~\cite{Grzadkowski:2021kgi}. In the context of cosmological fluids, these divergences are regulated by thermal corrections to the self-energy of $A_\psi$, which impart an imaginary contribution to the mass of $A_\psi$~\cite{Grzadkowski:2021kgi}. We have checked that even including Bose enhancement and regulating this divergence with loop contributions from a thermal distribution of $\psi$~\cite{Redondo:2008ec}, all elastic scattering processes are indeed subdominant to the leading inverse decay process discussed here.}

Our proposed model therefore satisfies the requirement that all distortions to the $A'_\text{osc}$ power law and blackbody spectrum are small, and is a viable model for explaining the ERB\@. We stress that many of the required conditions are invoked in order to simplify the analysis and to allow an unambiguous prediction of the ERB in this model. Some of the conditions on the distortions, for example, may be relaxed under a complete analysis, which would include numerically solving the Boltzmann equations for each mode. 

\section{Experimental constraints}\label{Sec:ExptConstraints}

\subsection{Dark matter lifetime constraint with stimulated decay}\label{Sec:DecayBound}

Cosmological constraints on the decay lifetime of dark matter have been obtained for decays without stimulated emission~\cite{Poulin:2016nat,Nygaard:2020sow,DES:2020mpv,Mau:2022sbf,Simon:2022ftd,Alvi:2022aam}. These results show that if a subcomponent $f_\text{dcdm}$ decays between recombination and today, \emph{Planck} 2018 and baryon acoustic oscillation (BAO) data limit $f_\text{dcdm} < 0.0216$~\cite{Simon:2022ftd} at the 95\% confidence level. This result is consistent with the uncertainty in the dark matter energy density reported by \emph{Planck} 2018~\cite{Planck:2018vyg}. From this, we can deduce a limit on the decay lifetime with stimulated emission, by adopting a limit of $f_\text{lim} = 0.0216$ for the proportion of dark matter that decays away by the present day. For a stimulated emission lifetime given by $\tau(z) = \tau_\text{vec}(1 + 2 f_{A'}^\text{BB})^{-1}$, which is the expression we obtain with our fiducial model, we require\footnote{This corresponds to an approximation of the effect of stimulation, we leave a more complete treatment for future work.}.
\begin{alignat*}{1}
    \int_0^t \frac{\dd t}{\tau_\text{vac}} \left(1 + \frac{2}{e^{\omega_{A'}/T'} - 1}\right) < f_\text{lim} \,.
\end{alignat*}

Numerically, we find 
\begin{alignat*}{1}
    \int_0^t \frac{\dd t}{\tau_\text{vac}} \left(1 + \frac{2}{e^{\omega_{A'}/T'} - 1}\right) \approx \frac{1}{\tau_\text{vac}} \times \begin{dcases}
        \SI{1.5e18}{\second} \left( \frac{2 T'_0}{\omega_{A'}} \right) \,, & \frac{\omega_{A'}}{2 T'_0} < 3.2 \,, \\
        H_0^{-1} \,, & \frac{\omega_{A'}}{2 T'_0} \geq 3.2 \,,
    \end{dcases}
\end{alignat*}
which leads to a limit on the vacuum decay lifetime of
\begin{alignat}{1}
    \tau_\text{vac} > \SI{e19}{\second} \times \max \left[2.1 \,, \,\, 6.3 \left(\frac{\SI{e-4}{\eV}}{\omega_{A'}}\right) \bigg( \frac{\gamma}{0.2} \bigg) \right] \,.
\end{alignat}
This limit is roughly an order of magnitude stronger than the lifetime limits on DM decays without stimulated emission for our fiducial choice of parameters~\cite{Simon:2022ftd}.  

\subsection{Constraints on relativistic degrees of freedom}
\label{Sec:Neff_constraint}

We complete our discussion of the particle physics model by discussing its effect on $N_\text{eff}$, and relevant constraints. $A_\text{osc}$, $A_\psi$ contribute to $N_\text{eff}$ as relativistic degrees of freedom. In addition, other fermions that couple to these bosons can contribute as well if they are sufficiently light. Assuming the existence of two such Dirac fermions on top of $A_\text{osc}$ and $A_\psi$ before recombination, the total energy density of these particles divided by the energy density of a neutrino is
\begin{alignat}{1}
    \Delta N_\text{eff} \approx \frac{2 \times 3 + 2 \times 4 \times \frac{7}{8}}{2 \times \frac{7}{8}}  \frac{\gamma^4}{\left(\frac{4}{11}\right)^{4/3}} = 0.04 \left(\frac{\gamma}{0.2}\right)^4 \,.
\end{alignat}
This is to be compared with the \emph{Planck} measurement of $N_\text{eff} = 2.99 \pm 0.17$~\cite{Planck:2018vyg}, which limits $\Delta N_\text{eff} < 0.34$ at the 95\% confidence level. Including the irreducible contribution from the thermal distribution of just one dark photon, which must be present in the class of models considered in this work, gives $\Delta N_\text{eff} \approx 0.01 (\gamma / 0.2)^4$.

\section{Anisotropy due to dark photon conversions}\label{Sec:Anisotropy}

In this appendix, we obtain the conversion anisotropy power spectrum, which originates from variations in electron density fluctuations along two lines-of-sight separated in the sky by some angle. Our discussion follows Ref.~\cite{Fornengo:2013rga} closely, but with some novel and peculiar features originating from the resonance conversion process; for clarity, we repeat many of the same calculations presented in that reference.

For dark photon conversions, the observed temperature in any direction in the sky $\hat{n}$ at a fixed frequency is directly proportional to the total probability of conversion $P(\hat{n})$ of dark photons into photons. We begin by defining the fluctuation of the conversion probability in some particular direction in the sky $\hat{n}$, defined explicitly by $\delta P(\hat{n}) \equiv P(\hat{n}) - \langle P \rangle$, where $\langle P \rangle$ is the sky-averaged conversion probability. In any particular direction $\hat{n}$, we have~\cite{Caputo:2020bdy,Caputo:2020rnx}
\begin{alignat}{1}
    P(\hat{n}) = \frac{\pi \epsilon^2 m_{A'}^4}{\omega_0} \int_0^{z_\star} \dd z \, \frac{\delta_\text{D}(m_\gamma^2(\vec{r}, z) - m_{A'}^2)}{H(z) (1+z)^2} \,,
\end{alignat}
where $m_\gamma^2(\vec{r}, z)$ is the effective plasma mass of photons along the particular line-of-sight, and $\omega_0$ is the present-day, observed angular frequency of the photons. $\vec{r} \equiv \hat{n} \chi(z)$, where $\chi(z)$ is the comoving distance traveled by light between $z$ and the present day. $\omega_0(1+z_\star)$ is the energy of the daughter particle from the DM decay; photons that have energy $\omega_0$ today were produced by decays at $z_\star$, which is the maximum redshift we integrate up to. 

We can compute the sky-averaged conversion probability by integrating over the one-point probability density function (PDF) $f_1(m_\gamma^2; z)$ of $m_\gamma^2$, to obtain~\cite{Caputo:2020bdy,Caputo:2020rnx}
\begin{alignat}{1}
    \langle P \rangle = \frac{\pi \epsilon^2 m_{A'}^4}{\omega_0} \int_0^{z_\star} \dd z  \int \dd m_\gamma^2 \, f_1(m_\gamma^2; z) \frac{\delta_\text{D} (m_\gamma^2 - m_{A'}^2)}{H(z) (1+z)^2} = \frac{\pi \epsilon^2 m_{A'}^4}{\omega_0} \int_0^{z_\star} \dd z \, \frac{f_1(m_\gamma^2 = m_{A'}^2 ; z)}{H(z) (1+z)^2} \,,
\end{alignat}
so that
\begin{alignat}{1}
    \delta P(\hat{n}) = \frac{\pi \epsilon^2 m_{A'}^4}{\omega_0} \int_0^{z_\star} \frac{\dd z}{H(z) (1+z)^2} \left[\delta_\text{D} (m_\gamma^2(\vec{r}, z) - m_{A'}^2) - f_1(m_\gamma^2 = m_{A'}^2 ; z)\right] \,.
\end{alignat}

We now decompose the observed conversion probability over the whole sky into spherical harmonics $Y_{\ell m}(\hat{n})$, with coefficients $a_{\ell m}$ given by
\begin{alignat*}{1}
    a_{\ell m} = \frac{1}{\langle P \rangle} \int \dd \hat{n} \, \delta P(\hat{n}) Y_{\ell m}^*(\hat{n}) \,.
\end{alignat*}
The anisotropy power spectrum $C_\ell$ for the conversion probability, defined as $C_\ell \equiv \langle a_{\ell m}^* a_{\ell m} \rangle$, can then be computed as
\begin{alignat*}{1}
    C_\ell = \frac{1}{\langle P \rangle^2}\int \dd \hat{n} \int \dd \hat{n}' \langle \delta P(\hat{n}) \delta P(\hat{n}') \rangle Y_{\ell m}^*(\hat{n}) Y_{\ell m} (\hat{n}') \,,
\end{alignat*}
where $\langle \cdots \rangle$ in the integral should be interpreted as an all-sky average. Note that defined in the following manner, $C_\ell$ is dimensionless, since it describes fluctuations in conversion probability. 

To make further progress, we define the quantity
\begin{alignat}{1}
    Q(\vec{r}, z; \vec{r}', z') \equiv \left< \left[\delta_\text{D} (m_\gamma^2(\vec{r}, z) - m_{A'}^2) - f_1(m_\gamma^2 = m_{A'}^2 ; z)\right] \left[\delta_\text{D} (m_\gamma^2(\vec{r}', z') - m_{A'}^2) - f_1(m_\gamma^2 = m_{A'}^2 ; z')\right] \right> \,,
    \label{eq:correlation_function_for_conversion_anisotropy}
\end{alignat}
a two-point correlation function along two different lines-of-sight, described by comoving coordinates $\vec{r}$ and $\vec{r}'$, with photons passing through each point at redshifts $z$ and $z'$ respectively. In fact, homogeneity and isotropy guarantee that this function does not depend on $\vec{r}$ and $\vec{r}'$ separately, but only on $|\vec{r} - \vec{r}'|$. Inserting this into the expression above for $C_\ell$, we find
\begin{alignat}{1}
    C_\ell =  \frac{1}{\langle P \rangle^2}\left[ \frac{\pi \epsilon^2 m_{A'}^4}{\omega_0} \right]^2 \int_0^{z_\star} \dd z \, W(z) \int_0^{z_\star} \dd z' \, W(z') \int \dd \hat{n} \int \dd \hat{n}' \, Q(|\vec{r} - \vec{r}'|, z, z') Y_{\ell m}^*(\hat{n}) Y_{\ell m} (\hat{n}') \,,
    \label{eq:Cl_intermediate}
\end{alignat}
where $W(z) \equiv [H(z) (1+z)^2]^{-1}$. We can now write $Q$ in terms of its Fourier transform over $\vec{r} - \vec{r}'$, $\tilde{Q}$, giving: 
\begin{alignat*}{1}
    Q(|\vec{r} - \vec{r}'|, z, z') &= \int \frac{\dd^3 \vec{k}}{(2\pi)^3} e^{i \vec{k} \cdot (\vec{r} - \vec{r}')}  \tilde{Q}(k, z, z') \\
    &= (4\pi)^2 \sum_{p,s=0}^\infty \sum_{q=-p}^p \sum_{t=-s}^s \int \frac{\dd^3 \vec{k}}{(2\pi)^3}  \tilde{Q}(\vec{k}, z, z') i^{p-s} j_p(k r) j_s (k r') Y_{pq}^*(\hat{k}) Y_{pq}(\hat{n}) Y_{st}(\hat{k}) Y_{st}^* (\hat{n}') \\
    &= \frac{2}{\pi} \sum_{p=0}^\infty \sum_{q=-p}^p \int \dd k \, k^2 \tilde{Q}(k, z, z') j_p(kr) j_p(kr') Y_{pq}(\hat{n}) Y_{pq}^*(\hat{n}') \,.
\end{alignat*}
In the second line, we have used the Rayleigh expansion for plane waves, and $j_p$ is the spherical Bessel function of order $p$. In the last line, we use the orthogonality of spherical harmonics to integrate over solid angle. Substituting this expression into Eq.~\eqref{eq:Cl_intermediate} and integrating over $\hat{n}$ and $\hat{n}'$, once again exploiting the orthogonality of spherical harmonics, gives
\begin{alignat*}{1}
    C_\ell = \frac{1}{\langle P \rangle^2} \left[ \frac{\pi \epsilon^2 m_{A'}^4}{\omega_0} \right]^2 \frac{2}{\pi} \int_0^{z_\star} \dd z \, W(z) \int_0^{z_\star} \dd z' \, W(z') \int \dd k \, k^2 \tilde{Q}(k, z, z') j_\ell(k r) j_\ell(k r') \,.
\end{alignat*}
Finally, we can simplify this integral further assuming the Limber approximation~\cite{Limber, Kaiser:1991qi, Kaiser:1996tp}, which is a high-$\ell$ expansion that works particularly well for the multipoles in which we are interested~\cite{LoVerde:2008re} and allows us to approximate $k^2 j_\ell(k r) j_\ell(k r') \approx (\pi / 2) \delta_\text{D}(k - \ell / r) \delta_\text{D}(r - r') / r^2$.  This finally gives 
\begin{alignat}{1}
    C_\ell = \frac{1}{\langle P \rangle^2} \left[ \frac{\pi \epsilon^2 m_{A'}^4}{\omega_0} \right]^2 \int_0^{z_\star} \frac{\dd z}{r^2(z)} \, W^2(z) H(z) \tilde{Q}(k = \ell / r, z, z) \,.
    \label{eq:C_ell_compact_expression}
\end{alignat}
This compact expression makes the calculation of the conversion anisotropy numerically tractable; in particular, the correlation function $Q$ defined in Eq.~\eqref{eq:correlation_function_for_conversion_anisotropy} need only be evaluated at points along two different lines-of-sight which have equal redshifts.  

We now need an expression for $\tilde{Q}(k = \ell/r, z, z)$. First, we define $f_2(\rho, m_\gamma^2(\vec{r}), m_\gamma^2(\vec{r}'); z)$ as the two-point PDF for $m_\gamma^2$ at two different points, $\vec{r}$ and $\vec{r}'$, at redshift $z$, with $\rho \equiv |\vec{r} - \vec{r}'|$. Note that homogeneity and isotropy guarantee that $f_2$ only depends on $\rho$. With this, the averaging in Eq.~\eqref{eq:correlation_function_for_conversion_anisotropy} is performed by integrating the quantity with respect to $m_\gamma^2(\vec{r})$ and $m_\gamma^2(\vec{r}')$, weighted by $f_2$, which gives
\begin{alignat}{1}
    Q(|\vec{r} - \vec{r}'|, z, z) = f_2(|\vec{r} - \vec{r}'|, m_\gamma^2(\vec{r}) = m_{A'}^2, m_\gamma^2(\vec{r}') = m_{A'}^2; z) - [f_1(m_\gamma^2 = m_{A'}^2 ; z)]^2 \,,
    \label{eq:Q_is_f2_minus_f1_squared}
\end{alignat}
with
\begin{alignat*}{1}
    \tilde{Q}(k, z, z) = 4 \pi \int \dd \rho \, \rho^2 j_0(k \rho)  Q(\rho, z, z)
\end{alignat*}
Given both the one- and two-point PDF of $m_\gamma^2$, we can now proceed to calculate the anisotropy power spectrum.

We demonstrate that the conversion anisotropy signal can satisfy existing experimental bounds by assuming that \textit{1)} the free-electron number density is exactly equal to the baryon number density, so that the one- and two-point PDFs of $m_\gamma^2$ are given by the one- and two-point PDFs of baryonic fluctuations $\delta_\text{b}$ divided by one and two powers of $\overline{m_\gamma^2}(z)$, the mean value of $m_\gamma^2$ at redshift $z$, respectively. This is a good assumption for $z \lesssim 6$, where resonant converions generally take place, since reionization has been completed, and all baryons are ionized; and \textit{2)} the baryonic fluctuations follow analytic one- and two-point PDFs. We consider two analytic forms for the one- and two-points PDFs: normally distributed and log-normally distributed fluctuations, to give an indication of the dependence on the PDF\@. In both cases, the PDFs are fully specified by the correlation function $\xi_\text{b}(|\vec{r} - \vec{r}'|;z)$, which is related to the power spectrum of baryons $P_\text{b}$ via
\begin{alignat*}{1}
    \xi_\text{b}(\rho; z) = \int \frac{\dd^3 \vec{k}}{(2\pi)^3} j_0(k \rho) P_\text{b} (k; z) \,,
\end{alignat*}
with $\sigma_\text{b}^2(z) = \xi_\text{b}(0; z)$ being the variance of baryonic fluctuations. For normally distributed fluctuations, 
\begin{alignat}{1}
    f_1(m_\gamma^2 = m_{A'}^2 ; z) &= \frac{1}{\overline{m_\gamma^2}(z)\sqrt{2 \pi \sigma_\text{b}^2(z)}} \exp \left[- \frac{g(m_{A'}^2; z)^2}{2 \sigma_\text{b}^2(z)}\right] \,, \nonumber \\
    f_2(\rho, m_\gamma^2(\vec{r}) = m_{A'}^2, m_\gamma^2(\vec{r}') = m_{A'}^2; z) &= \frac{1}{2 \pi \overline{m_\gamma^2}^2(z) \sqrt{\sigma_\text{b}^4(z) - \xi_\text{b}^2(\rho; z)}} \exp \left(- \frac{g(m_{A'}^2; z)^2}{\sigma_\text{b}^2(z) + \xi_\text{b}(\rho; z)} \right) \,,
    \label{eq:f_1_and_f_2_gaussian}
\end{alignat}
where $g(m_{A'}^2; z) \equiv m_{A'}^2 / \overline{m_\gamma^2}(z) - 1$, while for log-normally distributed fluctuations~\cite{Kayo:2001gu,Caputo:2020rnx}, 
\begin{alignat}{1}
    f_1(m_\gamma^2 = m_{A'}^2 ; z) &= \frac{1}{\overline{m_\gamma^2}(z)\sqrt{2 \pi \Sigma^2(z)}} \exp \left[- \frac{L^2(m_{A'}^2; z)}{2 \Sigma^2(z)}\right] \frac{1}{1 + g(m_{A'}^2; z)} \,, \nonumber \\
    f_2(\rho, m_\gamma^2(\vec{r}) = m_{A'}^2, m_\gamma^2(\vec{r}') = m_{A'}^2; z) &= \frac{1}{2 \pi \overline{m_\gamma^2}^2(z) \sqrt{\Sigma^4(z) - X^2(\rho; z)}} \exp \left(- \frac{L^2(m_{A'}^2; z)}{\Sigma^2(z) + X(\rho; z)} \right) \frac{1}{[1 + g(m_{A'}^2; z)]^2} \,,
    \label{eq:f_1_and_f_2_lognormal}
\end{alignat}
where $\Sigma^2(z) = \log[1 + \sigma_\text{b}^2(z)]$, $X(\rho; z) = \log [1 + \xi_\text{b}(\rho; z)]$, and $L(m_{A'}^2 ; z) = \log[1 + g(m_{A'}^2 ; z)] + \Sigma^2(z) / 2$. 

In the limit that $\rho \to \infty$, we obtain $\xi_\text{b} \to 0$ and $f_2 \to f_1^2$ for both distributions, indicating that the two-point PDF factorizes into the product of two, independent one-point PDFs at sufficiently large separations, when correlations are unimportant. 

Eqs.~\eqref{eq:Q_is_f2_minus_f1_squared} and~\eqref{eq:C_ell_compact_expression} give us the following general expression for $C_\ell$, which is the main result of this section:
\begin{alignat}{1}
    C_\ell = \frac{1}{\langle P \rangle^2} \int_0^{z_\star} \frac{\dd z}{r^2 (z)} H(z) \left[\frac{\dd \langle P \rangle}{\dd z}\right]^2 \int \dd \rho \, 4\pi \rho^2 j_0 (\ell \rho / r) \left[ \frac{f_2(\rho, m_\gamma^2(\vec{r}) = m_{A'}^2, m_\gamma^2(\vec{r}') = m_{A'}^2; z)}{f_1^2(m_\gamma^2 = m_{A'}^2 ; z)} - 1 \right] \,,
    \label{eq:C_ell_final_expression}
\end{alignat}
an expression that we can evaluate numerically for either choice of baryon fluctuation PDFs. Note that the integrand over $\rho$ is finite as $\rho \to 0$. Following the convention used in CMB anisotropy power spectrum analyses as well as Ref.~\cite{Offringa:2021rwp}, the temperature anisotropy power spectrum is defined to be $\ell (\ell + 1) C_\ell \langle T \rangle^2$, where $\langle T \rangle$ is the sky-averaged brightness temperature. 

For small angular scales ($\ell \gtrsim 3000$), upper limits on the anisotropy of the ERB have been obtained by the Very Large Array (VLA) at \SI{4.86}{\giga\hertz}~\cite{1988AJ.....96.1187F} and \SI{8.4}{\giga\hertz}~\cite{1997ApJ...483...38P}, as well as the Australia Telescope Compact Array (ATCA) at \SI{8.7}{\giga\hertz}~\cite{Subrahmanyan:2000df}. These measurements were made with the intention of looking for CMB anisotropies at the arcminute scales, and all set an approximate upper limit of $\Delta T / T_\text{exc} \lesssim 10^{-2}$ for the ERB~\cite{Holder_2013}. More recently, LOFAR and TGSS observations have been used to measure the anisotropy power spectrum at $\sim$\SI{140}{\mega\hertz}~\cite{Choudhuri:2020dgd,Offringa:2021rwp}. This latest result confirms the fact that a currently unknown population of dim but numerous synchrotron sources is required to explain the ERB; they also find that these sources must exhibit some nontrivial clustering to produce the right power spectrum. 

\begin{figure}
\centering
\includegraphics[width=0.5\textwidth]{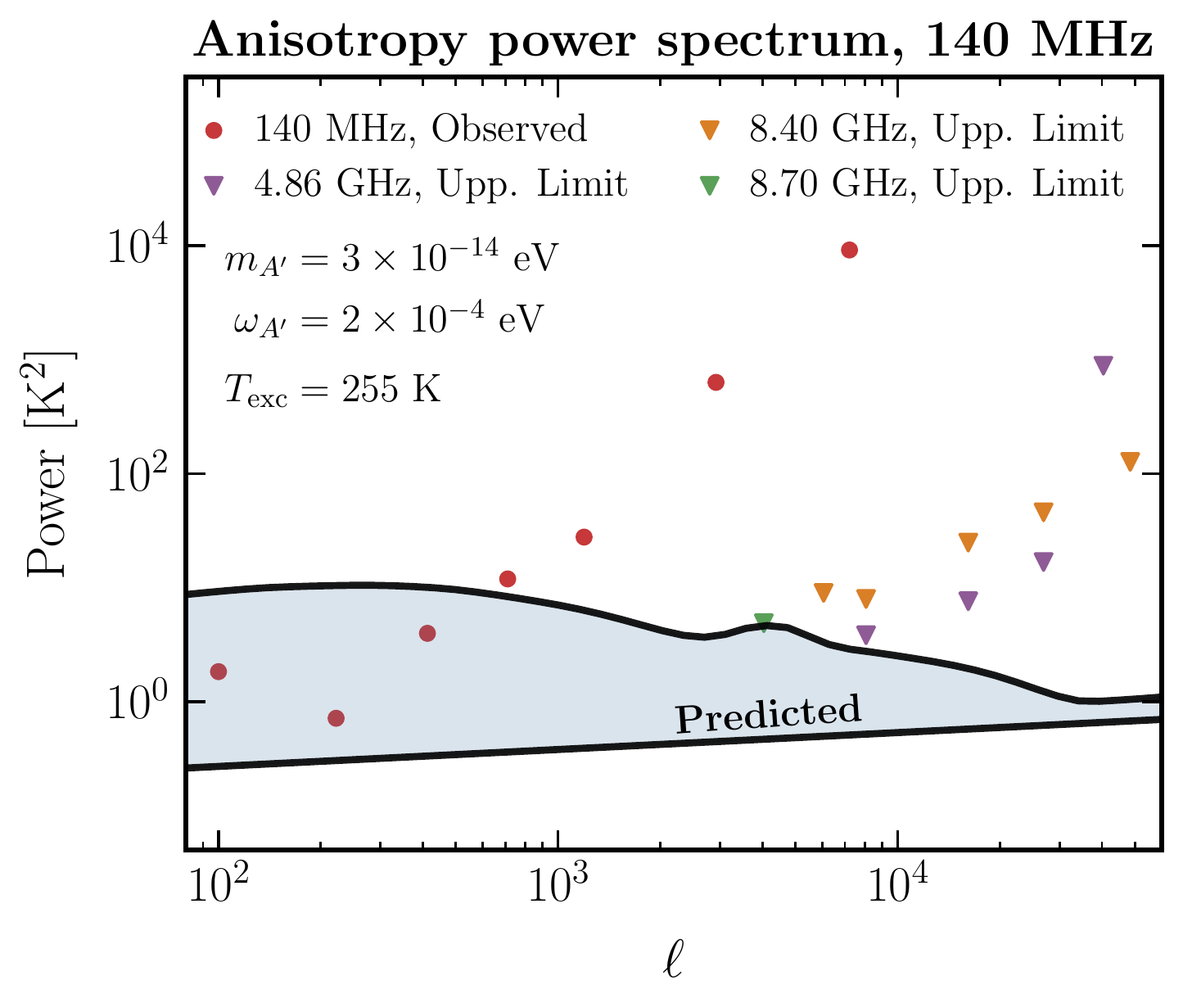}
\caption{Predicted anisotropy power spectrum with normally (solid black line, below) and log-normally (solid black line, above) distributed baryon fluctuations. The model parameters are $m_{A'} = \SI{3e-14}{\eV}$ and $\omega_{A'} = \SI{2e-4}{\eV}$. Upper limits from VLA at \SI{4.86}{\giga\hertz}~\cite{1988AJ.....96.1187F} (purple triangles), \SI{8.4}{\giga\hertz}~\cite{1997ApJ...483...38P} (orange triangles) and ATCA at \SI{8.7}{\giga\hertz}~\cite{Subrahmanyan:2000df} (green triangle) are shown, and have been rescaled to $T_\text{exc} = \SI{255}{\kelvin}$, the expected value at \SI{140}{\mega\hertz}. Representative data points of the power spectrum measured by LOFAR are shown in red~\cite{Offringa:2021rwp}.}
\label{fig:anisotropy_power_spec}
\end{figure}

Fig.~\ref{fig:anisotropy_power_spec} shows the predicted anisotropy power spectrum, $[\ell(\ell + 1) / (2 \pi)] C_\ell \langle T \rangle^2$, as a function of $\ell$, obtained by integrating Eq.~\eqref{eq:C_ell_final_expression} numerically for both normally (lower line) and log-normally (upper line) distributed baryon fluctuations.\footnote{The theoretical prediction from our model only has a weak dependence on the observational frequency, coming from the integration limit $z_\star$; we have only included the largest fluctuations here.} The range between the two lines, shaded in blue, gives some indication of the uncertainty of our prediction. Representative values of $m_{A'} = \SI{3e-14}{\eV}$ and $\omega_{A'} = \SI{2e-4}{\eV}$ have been chosen, but the results are qualitatively similar for other values of these parameters. This result should be compared with existing measurements of the anisotropy power, including \textit{1)} observed upper limits of the CMB anisotropy power spectrum at small scales between \SIrange{4}{9}{\giga\hertz} by the VLA~\cite{1988AJ.....96.1187F,1997ApJ...483...38P} and the ATCA, subsequently reinterpreted in Ref.~\cite{Holder_2013} as an upper limit on the anisotropy power of the excess radio power above the CMB and known point sources, and \textit{2)} the LOFAR anisotropy power spectrum of the radio background at \SI{140}{\mega\hertz}~\cite{Offringa:2021rwp}. We take $\langle T \rangle = \SI{255}{\kelvin}$, the expected radio temperature at \SI{140}{\mega\hertz} based on the power-law fit in Eq.~(1) of the \textit{Letter}. The anisotropy power reported by the high frequency measurements at frequency $\nu_\text{high}$ has been rescaled by $(\SI{140}{\mega\hertz} / \nu_\text{high})^{2\beta}$, where $\beta = 2.6$; this assumes that the relative size of fluctuations $\delta T / T$ is independent of frequency, and can be rescaled by $\langle T \rangle^2$ at each frequency. Note that the observed upper limits in the \SIrange{4}{9}{\giga\hertz} range are not necessarily in tension with the $\ell \gtrsim 10^3$ results at \SI{140}{\mega\hertz}, since the anisotropy can in fact be frequency dependent. 

The experimental results for the anisotropy power spectrum shown in Fig.~\ref{fig:anisotropy_power_spec} cover $10^2 \lesssim \ell \lesssim 6 \times 10^4$, finding a power of \SIrange{1}{e2}{\kelvin\squared} across this range; this represents a temperature fluctuation of $0.004 \lesssim \Delta T / T_\text{exc} \lesssim 0.04$, a result that seems unusually smooth if radio emission were correlated with large scale structure at low redshifts~\cite{Holder_2013}. Our predicted conversion anisotropy power spectrum for the fiducial values shown here lies mostly below the measured anisotropy power, and is therefore consistent with these experimental results, except for $\ell \lesssim 800$ at \SI{140}{\mega\hertz}. However, as mentioned in the \textit{Letter}, lower-$\ell$ measurements of the anisotropy are difficult, and within the LOFAR dataset, significant scatter can be observed in the anisotropy power between adjacent frequency bands, which are subsequently combined to produce the final result shown here~\cite{Offringa:2021rwp}. We have thus demonstrated that our model can in principle produce a sufficiently smooth conversion anisotropy power spectrum; more detailed studies involving more realistic one- and two-point PDFs, as well as potentially a more careful analysis of the \SIrange{4}{9}{\giga\hertz} radio data may be of interest in future work.

\section{Data analysis and statistical methodology}\label{Sec:DataAnalysis}

\subsection{Likelihood and posterior sampling}

We assume the likelihood of the observed brightness temperatures $T_{\mathrm{obs},i}(\nu_i)$ given a model for the emission $T(\nu;\theta)$ to be Gaussian, where $i$ indexes individual data points and $\theta = \{m_a, m_{A'}, \tau_\mathrm{vac}, \epsilon, T_0', T_0\}$ are the parameters of interest describing the fiducial new-physics model. Assuming uncorrelated measurements, the total likelihood for a dataset is given by $p\left(T_\mathrm{obs}\mid\theta\right) = \prod_i \mathcal N\left(T_{\mathrm{obs},i}(\nu_i)\mid T(\nu_i;\theta) ,\sigma_{T, i}\right)$, where $\sigma_{T, i}$ are the corresponding standard deviation uncertainties describing the noise model. 

Given a prior $p(\theta)$ as defined in Sec.~\ref{sec:prior} below, we infer an approximation for the posterior distribution $p\left(\theta\mid T_\mathrm{obs}\right) = p\left(T_\mathrm{obs}\mid\theta\right)p(\theta)/\mathcal Z$, where $\mathcal Z$ is the marginal evidence $\mathcal Z = \int \dd \theta\,p(T_\mathrm{obs}\mid\theta)\,p(\theta)$. This is done through Monte Carlo nested sampling implemented in \texttt{dynesty}~\cite{Speagle_2020}. $1000$ live points are used to model the posterior and sampling is performed until the estimated expected contribution to the log-evidence is less than $\Delta\log\mathcal Z=0.05$, with an otherwise default configuration of the static nested sampler. 

\subsection{Constrained prior definition}
\label{sec:prior}

A baseline prior is assumed such that $\log_{10} m_a\sim \mathcal U(-7, -1)$, $\log_{10} m_{A'}\sim \mathcal U(-17, -10)$, $T_0'/T_0\sim\mathcal U(0, 0.4)$, $\log_{10}\tau_\text{vac}\sim\mathcal U(16, 25)$, $\log_{10} \epsilon\sim \mathcal U(-11, -4)$, and $T_0\sim\mathcal N(2.7255, 0.00086)$, with the prior on $T_0$ motivated by an analysis of COBE/FIRAS CMB data in the $2.27$--$21.33$\,cm$^{-1}$ frequency range~\cite{Fixsen:1996nj,Fixsen:2009ug}, and the prior on $T_0' / T_0$ motivated by the constraints derived in Sec.~\ref{Sec:Neff_constraint}. Furthermore, for our fiducial analysis we wish to incorporate external constraints on the considered parameter space as mentioned in the main body of this  \emph{Letter}. Practical implementations of nested sampling require that the prior distribution be provided as a transformation from the unit hypercube to the target prior density. Although such a transformation is readily available for commonly used prior specifications \emph{e.g.}, uniform- or Gaussian-distributed, transformations from an arbitrary prior such as that respecting the constraints laid out in the main \emph{Letter} through combinations of multiple parameters are not always readily available or analytically tractable.

In order to construct an implicit prior for our purposes we follow a procedure similar to that outlined in Ref.~\cite{2021arXiv210212478A} and use normalizing flows~\cite{pmlr-v37-rezende15,papamakarios2019normalizing}, which use a series of bijections parameterized by neural networks with tractable Jacobians in order to define transformations from a simple base distribution to complex, expressive target distributions. Specifically, we generate a large number $10^6$ of samples from the baseline prior, remove the samples incompatible with the external constraints, and use a normalizing flow consisting of 8 neural spline flows~\cite{NEURIPS2019_7ac71d43} to learn a transformation from a standard Gaussian $\mathcal N(0, \mathbbm{1})$ into the constrained target prior density. Samples $u$ from the unit hypercube proposal can then be transformed to those corresponding to a standard Gaussian through an affine transformation $u\rightarrow 0 + \mathbbm{1}\cdot F^{-1}_\mathcal{N}(u)$, where $F^{-1}_\mathcal{N}$ is the inverse cumulative distribution function of the Gaussian distribution. These samples can then be further transformed to those on the constrained prior density by passing them through the learned normalizing flow. 

\subsection{Prior-predictive check}

Generally in Bayesian analyses and in particular when non-trivial priors are utilized such as in this work, it is useful to conduct prior-predictive checks in order to instill confidence that the prior does not bias the analysis towards ``favorable'' regions of the parameter space. In that case, formally low-likelihood parameter points could still correspond to high posterior density regions.

\begin{figure*}
    \centering
    \includegraphics[width=0.6\textwidth]{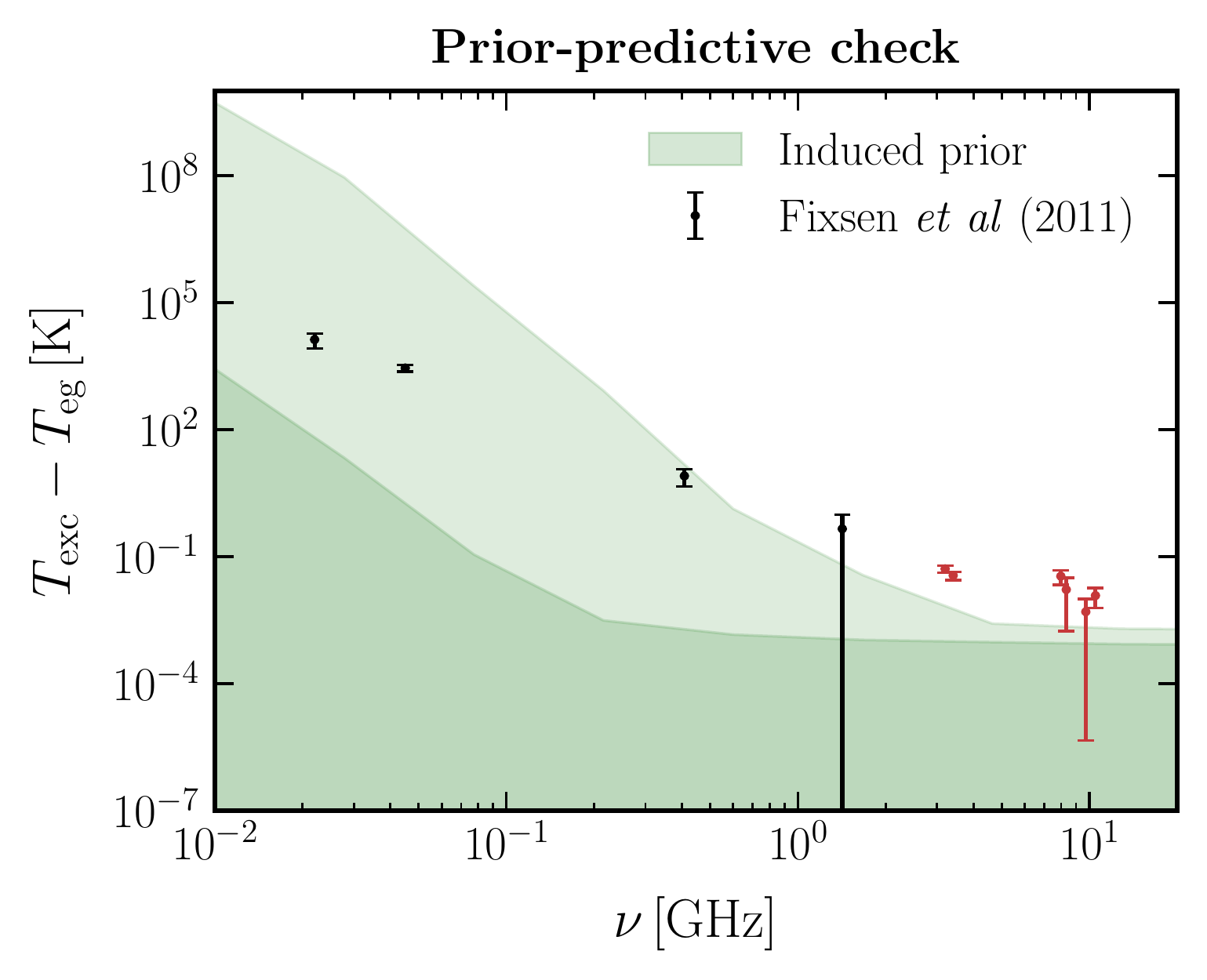}
    \caption{68\% (dark green) and 95\% (light green) containment of the induced prior on the excess temperature above the modeled extragalactic background, $T_\text{eg} = \SI{0.23}{\kelvin} (\nu / \SI{}{\giga\hertz})^{-2.7}$~\cite{Gervasi:2008rr}. Measured data points from Ref.~\cite{Fixsen:2009xn} are shown for comparison. The induced prior covers a large range of possible excess temperatures and does not bias the analysis in favour of the observed data points.}
    \label{fig:pp_check}
\end{figure*}

Figure~\ref{fig:pp_check} shows the 68 and 95\% contours of the prior induced on the excess temperature above the modeled extragalactic background, defined as $T_\text{eg} = \SI{0.23}{\kelvin} (\nu / \SI{}{\giga\hertz})^{-2.7}$~\cite{Gervasi:2008rr,Holder_2013}. It can be seen that the induced prior covers a large range of possible excess temperatures and does not bias the current analysis in favour of the observed data points (shown in black and red).

\section{Extended results}\label{Sec:ExtendedResults}

\subsection{Parameter posteriors}

\begin{figure*}
    \centering
    \includegraphics[width=0.9\textwidth]{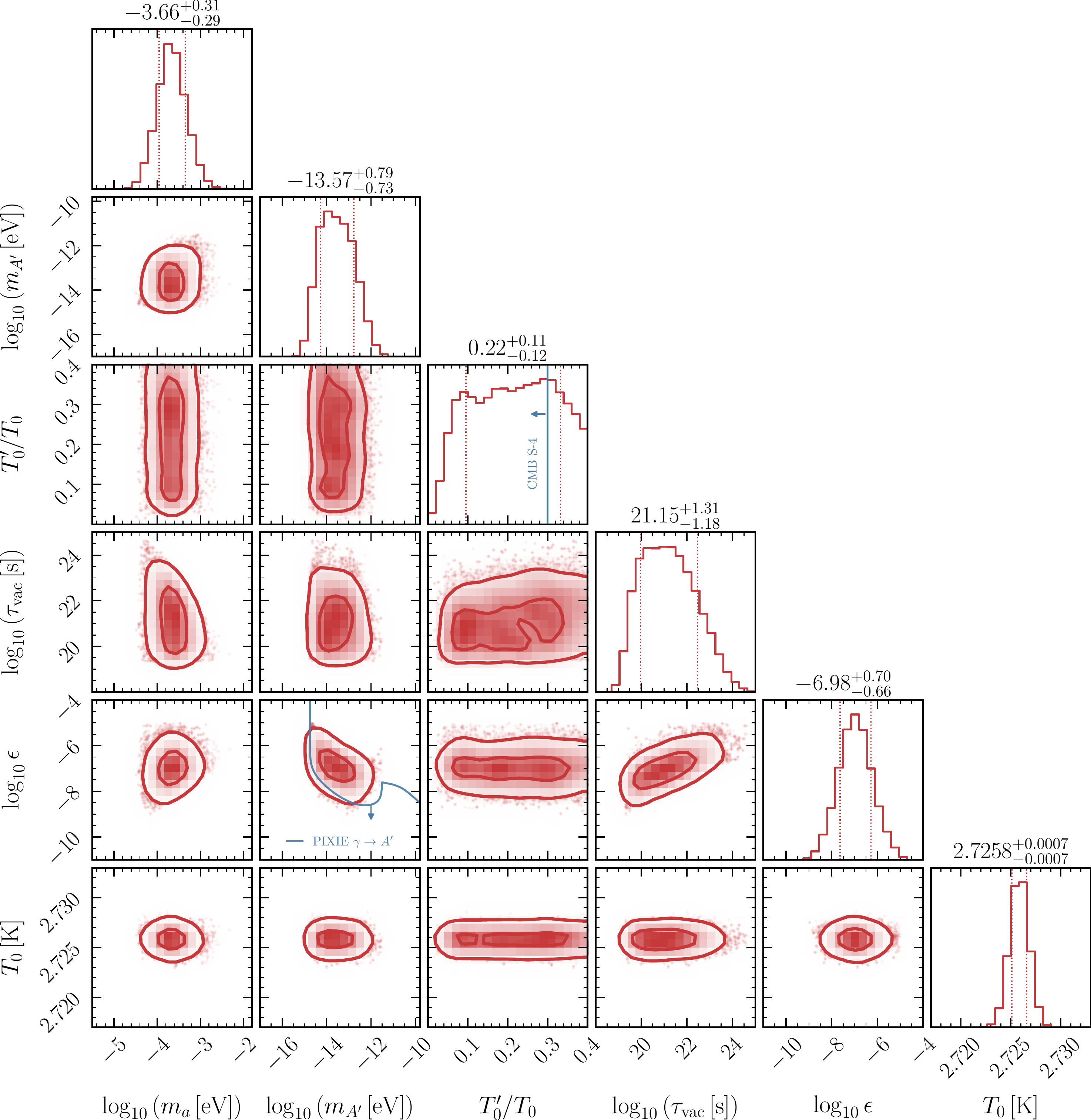}
    \caption{Joint and individual marginal posterior obtained in the fiducial model, same format as Fig.~\ref{fig:T_exc_fiducial}.}
    \label{fig:corner_fiducial}
\end{figure*}

Figure~\ref{fig:corner_fiducial} shows the individual and pair-wise marginal posteriors on all the modeled parameters obtained for the baseline analysis.

\subsection{Systematic variations on the analysis}

In this section we consider various systematic variations of our baseline analysis. In particular we consider: 
\begin{enumerate}
    \item results obtained using the data from Ref.~\cite{Dowell:2018mdb} instead of those from Ref.~\cite{Fixsen:2009xn}. Ref.~\cite{Dowell:2018mdb} includes recent data points from the LWA1 Low Frequency Sky Survey (LLFSS), which spans a frequency range of \SIrange{40}{80}{\mega\hertz}. These new data points are consistent with the power-law fit in Ref.~\cite{Fixsen:2009xn}. Ref.~\cite{Dowell:2018mdb} also independently reanalyzed the data from ARCADE 2 and the other experiments used in Ref.~\cite{Fixsen:2009xn}, finding broad agreement, but with small changes to the central values and more significant changes to the error bars, despite using a similar data analysis method. Ref.~\cite{Dowell:2018mdb} therefore provides a useful systematics check on how the ERB spectrum is extracted; 
    
    \item results obtained by fitting only the ARCADE 2 data $ \gtrsim 3$\,GHz and ignoring the lower-frequency radio measurements. In this case, we neglect stimulated emission, since this is not necessary to produce a good fit, simplifying the model and reducing to that first introduced in Ref.~\cite{Pospelov:2018kdh}. ARCADE 2 is the only experiment in our list that is designed primarily with absolute zero-level calibration in mind, in order to produce an accurate determination of the sky-brightness~\cite{Singal:2017jlh}, which warrants a separate investigation; and
    
    \item results obtained without the modeled contribution from unresolved extragalactic radio sources~\cite{Gervasi:2008rr,Holder_2013}. This should concretely demonstrate that our model does not rely on unresolved extragalactic radio sources for a good fit, which can be anticipated from the fact that these sources are at least 3 times less bright than the measured $T_\text{exc}$. 
\end{enumerate}
    
\begin{figure*}
    \centering
    \includegraphics[width=0.46\textwidth]{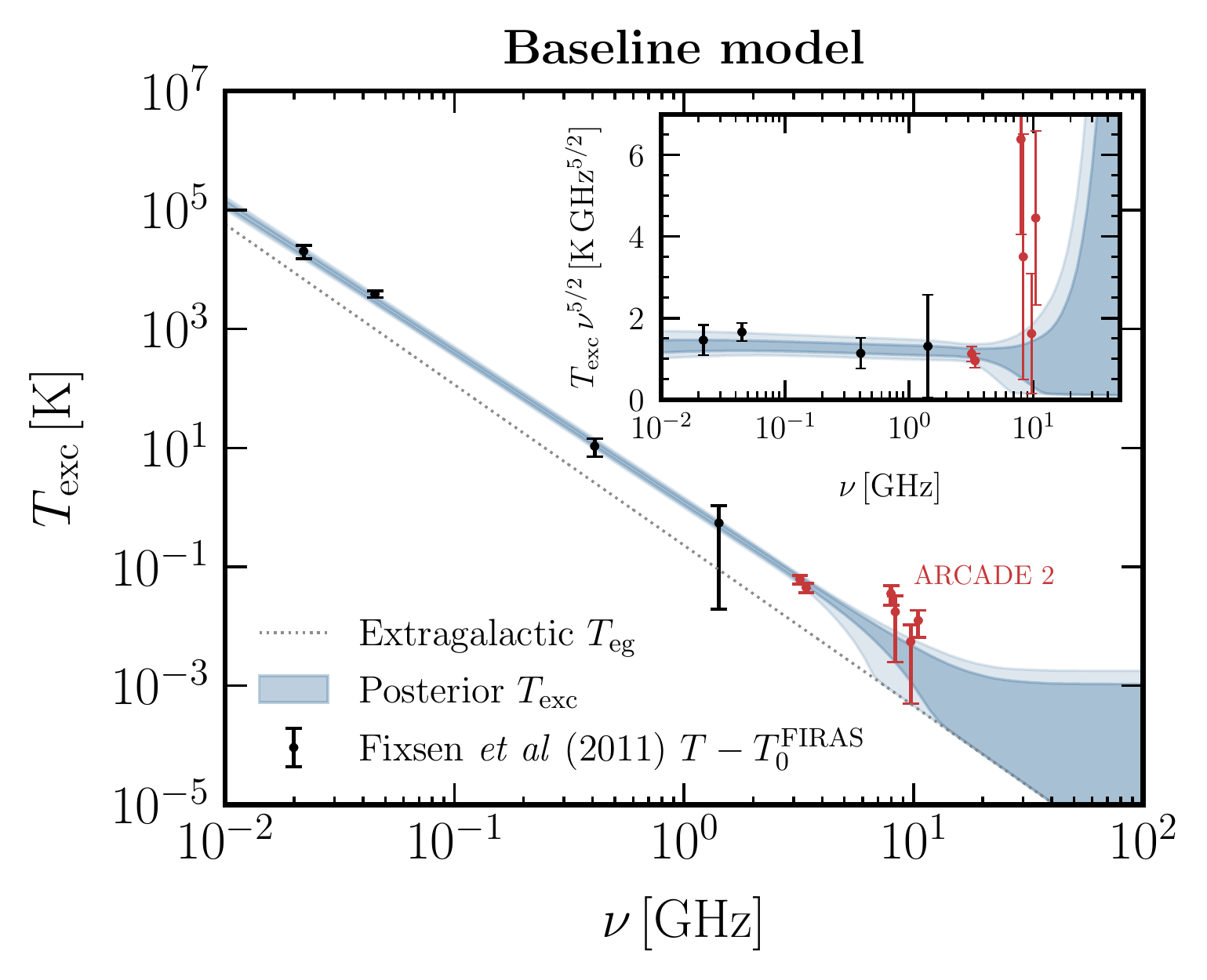}
    \includegraphics[width=0.46\textwidth]{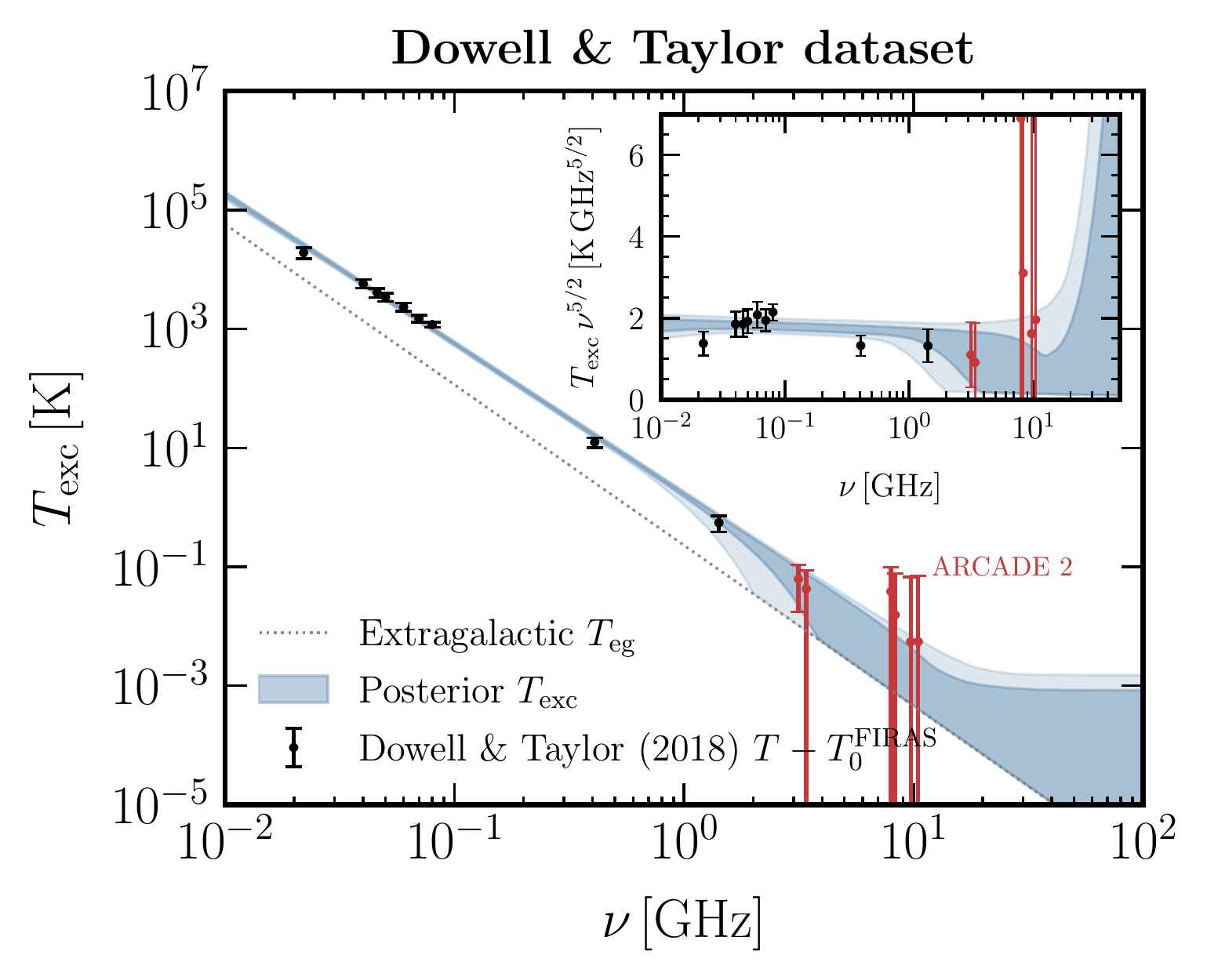}
    \includegraphics[width=0.46\textwidth]{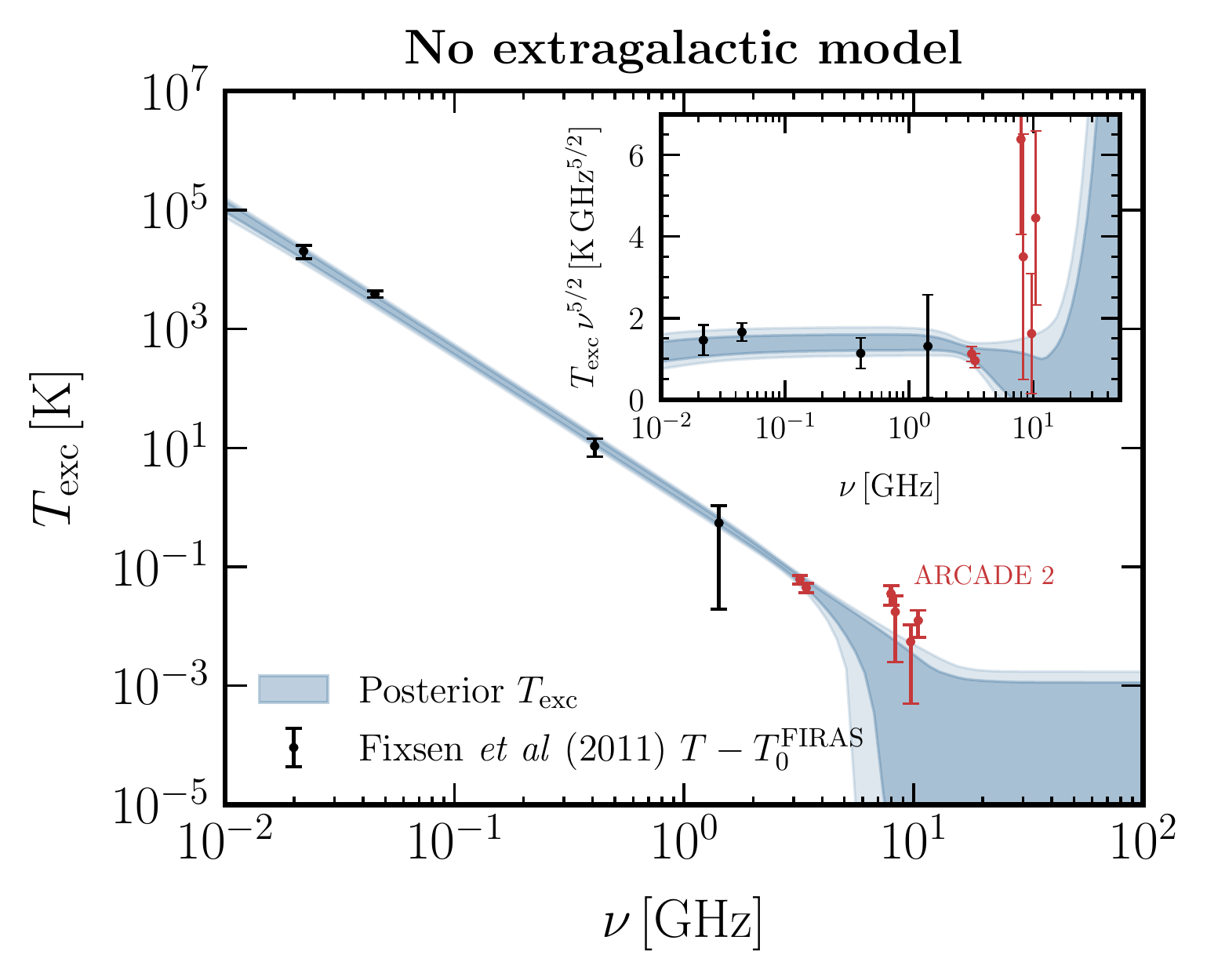}
    \includegraphics[width=0.46\textwidth]{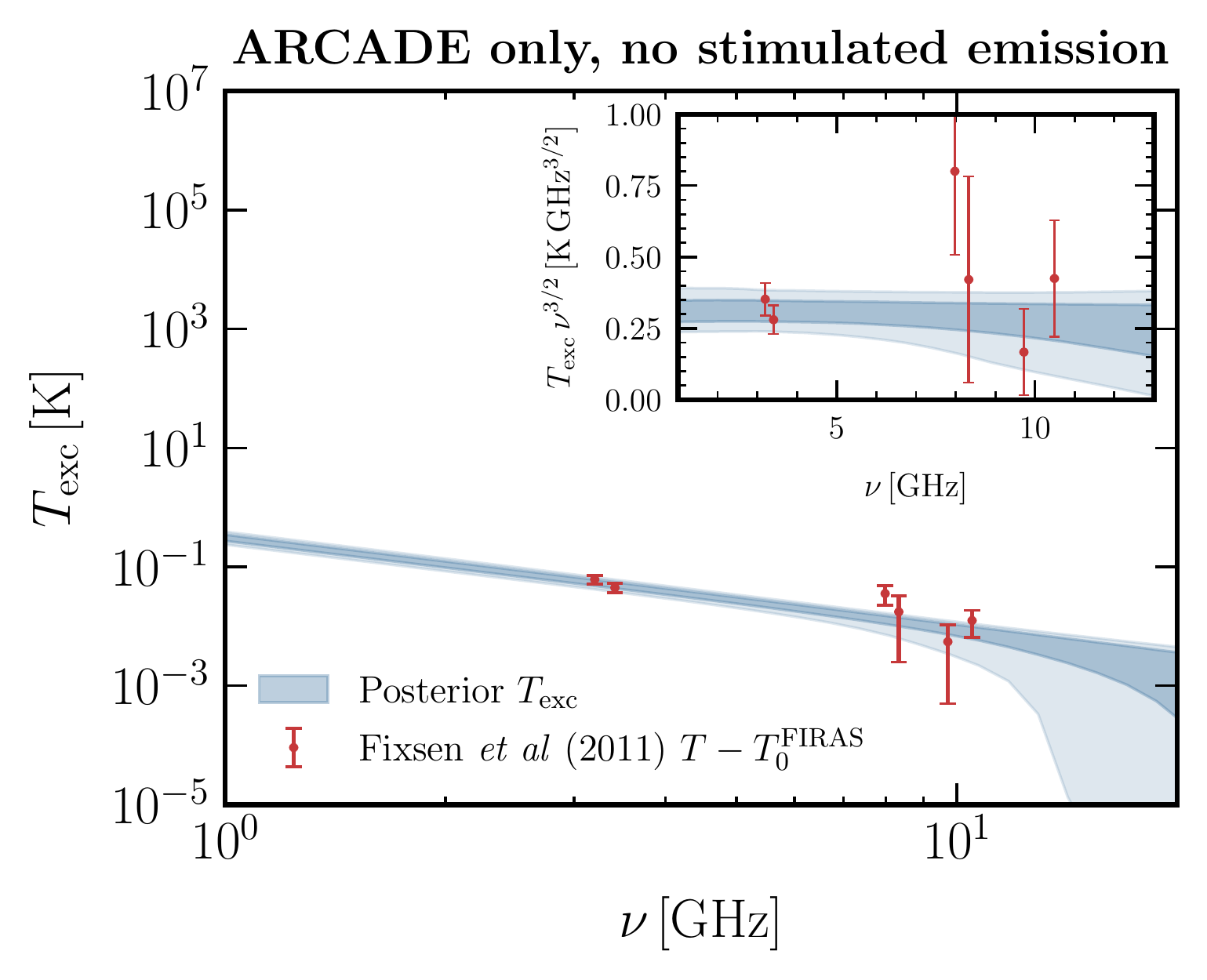}
    \caption{Systematic variations on the analysis, same format as Fig.~\ref{fig:T_exc_fiducial}.}
    \label{fig:systematics}
\end{figure*}

Figure~\ref{fig:systematics} shows the excess temperature posterior for all these systematic variations, while Tab.~\ref{tab:results} shows the posterior summaries for each case considered. The parameters ranges compatible with observations show minor variations from case to case, while providing a visually good fit to the data in all cases.

We summarize the inferred posteriors in Tab.~\ref{tab:results} through the median and 68\% containment of the individual marginal parameter posteriors (labeled ``Marginal'') and the 68\% highest posterior density intervals, describing the shortest interval in the higher-dimensional joint parameter space where 68\% of the posterior mass is contained (labeled ``HDPI'').

\subsection{Power-law fit and model comparison}

\begin{figure*}
    \centering
    \includegraphics[width=0.46\textwidth]{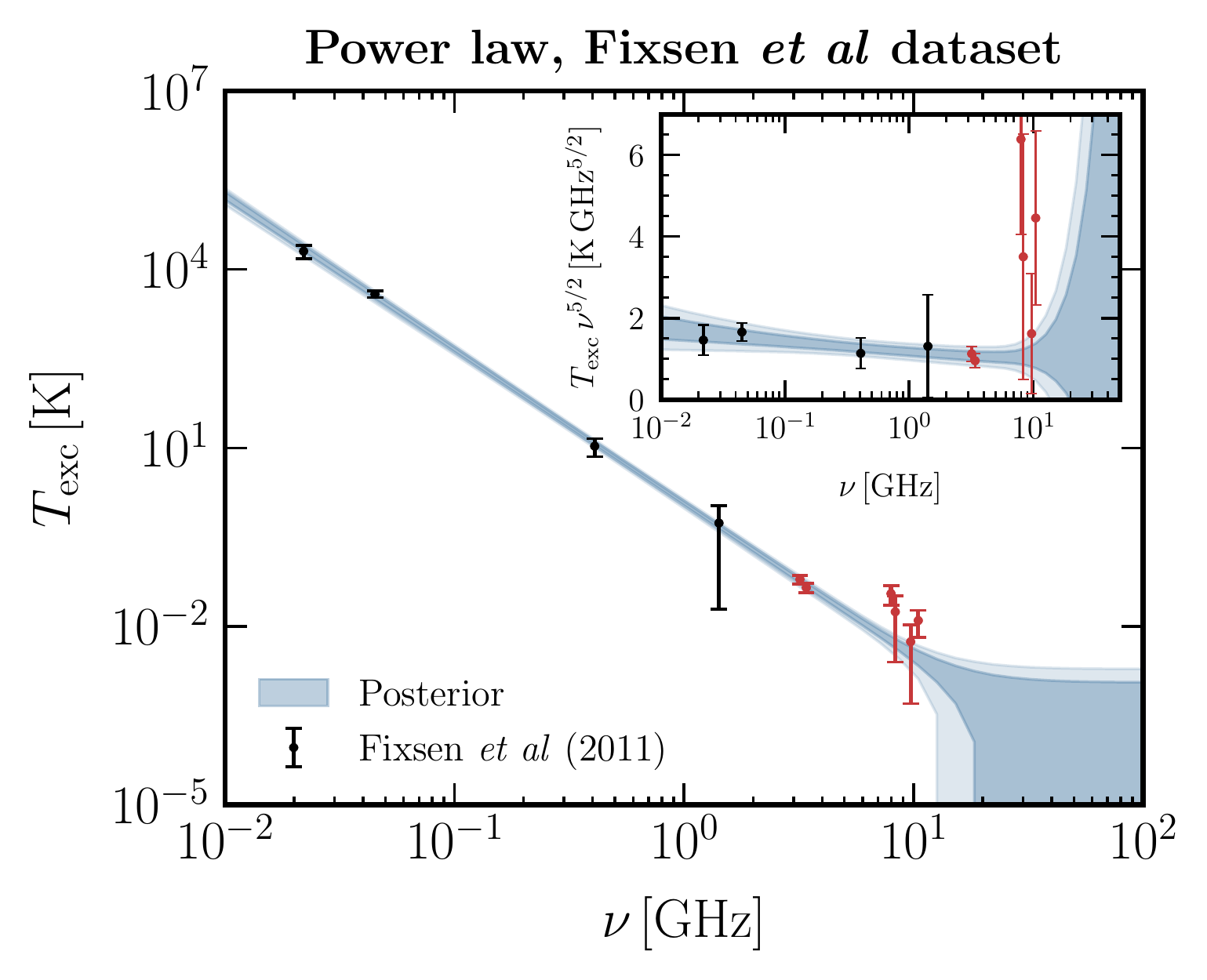}
    \includegraphics[width=0.46\textwidth]{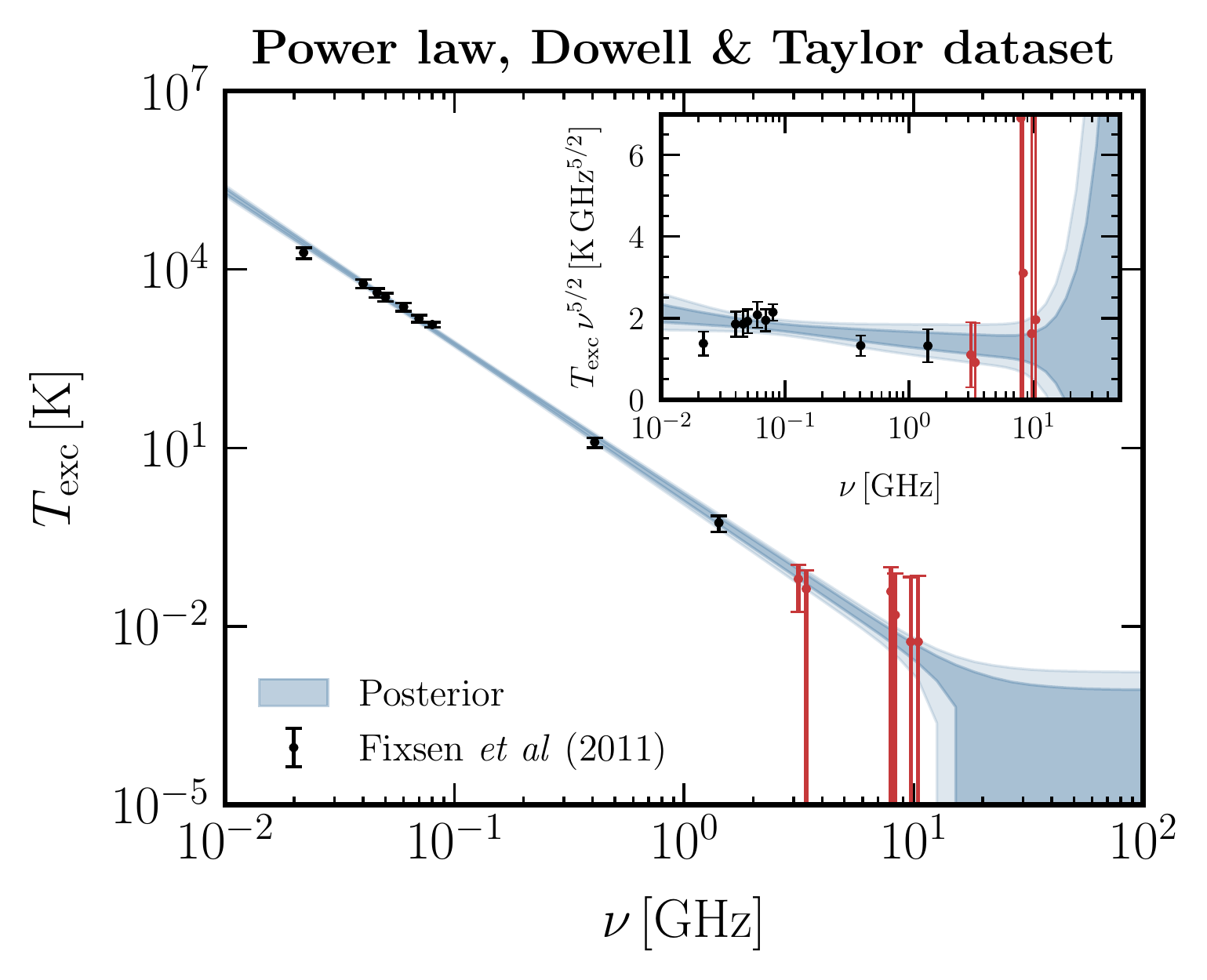}
    \caption{Excess temperature for the power-law fit, same format as Fig.~\ref{fig:T_exc_fiducial}}
    \label{fig:systematics-pl}
\end{figure*}

As a point of comparison, we also show results using a variant of the commonly-used power-law ansatz (equivalent to Eq.~\ref{eq:T_power_law}) for the excess temperature over $T_0$,
\begin{equation}
    T(\nu) = T_0\left[1 + A_\mathrm{GHz}\left(\frac{\nu}{1\,\mathrm{GHz}}\right)^\beta\right]
\end{equation}
which is parameterized through the CMB black-body temperature $T_0$, excess temperature at 1 GHz $A_\mathrm{GHz}$, and spectral index $\beta$. The results of these fits for the Fixsen \emph{et al} and Dowell \& Taylor datasets are shown in the left and right columns of Fig.~\ref{fig:systematics-pl}, respectively. As previously reported in the literature, these parameterizations provide a formally good fit to the low-frequency radio data~\cite{Fialkov:2019vnb,Dowell:2018mdb}. Posterior summaries for these fits are given in Tab.~\ref{tab:results-pl} in the same format as Tab.~\ref{tab:results}.

We show a quantitative comparison between the power law and new physics models as follows. For a given model $\mathcal M$ (in our case corresponding to either the new-physics model or the power-law ansatz) the model evidence, also known as the marginal likelihood, is defined for data $x$ and parameters $\theta$ as $\mathcal Z_{\mathcal M} \equiv \int\mathrm{d}\theta \, p_{\mathcal M}(x\mid\theta)p_{\mathcal M}(\theta)$ where $p_{\mathcal M}(x\mid\theta)$ and $p_{\mathcal M}(\theta)$ are the model likelihood and prior, respectively. The model evidence is a measure of compatibility between the model and data, and implicitly penalizes excessive model complexity---this is relevant in our case, since the new physics and power-law models have different number of parameters (3 and 6 respectively). The ratio of evidences between two models $\mathcal M_1$ and $\mathcal M_2$ is known as the Bayes factor, and can be formally used in a model comparison setting to quantify how much better of a fit $\mathcal M_1$ is compared to $\mathcal M_2$.

We show the log-Bayes factors in favor of the power-law model in the last column of Tab.~\ref{tab:results-pl}. The power-law model tends to provide a formally better fit using both the Fixsen \emph{et al} and Dowell \& Taylor datasets, corresponding to log-Bayes factors in favor of the power-law model of $\Delta\log\mathcal Z = 1.89$ and $\Delta\log\mathcal Z = 0.76$ respectively. This reflects the fact that a power law with $\beta \approx -2.6$ is a slightly better fit to the data points, rather than $\beta \approx -5/2$ as predicted by our model. We note however that given significant differences between models considered, a principled model comparison is difficult. Specifically, $\mathcal O (1)$ Bayes factors should not be taken at face value, since they can depend sensitively on chosen prior specifications and ranges; see Ref.~\cite{Isi:2022cii} for a detailed discussion of this point. Finally, we note that even though synchrotron emission from unknown sources can plausibly produce a power law close to $\beta \approx -2.6$ (see App.~\ref{Sec:Synchrotron}), this is highly dependent on the electron distribution within these individual sources as a function of source brightness and counts; it is therefore not immediately obvious that any future model ascribing $T_\text{exc}$ to currently unknown synchrotron sources would provide a better fit than our model does. 

\begin{table*}[!t]
\footnotesize
\begin{center}
\begin{tabular}{C{2cm}C{1.9cm}|C{2.cm}C{2.0cm}C{2.0cm}C{2.0cm}C{2.0cm}C{2cm}}
\toprule
\textbf{Config.} & \textbf{Summary}  & $\log_{10}m_{a}$	 & $\log_{10}m_{A'}$  & $T_0'/T_0$ & $\log_{10}\tau_a$ &  $\log_{10}\epsilon$ & $T_0$ \\
- & - & \scriptsize{$\log_{10}[\mathrm{eV}]$}	 & \scriptsize{$\log_{10}[\mathrm{eV}]$}  & - & \scriptsize{$\log_{10}[\mathrm{s}]$} &  - & [K] \\
\Xhline{1\arrayrulewidth}
\footnotesize
\multirow{2}{*}{\makecell{Fixsen \\ (Baseline)}}  & Marginals&$-3.66^{+0.10}_{-0.26}$&$-13.57^{+0.32}_{-0.68}$&$0.22^{+0.05}_{-0.12}$&$21.15^{+0.47}_{-1.10}$&$-6.98^{+0.25}_{-0.60}$&$2.7258^{+0.0003}_{-0.0007}$\\
& HPDI & $[-3.97, -3.37]$&$[-14.52, -13.04]$&$[0.08, 0.32]$&$[19.58, 21.98]$&$[-7.65, -6.30]$&$[2.7251, 2.7266]$\\ \hline
\multirow{2}{*}{No EG model}  & Marginals&$-3.96^{+0.11}_{-0.26}$&$-13.22^{+0.28}_{-0.75}$&$0.22^{+0.05}_{-0.11}$&$21.48^{+0.58}_{-1.27}$&$-7.23^{+0.27}_{-0.71}$&$2.7259^{+0.0003}_{-0.0007}$\\
& HPDI & $[-4.27, -3.75]$&$[-13.85, -12.47]$&$[0.09, 0.33]$&$[19.96, 22.71]$&$[-7.87, -6.43]$&$[2.7252, 2.7267]$\\ \hline
\multirow{2}{*}{Dowell}  & Marginals&$-4.12^{+0.15}_{-0.33}$&$-13.28^{+0.30}_{-0.74}$&$0.22^{+0.05}_{-0.11}$&$21.61^{+0.60}_{-1.26}$&$-7.25^{+0.27}_{-0.78}$&$2.7256^{+0.0003}_{-0.0007}$\\
& HPDI & $[-4.54, -3.76]$&$[-14.01, -12.50]$&$[0.15, 0.40]$&$[20.10, 22.98]$&$[-8.06, -6.50]$&$[2.7249, 2.7264]$\\ \hline
\multirow{2}{*}{\makecell{Fixsen, no stim. \\ ARCADE only}}  & Marginals&$-3.38^{+0.13}_{-0.30}$&$-13.59^{+0.32}_{-0.64}$&-&$20.33^{+0.34}_{-0.65}$&$-6.61^{+0.20}_{-0.45}$&$2.7256^{+0.0002}_{-0.0007}$\\
& HPDI & $[-3.78, -3.10]$&$[-14.54, -13.06]$&-&$[19.36, 20.80]$&$[-7.15, -6.08]$&$[2.7249, 2.7264]$\\
\botrule
\end{tabular}
\end{center}
\caption{Posterior summaries for all the systematic variations as compared to our baseline analysis.}
\label{tab:results}
\end{table*}    
\begin{table*}[!t]
\footnotesize
\begin{center}
\begin{tabular}{C{2.2cm}C{2.0cm}|C{2.cm}C{2cm}C{2cm}|C{1.3cm}}
\toprule
\textbf{Config.} & \textbf{Summary}  & $A_\mathrm{GHz}$	 & $\beta$  & $T_0$ & $\Delta\log\mathcal Z$ \\
- & - & - & - & [K] & [nat]\\
\Xhline{1\arrayrulewidth}
\multirow{2}{*}{\makecell{Fixsen \\ Power law}}  & Marginals&$0.43^{+0.01}_{-0.04}$&$-2.58^{+0.01}_{-0.04}$&$2.7258^{+0.0003}_{-0.0008}$&\multirow{2}{*}{1.89}\\
& HPDI & $[0.40, 0.47]$&$[-2.62, -2.55]$&$[2.7250, 2.7266]$&\\ \hline
\multirow{2}{*}{\makecell{Dowell \\ Power law}}  & Marginals&$0.54^{+0.03}_{-0.06}$&$-2.58^{+0.02}_{-0.05}$&$2.7255^{+0.0003}_{-0.0008}$&\multirow{2}{*}{0.76}\\
& HPDI & $[0.47, 0.61]$&$[-2.63, -2.53]$&$[2.7247, 2.7264]$&\\
\botrule
\end{tabular}
\end{center}
\caption{Posterior summaries for the power-law fits. The last column additionally shows the log-Bayes factor $\Delta \log\mathcal Z$ in favor of the power-law fit, with positive values corresponding to a preference for the power-law model.}
\label{tab:results-pl}
\end{table*}    

\section{A recap of synchrotron radiation}
\label{Sec:Synchrotron}

As extensively discussed in the \textit{Letter}, many astrophysical solutions to the ERB advocate for the presence of synchrotron radiation from unresolved astrophysical sources. In this appendix, we briefly review synchrotron emission with a particular focus on its spectral features. The basic physics behind synchrotron radiation is straightforward: a charged particle accelerated in a magnetic field will radiate. Nonrelativistic particles emit cyclotron radiation, with a frequency given by the cyclotron frequency $\omega_{\rm cyc} = q B / m$, where $q$ is the particle charge, $B$ the magnetic field in which it moves and $m$ its mass. Relativistic particles emit what we call synchrotron radiation: the frequency spectrum is far richer, covering a range well away from the cyclotron frequency.

Taking into account strong relativistic effects (such as forward beaming), one finds that the characteristic frequency of the synchrotron radiation of a \textit{single} charged particle is $\omega_{\rm syn} \sim \gamma^2 \omega_{\rm cyc}$~\cite{1979rpa..book.....R}. The spectrum of the radiation is broad: $\omega < \omega_{\rm syn}$ the spectrum falls off like a power law $\propto \omega^{1/3}$, while for $\omega > \omega_{\rm syn}$, the spectrum falls off exponentially. 

This is, however, very different from what we observe in astrophysical systems. The reason is that we need to consider what happens to the total synchrotron radiation emitted by a \textit{population} of charged particles, with a given energy distribution. Let us consider an ensemble of charged particles with number density distribution per unit energy following a power law $\dd N/\dd E \propto E^p$. The power per unit energy emitted by this ensemble of charges reads
\begin{equation*}
\frac{\dd P}{\dd E} \propto \frac{\dd N}{\dd E} P_E \propto E^p E^2 B^2,
\end{equation*} 
where $P_E$ is the power emitted by a single particle, given by the Larmor formula $P_E \propto B^2 \gamma^2 \propto B^2 E^2$ (we drop here all the irrelevant constants given that we are only interested in the final spectral shape of the synchrotron radiation power). Let us now assume that the \textit{single}-charge synchrotron spectrum is peaked enough to assume that all the power is emitted at the synchrotron frequency, $\omega_{\rm syn}$. This implies a one-to-one relation between the observed frequency and the energy of the single particle $\nu \simeq \omega_{\rm syn} / (2\pi)  \propto E^2 B$. It then follows
\begin{equation*}
\frac{\dd P}{\dd \nu} = \frac{\dd P}{\dd E}\frac{\dd E}{\dd \nu} \propto E^{p+2} B \times \nu^{-1/2} B^{-1/2} \propto B \Big(\frac{\nu}{B}\Big)^{\frac{1+p}{2}},
\end{equation*}
and therefore the spectral index reads $\alpha = (1+p) / 2$. In terms of the spectral index of brightness temperature, $\beta$, this corresponds to $\beta = (p - 3) / 2$, since the brightness temperature is related to the intensity via $T_\nu \propto I_\nu/\nu^2$. Therefore, the spectral index for synchrotron radiation is intrinsically related to the energy distribution of the charged particles which generate the radiation. In particular, a power law is expected if the energy distribution is also a power law as assumed above. 

One well-known example for which a power-law distribution is expected comes from particle acceleration by astrophysical shocks~\cite{Bell:1978fj, BellI, Fermi49, Blandford:1980hv, Blandford:1978ky}. For example, if one considers shock waves which propagate in the interstellar medium outside a supernova remnant, it can be shown that at linear level $p = - 2$~\cite{Bell:1978fj, BellI}, which implies $\alpha = -0.5$ and $\beta = -2.5$. The spectral index predicted by the simplest scenario of shock acceleration coincides with the spectral index expected in our model, and is therefore slightly different than the power-law best-fit obtained in Ref.~\cite{Fixsen:2009xn}. Nevertheless, steeper spectra are possible taking into account non-linear effects in the theory of diffusive shock acceleration~\cite{Malkov:2001kya}. As a matter of fact, these effects are crucial to match observations of the supernova remnant Cassiopeia A, which exhibits a steeper spectral index, $\alpha \simeq -0.77 $ or $\beta \simeq -2.77$~\cite{Domcek:2020pfz}.

%% file: low-radio.bbl
\begin{thebibliography}{110}%
\makeatletter
\providecommand \@ifxundefined [1]{%
 \@ifx{#1\undefined}
}%
\providecommand \@ifnum [1]{%
 \ifnum #1\expandafter \@firstoftwo
 \else \expandafter \@secondoftwo
 \fi
}%
\providecommand \@ifx [1]{%
 \ifx #1\expandafter \@firstoftwo
 \else \expandafter \@secondoftwo
 \fi
}%
\providecommand \natexlab [1]{#1}%
\providecommand \enquote  [1]{``#1''}%
\providecommand \bibnamefont  [1]{#1}%
\providecommand \bibfnamefont [1]{#1}%
\providecommand \citenamefont [1]{#1}%
\providecommand \href@noop [0]{\@secondoftwo}%
\providecommand \href [0]{\begingroup \@sanitize@url \@href}%
\providecommand \@href[1]{\@@startlink{#1}\@@href}%
\providecommand \@@href[1]{\endgroup#1\@@endlink}%
\providecommand \@sanitize@url [0]{\catcode `\\12\catcode `\$12\catcode
  `\&12\catcode `\#12\catcode `\^12\catcode `\_12\catcode `\%12\relax}%
\providecommand \@@startlink[1]{}%
\providecommand \@@endlink[0]{}%
\providecommand \url  [0]{\begingroup\@sanitize@url \@url }%
\providecommand \@url [1]{\endgroup\@href {#1}{\urlprefix }}%
\providecommand \urlprefix  [0]{URL }%
\providecommand \Eprint [0]{\href }%
\providecommand \doibase [0]{https://doi.org/}%
\providecommand \selectlanguage [0]{\@gobble}%
\providecommand \bibinfo  [0]{\@secondoftwo}%
\providecommand \bibfield  [0]{\@secondoftwo}%
\providecommand \translation [1]{[#1]}%
\providecommand \BibitemOpen [0]{}%
\providecommand \bibitemStop [0]{}%
\providecommand \bibitemNoStop [0]{.\EOS\space}%
\providecommand \EOS [0]{\spacefactor3000\relax}%
\providecommand \BibitemShut  [1]{\csname bibitem#1\endcsname}%
\let\auto@bib@innerbib\@empty
%</preamble>
\bibitem [{\citenamefont {Fixsen}\ \emph {et~al.}(1996)\citenamefont {Fixsen},
  \citenamefont {Cheng}, \citenamefont {Gales}, \citenamefont {Mather},
  \citenamefont {Shafer},\ and\ \citenamefont {Wright}}]{Fixsen:1996nj}%
  \BibitemOpen
  \bibfield  {author} {\bibinfo {author} {\bibfnamefont {D.~J.}\ \bibnamefont
  {Fixsen}}, \bibinfo {author} {\bibfnamefont {E.~S.}\ \bibnamefont {Cheng}},
  \bibinfo {author} {\bibfnamefont {J.~M.}\ \bibnamefont {Gales}}, \bibinfo
  {author} {\bibfnamefont {J.~C.}\ \bibnamefont {Mather}}, \bibinfo {author}
  {\bibfnamefont {R.~A.}\ \bibnamefont {Shafer}},\ and\ \bibinfo {author}
  {\bibfnamefont {E.~L.}\ \bibnamefont {Wright}},\ }\bibfield  {title}
  {\bibinfo {title} {{The Cosmic Microwave Background spectrum from the full
  COBE FIRAS data set}},\ }\href {https://doi.org/10.1086/178173} {\bibfield
  {journal} {\bibinfo  {journal} {Astrophys. J.}\ }\textbf {\bibinfo {volume}
  {473}},\ \bibinfo {pages} {576} (\bibinfo {year} {1996})},\ \Eprint
  {https://arxiv.org/abs/astro-ph/9605054} {arXiv:astro-ph/9605054 [astro-ph]}
  \BibitemShut {NoStop}%
%%CITATION = ASTRO-PH/9605054;%%
\bibitem [{\citenamefont {Aghanim}\ \emph {et~al.}(2020)\citenamefont {Aghanim}
  \emph {et~al.}}]{Planck:2018vyg}%
  \BibitemOpen
  \bibfield  {author} {\bibinfo {author} {\bibfnamefont {N.}~\bibnamefont
  {Aghanim}} \emph {et~al.} (\bibinfo {collaboration} {Planck}),\ }\bibfield
  {title} {\bibinfo {title} {{Planck 2018 results. VI. Cosmological
  parameters}},\ }\href {https://doi.org/10.1051/0004-6361/201833910}
  {\bibfield  {journal} {\bibinfo  {journal} {Astron. Astrophys.}\ }\textbf
  {\bibinfo {volume} {641}},\ \bibinfo {pages} {A6} (\bibinfo {year} {2020})},\
  \bibinfo {note} {[Erratum: Astron.Astrophys. 652, C4 (2021)]},\ \Eprint
  {https://arxiv.org/abs/1807.06209} {arXiv:1807.06209 [astro-ph.CO]}
  \BibitemShut {NoStop}%
\bibitem [{\citenamefont {Fixsen}\ \emph {et~al.}(2011)\citenamefont {Fixsen}
  \emph {et~al.}}]{Fixsen:2009xn}%
  \BibitemOpen
  \bibfield  {author} {\bibinfo {author} {\bibfnamefont {D.~J.}\ \bibnamefont
  {Fixsen}} \emph {et~al.},\ }\bibfield  {title} {\bibinfo {title} {{ARCADE 2
  Measurement of the Extra-Galactic Sky Temperature at 3-90 GHz}},\ }\href
  {https://doi.org/10.1088/0004-637X/734/1/5} {\bibfield  {journal} {\bibinfo
  {journal} {Astrophys. J.}\ }\textbf {\bibinfo {volume} {734}},\ \bibinfo
  {pages} {5} (\bibinfo {year} {2011})},\ \Eprint
  {https://arxiv.org/abs/0901.0555} {arXiv:0901.0555 [astro-ph.CO]}
  \BibitemShut {NoStop}%
\bibitem [{\citenamefont {Roger}\ \emph {et~al.}(1999)\citenamefont {Roger},
  \citenamefont {Costain}, \citenamefont {Landecker},\ and\ \citenamefont
  {Swerdlyk}}]{Roger:1999jy}%
  \BibitemOpen
  \bibfield  {author} {\bibinfo {author} {\bibfnamefont {R.~S.}\ \bibnamefont
  {Roger}}, \bibinfo {author} {\bibfnamefont {C.~H.}\ \bibnamefont {Costain}},
  \bibinfo {author} {\bibfnamefont {T.~L.}\ \bibnamefont {Landecker}},\ and\
  \bibinfo {author} {\bibfnamefont {C.~M.}\ \bibnamefont {Swerdlyk}},\
  }\bibfield  {title} {\bibinfo {title} {{The radio emission from the galaxy at
  22 mhz}},\ }\href {https://doi.org/10.1051/aas:1999239} {\bibfield  {journal}
  {\bibinfo  {journal} {Astron. Astrophys. Suppl. Ser.}\ }\textbf {\bibinfo
  {volume} {137}},\ \bibinfo {pages} {7} (\bibinfo {year} {1999})},\ \Eprint
  {https://arxiv.org/abs/astro-ph/9902213} {arXiv:astro-ph/9902213}
  \BibitemShut {NoStop}%
\bibitem [{\citenamefont {{Maeda}}\ \emph {et~al.}(1999)\citenamefont
  {{Maeda}}, \citenamefont {{Alvarez}}, \citenamefont {{Aparici}},
  \citenamefont {{May}},\ and\ \citenamefont {{Reich}}}]{1999A&AS..140..145M}%
  \BibitemOpen
  \bibfield  {author} {\bibinfo {author} {\bibfnamefont {K.}~\bibnamefont
  {{Maeda}}}, \bibinfo {author} {\bibfnamefont {H.}~\bibnamefont {{Alvarez}}},
  \bibinfo {author} {\bibfnamefont {J.}~\bibnamefont {{Aparici}}}, \bibinfo
  {author} {\bibfnamefont {J.}~\bibnamefont {{May}}},\ and\ \bibinfo {author}
  {\bibfnamefont {P.}~\bibnamefont {{Reich}}},\ }\bibfield  {title} {\bibinfo
  {title} {{A 45-MHz continuum survey of the northern hemisphere}},\ }\href
  {https://doi.org/10.1051/aas:1999413} {\bibfield  {journal} {\bibinfo
  {journal} {\aaps}\ }\textbf {\bibinfo {volume} {140}},\ \bibinfo {pages}
  {145} (\bibinfo {year} {1999})}\BibitemShut {NoStop}%
\bibitem [{\citenamefont {{Haslam}}\ \emph {et~al.}(1981)\citenamefont
  {{Haslam}}, \citenamefont {{Klein}}, \citenamefont {{Salter}}, \citenamefont
  {{Stoffel}}, \citenamefont {{Wilson}}, \citenamefont {{Cleary}},
  \citenamefont {{Cooke}},\ and\ \citenamefont
  {{Thomasson}}}]{1981A&A...100..209H}%
  \BibitemOpen
  \bibfield  {author} {\bibinfo {author} {\bibfnamefont {C.~G.~T.}\
  \bibnamefont {{Haslam}}}, \bibinfo {author} {\bibfnamefont {U.}~\bibnamefont
  {{Klein}}}, \bibinfo {author} {\bibfnamefont {C.~J.}\ \bibnamefont
  {{Salter}}}, \bibinfo {author} {\bibfnamefont {H.}~\bibnamefont {{Stoffel}}},
  \bibinfo {author} {\bibfnamefont {W.~E.}\ \bibnamefont {{Wilson}}}, \bibinfo
  {author} {\bibfnamefont {M.~N.}\ \bibnamefont {{Cleary}}}, \bibinfo {author}
  {\bibfnamefont {D.~J.}\ \bibnamefont {{Cooke}}},\ and\ \bibinfo {author}
  {\bibfnamefont {P.}~\bibnamefont {{Thomasson}}},\ }\bibfield  {title}
  {\bibinfo {title} {{A 408 MHz all-sky continuum survey. I - Observations at
  southern declinations and for the North Polar region.}},\ }\href@noop {}
  {\bibfield  {journal} {\bibinfo  {journal} {\aap}\ }\textbf {\bibinfo
  {volume} {100}},\ \bibinfo {pages} {209} (\bibinfo {year}
  {1981})}\BibitemShut {NoStop}%
\bibitem [{\citenamefont {{Reich}}\ and\ \citenamefont
  {{Reich}}(1986)}]{1986A&AS...63..205R}%
  \BibitemOpen
  \bibfield  {author} {\bibinfo {author} {\bibfnamefont {P.}~\bibnamefont
  {{Reich}}}\ and\ \bibinfo {author} {\bibfnamefont {W.}~\bibnamefont
  {{Reich}}},\ }\bibfield  {title} {\bibinfo {title} {{A radio continuum survey
  of the northern sky at 1420 MHz. II}},\ }\href@noop {} {\bibfield  {journal}
  {\bibinfo  {journal} {\aaps}\ }\textbf {\bibinfo {volume} {63}},\ \bibinfo
  {pages} {205} (\bibinfo {year} {1986})}\BibitemShut {NoStop}%
\bibitem [{\citenamefont {Fixsen}(2009)}]{Fixsen:2009ug}%
  \BibitemOpen
  \bibfield  {author} {\bibinfo {author} {\bibfnamefont {D.~J.}\ \bibnamefont
  {Fixsen}},\ }\bibfield  {title} {\bibinfo {title} {{The Temperature of the
  Cosmic Microwave Background}},\ }\href
  {https://doi.org/10.1088/0004-637X/707/2/916} {\bibfield  {journal} {\bibinfo
   {journal} {Astrophys. J.}\ }\textbf {\bibinfo {volume} {707}},\ \bibinfo
  {pages} {916} (\bibinfo {year} {2009})},\ \Eprint
  {https://arxiv.org/abs/0911.1955} {arXiv:0911.1955 [astro-ph.CO]}
  \BibitemShut {NoStop}%
\bibitem [{\citenamefont {Dowell}\ and\ \citenamefont
  {Taylor}(2018)}]{Dowell:2018mdb}%
  \BibitemOpen
  \bibfield  {author} {\bibinfo {author} {\bibfnamefont {J.}~\bibnamefont
  {Dowell}}\ and\ \bibinfo {author} {\bibfnamefont {G.~B.}\ \bibnamefont
  {Taylor}},\ }\bibfield  {title} {\bibinfo {title} {{The Radio Background
  Below 100 MHz}},\ }\href {https://doi.org/10.3847/2041-8213/aabf86}
  {\bibfield  {journal} {\bibinfo  {journal} {Astrophys. J. Lett.}\ }\textbf
  {\bibinfo {volume} {858}},\ \bibinfo {pages} {L9} (\bibinfo {year} {2018})},\
  \Eprint {https://arxiv.org/abs/1804.08581} {arXiv:1804.08581 [astro-ph.CO]}
  \BibitemShut {NoStop}%
\bibitem [{\citenamefont {Singal}\ \emph {et~al.}(2018)\citenamefont {Singal}
  \emph {et~al.}}]{Singal:2017jlh}%
  \BibitemOpen
  \bibfield  {author} {\bibinfo {author} {\bibfnamefont {J.}~\bibnamefont
  {Singal}} \emph {et~al.},\ }\bibfield  {title} {\bibinfo {title} {{The Radio
  Synchrotron Background: Conference Summary and Report}},\ }\href
  {https://doi.org/10.1088/1538-3873/aaa6b0} {\bibfield  {journal} {\bibinfo
  {journal} {Publ. Astron. Soc. Pac.}\ }\textbf {\bibinfo {volume} {130}},\
  \bibinfo {pages} {036001} (\bibinfo {year} {2018})},\ \Eprint
  {https://arxiv.org/abs/1711.09979} {arXiv:1711.09979 [astro-ph.HE]}
  \BibitemShut {NoStop}%
\bibitem [{\citenamefont {{Wright}}\ \emph {et~al.}(1999)\citenamefont
  {{Wright}}, \citenamefont {{Dickel}}, \citenamefont {{Koralesky}},\ and\
  \citenamefont {{Rudnick}}}]{Cas1999ApJ}%
  \BibitemOpen
  \bibfield  {author} {\bibinfo {author} {\bibfnamefont {M.}~\bibnamefont
  {{Wright}}}, \bibinfo {author} {\bibfnamefont {J.}~\bibnamefont {{Dickel}}},
  \bibinfo {author} {\bibfnamefont {B.}~\bibnamefont {{Koralesky}}},\ and\
  \bibinfo {author} {\bibfnamefont {L.}~\bibnamefont {{Rudnick}}},\ }\bibfield
  {title} {\bibinfo {title} {{The Supernova Remnant Cassiopeia A at Millimeter
  Wavelengths}},\ }\href {https://doi.org/10.1086/307270} {\bibfield  {journal}
  {\bibinfo  {journal} {\apj}\ }\textbf {\bibinfo {volume} {518}},\ \bibinfo
  {pages} {284} (\bibinfo {year} {1999})}\BibitemShut {NoStop}%
\bibitem [{\citenamefont {{Scheuer}}\ and\ \citenamefont
  {{Williams}}(1968)}]{1968ARA&A...6..321S}%
  \BibitemOpen
  \bibfield  {author} {\bibinfo {author} {\bibfnamefont {P.~A.~G.}\
  \bibnamefont {{Scheuer}}}\ and\ \bibinfo {author} {\bibfnamefont {P.~J.~S.}\
  \bibnamefont {{Williams}}},\ }\bibfield  {title} {\bibinfo {title} {{Radio
  Spectra}},\ }\href {https://doi.org/10.1146/annurev.aa.06.090168.001541}
  {\bibfield  {journal} {\bibinfo  {journal} {The Annual Review of Astronomy
  and Astrophysics}\ }\textbf {\bibinfo {volume} {6}},\ \bibinfo {pages} {321}
  (\bibinfo {year} {1968})}\BibitemShut {NoStop}%
\bibitem [{\citenamefont {Platania}\ \emph {et~al.}(1998)\citenamefont
  {Platania}, \citenamefont {Bensadoun}, \citenamefont {Bersanelli},
  \citenamefont {De~Amici}, \citenamefont {Kogut}, \citenamefont {Levin},
  \citenamefont {Maino},\ and\ \citenamefont {Smoot}}]{Platania:1997zn}%
  \BibitemOpen
  \bibfield  {author} {\bibinfo {author} {\bibfnamefont {P.}~\bibnamefont
  {Platania}}, \bibinfo {author} {\bibfnamefont {M.}~\bibnamefont {Bensadoun}},
  \bibinfo {author} {\bibfnamefont {M.}~\bibnamefont {Bersanelli}}, \bibinfo
  {author} {\bibfnamefont {G.}~\bibnamefont {De~Amici}}, \bibinfo {author}
  {\bibfnamefont {A.}~\bibnamefont {Kogut}}, \bibinfo {author} {\bibfnamefont
  {S.}~\bibnamefont {Levin}}, \bibinfo {author} {\bibfnamefont
  {D.}~\bibnamefont {Maino}},\ and\ \bibinfo {author} {\bibfnamefont {G.~F.}\
  \bibnamefont {Smoot}},\ }\bibfield  {title} {\bibinfo {title} {{A
  determination of the spectral index of galactic synchrotron emission in the
  1-10 GHz range}},\ }\href {https://doi.org/10.1086/306175} {\bibfield
  {journal} {\bibinfo  {journal} {Astrophys. J.}\ }\textbf {\bibinfo {volume}
  {505}},\ \bibinfo {pages} {473} (\bibinfo {year} {1998})},\ \Eprint
  {https://arxiv.org/abs/astro-ph/9707252} {arXiv:astro-ph/9707252}
  \BibitemShut {NoStop}%
\bibitem [{\citenamefont {Carilli}\ \emph {et~al.}(1991)\citenamefont
  {Carilli}, \citenamefont {Perley}, \citenamefont {Dreher},\ and\
  \citenamefont {Leahy}}]{Carilli:1991zz}%
  \BibitemOpen
  \bibfield  {author} {\bibinfo {author} {\bibfnamefont {C.~L.}\ \bibnamefont
  {Carilli}}, \bibinfo {author} {\bibfnamefont {R.~A.}\ \bibnamefont {Perley}},
  \bibinfo {author} {\bibfnamefont {J.~W.}\ \bibnamefont {Dreher}},\ and\
  \bibinfo {author} {\bibfnamefont {J.~P.}\ \bibnamefont {Leahy}},\ }\bibfield
  {title} {\bibinfo {title} {{Multifrequency radio observations of Cygnus A -
  Spectral aging in powerful radio galaxies}},\ }\href
  {https://doi.org/10.1086/170813} {\bibfield  {journal} {\bibinfo  {journal}
  {Astrophys. J.}\ }\textbf {\bibinfo {volume} {383}},\ \bibinfo {pages} {554}
  (\bibinfo {year} {1991})}\BibitemShut {NoStop}%
\bibitem [{\citenamefont {Protheroe}\ and\ \citenamefont
  {Biermann}(1996)}]{Protheroe:1996si}%
  \BibitemOpen
  \bibfield  {author} {\bibinfo {author} {\bibfnamefont {R.~J.}\ \bibnamefont
  {Protheroe}}\ and\ \bibinfo {author} {\bibfnamefont {P.~L.}\ \bibnamefont
  {Biermann}},\ }\bibfield  {title} {\bibinfo {title} {{A New estimate of the
  extragalactic radio background and implications for ultrahigh-energy
  gamma-ray propagation}},\ }\href
  {https://doi.org/10.1016/S0927-6505(96)00041-2} {\bibfield  {journal}
  {\bibinfo  {journal} {Astropart. Phys.}\ }\textbf {\bibinfo {volume} {6}},\
  \bibinfo {pages} {45} (\bibinfo {year} {1996})},\ \bibinfo {note} {[Erratum:
  Astropart.Phys. 7, 181 (1997)]},\ \Eprint
  {https://arxiv.org/abs/astro-ph/9605119} {arXiv:astro-ph/9605119}
  \BibitemShut {NoStop}%
\bibitem [{\citenamefont {Gervasi}\ \emph {et~al.}(2008)\citenamefont
  {Gervasi}, \citenamefont {Tartari}, \citenamefont {Zannoni}, \citenamefont
  {Boella},\ and\ \citenamefont {Sironi}}]{Gervasi:2008rr}%
  \BibitemOpen
  \bibfield  {author} {\bibinfo {author} {\bibfnamefont {M.}~\bibnamefont
  {Gervasi}}, \bibinfo {author} {\bibfnamefont {A.}~\bibnamefont {Tartari}},
  \bibinfo {author} {\bibfnamefont {M.}~\bibnamefont {Zannoni}}, \bibinfo
  {author} {\bibfnamefont {G.}~\bibnamefont {Boella}},\ and\ \bibinfo {author}
  {\bibfnamefont {G.}~\bibnamefont {Sironi}},\ }\bibfield  {title} {\bibinfo
  {title} {{The contribution of the Unresolved Extragalactic Radio Sources to
  the Brightness Temperature of the sky}},\ }\href
  {https://doi.org/10.1086/588628} {\bibfield  {journal} {\bibinfo  {journal}
  {Astrophys. J.}\ }\textbf {\bibinfo {volume} {682}},\ \bibinfo {pages} {223}
  (\bibinfo {year} {2008})},\ \Eprint {https://arxiv.org/abs/0803.4138}
  {arXiv:0803.4138 [astro-ph]} \BibitemShut {NoStop}%
\bibitem [{\citenamefont {Ni\c{t}u}\ \emph {et~al.}(2021)\citenamefont
  {Ni\c{t}u}, \citenamefont {Bevins}, \citenamefont {Bray},\ and\ \citenamefont
  {Scaife}}]{Nitu:2020vzn}%
  \BibitemOpen
  \bibfield  {author} {\bibinfo {author} {\bibfnamefont {I.~C.}\ \bibnamefont
  {Ni\c{t}u}}, \bibinfo {author} {\bibfnamefont {H.~T.~J.}\ \bibnamefont
  {Bevins}}, \bibinfo {author} {\bibfnamefont {J.~D.}\ \bibnamefont {Bray}},\
  and\ \bibinfo {author} {\bibfnamefont {A.~M.~M.}\ \bibnamefont {Scaife}},\
  }\bibfield  {title} {\bibinfo {title} {{An updated estimate of the cosmic
  radio background and implications for ultra-high-energy photon
  propagation}},\ }\href {https://doi.org/10.1016/j.astropartphys.2020.102532}
  {\bibfield  {journal} {\bibinfo  {journal} {Astropart. Phys.}\ }\textbf
  {\bibinfo {volume} {126}},\ \bibinfo {pages} {102532} (\bibinfo {year}
  {2021})},\ \Eprint {https://arxiv.org/abs/2004.13596} {arXiv:2004.13596
  [astro-ph.HE]} \BibitemShut {NoStop}%
\bibitem [{\citenamefont {Vernstrom}\ \emph {et~al.}(2011)\citenamefont
  {Vernstrom}, \citenamefont {Scott},\ and\ \citenamefont
  {Wall}}]{Vernstrom_2011}%
  \BibitemOpen
  \bibfield  {author} {\bibinfo {author} {\bibfnamefont {T.}~\bibnamefont
  {Vernstrom}}, \bibinfo {author} {\bibfnamefont {D.}~\bibnamefont {Scott}},\
  and\ \bibinfo {author} {\bibfnamefont {J.~V.}\ \bibnamefont {Wall}},\
  }\bibfield  {title} {\bibinfo {title} {Contribution to the diffuse radio
  background from extragalactic radio sources},\ }\href
  {https://doi.org/10.1111/j.1365-2966.2011.18990.x} {\bibfield  {journal}
  {\bibinfo  {journal} {Monthly Notices of the Royal Astronomical Society}\
  }\textbf {\bibinfo {volume} {415}},\ \bibinfo {pages} {3641} (\bibinfo {year}
  {2011})}\BibitemShut {NoStop}%
\bibitem [{\citenamefont {Condon}\ \emph {et~al.}(2012)\citenamefont {Condon},
  \citenamefont {Cotton}, \citenamefont {Fomalont}, \citenamefont {Kellermann},
  \citenamefont {Miller}, \citenamefont {Perley}, \citenamefont {Scott},
  \citenamefont {Vernstrom},\ and\ \citenamefont {Wall}}]{Condon_2012}%
  \BibitemOpen
  \bibfield  {author} {\bibinfo {author} {\bibfnamefont {J.~J.}\ \bibnamefont
  {Condon}}, \bibinfo {author} {\bibfnamefont {W.~D.}\ \bibnamefont {Cotton}},
  \bibinfo {author} {\bibfnamefont {E.~B.}\ \bibnamefont {Fomalont}}, \bibinfo
  {author} {\bibfnamefont {K.~I.}\ \bibnamefont {Kellermann}}, \bibinfo
  {author} {\bibfnamefont {N.}~\bibnamefont {Miller}}, \bibinfo {author}
  {\bibfnamefont {R.~A.}\ \bibnamefont {Perley}}, \bibinfo {author}
  {\bibfnamefont {D.}~\bibnamefont {Scott}}, \bibinfo {author} {\bibfnamefont
  {T.}~\bibnamefont {Vernstrom}},\ and\ \bibinfo {author} {\bibfnamefont
  {J.~V.}\ \bibnamefont {Wall}},\ }\bibfield  {title} {\bibinfo {title}
  {Resolving the radio source background: Deeper understanding through
  confusion},\ }\href {https://doi.org/10.1088/0004-637x/758/1/23} {\bibfield
  {journal} {\bibinfo  {journal} {The Astrophysical Journal}\ }\textbf
  {\bibinfo {volume} {758}},\ \bibinfo {pages} {23} (\bibinfo {year}
  {2012})}\BibitemShut {NoStop}%
\bibitem [{\citenamefont {Fornengo}\ \emph {et~al.}(2014)\citenamefont
  {Fornengo}, \citenamefont {Lineros}, \citenamefont {Regis},\ and\
  \citenamefont {Taoso}}]{Fornengo:2014mna}%
  \BibitemOpen
  \bibfield  {author} {\bibinfo {author} {\bibfnamefont {N.}~\bibnamefont
  {Fornengo}}, \bibinfo {author} {\bibfnamefont {R.~A.}\ \bibnamefont
  {Lineros}}, \bibinfo {author} {\bibfnamefont {M.}~\bibnamefont {Regis}},\
  and\ \bibinfo {author} {\bibfnamefont {M.}~\bibnamefont {Taoso}},\ }\bibfield
   {title} {\bibinfo {title} {{The isotropic radio background revisited}},\
  }\href {https://doi.org/10.1088/1475-7516/2014/04/008} {\bibfield  {journal}
  {\bibinfo  {journal} {JCAP}\ }\textbf {\bibinfo {volume} {04}},\ \bibinfo
  {pages} {008}},\ \Eprint {https://arxiv.org/abs/1402.2218} {arXiv:1402.2218
  [astro-ph.CO]} \BibitemShut {NoStop}%
\bibitem [{\citenamefont {{Vernstrom}}\ \emph {et~al.}(2015)\citenamefont
  {{Vernstrom}}, \citenamefont {{Norris}}, \citenamefont {{Scott}},\ and\
  \citenamefont {{Wall}}}]{2015MNRAS.447.2243V}%
  \BibitemOpen
  \bibfield  {author} {\bibinfo {author} {\bibfnamefont {T.}~\bibnamefont
  {{Vernstrom}}}, \bibinfo {author} {\bibfnamefont {R.~P.}\ \bibnamefont
  {{Norris}}}, \bibinfo {author} {\bibfnamefont {D.}~\bibnamefont {{Scott}}},\
  and\ \bibinfo {author} {\bibfnamefont {J.~V.}\ \bibnamefont {{Wall}}},\
  }\bibfield  {title} {\bibinfo {title} {{The deep diffuse extragalactic radio
  sky at 1.75 GHz}},\ }\href {https://doi.org/10.1093/mnras/stu2595} {\bibfield
   {journal} {\bibinfo  {journal} {\mnras}\ }\textbf {\bibinfo {volume}
  {447}},\ \bibinfo {pages} {2243} (\bibinfo {year} {2015})},\ \Eprint
  {https://arxiv.org/abs/1408.4160} {arXiv:1408.4160 [astro-ph.GA]}
  \BibitemShut {NoStop}%
\bibitem [{\citenamefont {Hardcastle}\ \emph {et~al.}(2021)\citenamefont
  {Hardcastle}, \citenamefont {Shimwell}, \citenamefont {Tasse}, \citenamefont
  {Best}, \citenamefont {Drabent}, \citenamefont {Jarvis}, \citenamefont
  {Prandoni}, \citenamefont {R\"ottgering}, \citenamefont {Sabater},\ and\
  \citenamefont {Schwarz}}]{Hardcastle:2020dfj}%
  \BibitemOpen
  \bibfield  {author} {\bibinfo {author} {\bibfnamefont {M.~J.}\ \bibnamefont
  {Hardcastle}}, \bibinfo {author} {\bibfnamefont {T.~W.}\ \bibnamefont
  {Shimwell}}, \bibinfo {author} {\bibfnamefont {C.}~\bibnamefont {Tasse}},
  \bibinfo {author} {\bibfnamefont {P.~N.}\ \bibnamefont {Best}}, \bibinfo
  {author} {\bibfnamefont {A.}~\bibnamefont {Drabent}}, \bibinfo {author}
  {\bibfnamefont {M.~J.}\ \bibnamefont {Jarvis}}, \bibinfo {author}
  {\bibfnamefont {I.}~\bibnamefont {Prandoni}}, \bibinfo {author}
  {\bibfnamefont {H.~J.~A.}\ \bibnamefont {R\"ottgering}}, \bibinfo {author}
  {\bibfnamefont {J.}~\bibnamefont {Sabater}},\ and\ \bibinfo {author}
  {\bibfnamefont {D.~J.}\ \bibnamefont {Schwarz}},\ }\bibfield  {title}
  {\bibinfo {title} {{The contribution of discrete sources to the sky
  temperature at 144 MHz}},\ }\href
  {https://doi.org/10.1051/0004-6361/202038814} {\bibfield  {journal} {\bibinfo
   {journal} {Astron. Astrophys.}\ }\textbf {\bibinfo {volume} {648}},\
  \bibinfo {pages} {A10} (\bibinfo {year} {2021})},\ \Eprint
  {https://arxiv.org/abs/2011.08294} {arXiv:2011.08294 [astro-ph.CO]}
  \BibitemShut {NoStop}%
\bibitem [{\citenamefont {Fang}\ and\ \citenamefont
  {Linden}(2016)}]{Fang:2015dga}%
  \BibitemOpen
  \bibfield  {author} {\bibinfo {author} {\bibfnamefont {K.}~\bibnamefont
  {Fang}}\ and\ \bibinfo {author} {\bibfnamefont {T.}~\bibnamefont {Linden}},\
  }\bibfield  {title} {\bibinfo {title} {{Cluster Mergers and the Origin of the
  ARCADE-2 Excess}},\ }\href {https://doi.org/10.1088/1475-7516/2016/10/004}
  {\bibfield  {journal} {\bibinfo  {journal} {JCAP}\ }\textbf {\bibinfo
  {volume} {10}},\ \bibinfo {pages} {004}},\ \Eprint
  {https://arxiv.org/abs/1506.05807} {arXiv:1506.05807 [astro-ph.HE]}
  \BibitemShut {NoStop}%
\bibitem [{\citenamefont {Subrahmanyan}\ and\ \citenamefont
  {Cowsik}(2013)}]{Subrahmanyan:2013eqa}%
  \BibitemOpen
  \bibfield  {author} {\bibinfo {author} {\bibfnamefont {R.}~\bibnamefont
  {Subrahmanyan}}\ and\ \bibinfo {author} {\bibfnamefont {R.}~\bibnamefont
  {Cowsik}},\ }\bibfield  {title} {\bibinfo {title} {{Is there an Unaccounted
  for Excess in the Extragalactic Cosmic Radio Background?}},\ }\href
  {https://doi.org/10.1088/0004-637X/776/1/42} {\bibfield  {journal} {\bibinfo
  {journal} {Astrophys. J.}\ }\textbf {\bibinfo {volume} {776}},\ \bibinfo
  {pages} {42} (\bibinfo {year} {2013})},\ \Eprint
  {https://arxiv.org/abs/1305.7060} {arXiv:1305.7060 [astro-ph.CO]}
  \BibitemShut {NoStop}%
\bibitem [{\citenamefont {Orlando}\ and\ \citenamefont
  {Strong}(2013)}]{Orlando:2013ysa}%
  \BibitemOpen
  \bibfield  {author} {\bibinfo {author} {\bibfnamefont {E.}~\bibnamefont
  {Orlando}}\ and\ \bibinfo {author} {\bibfnamefont {A.}~\bibnamefont
  {Strong}},\ }\bibfield  {title} {\bibinfo {title} {{Galactic synchrotron
  emission with cosmic ray propagation models}},\ }\href
  {https://doi.org/10.1093/mnras/stt1718} {\bibfield  {journal} {\bibinfo
  {journal} {Mon. Not. Roy. Astron. Soc.}\ }\textbf {\bibinfo {volume} {436}},\
  \bibinfo {pages} {2127} (\bibinfo {year} {2013})},\ \Eprint
  {https://arxiv.org/abs/1309.2947} {arXiv:1309.2947 [astro-ph.GA]}
  \BibitemShut {NoStop}%
\bibitem [{\citenamefont {Krause}\ and\ \citenamefont
  {Hardcastle}(2021)}]{Krause:2021xav}%
  \BibitemOpen
  \bibfield  {author} {\bibinfo {author} {\bibfnamefont {M.~G.~H.}\
  \bibnamefont {Krause}}\ and\ \bibinfo {author} {\bibfnamefont {M.~J.}\
  \bibnamefont {Hardcastle}},\ }\bibfield  {title} {\bibinfo {title} {{Can the
  Local Bubble explain the radio background?}},\ }\href
  {https://doi.org/10.1093/mnras/stab131} {\bibfield  {journal} {\bibinfo
  {journal} {Mon. Not. Roy. Astron. Soc.}\ }\textbf {\bibinfo {volume} {502}},\
  \bibinfo {pages} {2807} (\bibinfo {year} {2021})},\ \Eprint
  {https://arxiv.org/abs/2101.05255} {arXiv:2101.05255 [astro-ph.HE]}
  \BibitemShut {NoStop}%
\bibitem [{\citenamefont {Singal}\ \emph {et~al.}(2010)\citenamefont {Singal},
  \citenamefont {Stawarz}, \citenamefont {Lawrence},\ and\ \citenamefont
  {Petrosian}}]{Singal_2010}%
  \BibitemOpen
  \bibfield  {author} {\bibinfo {author} {\bibfnamefont {J.}~\bibnamefont
  {Singal}}, \bibinfo {author} {\bibfnamefont {L.}~\bibnamefont {Stawarz}},
  \bibinfo {author} {\bibfnamefont {A.}~\bibnamefont {Lawrence}},\ and\
  \bibinfo {author} {\bibfnamefont {V.}~\bibnamefont {Petrosian}},\ }\bibfield
  {title} {\bibinfo {title} {Sources of the radio background considered},\
  }\href {https://doi.org/10.1111/j.1365-2966.2010.17382.x} {\bibfield
  {journal} {\bibinfo  {journal} {Monthly Notices of the Royal Astronomical
  Society}\ }\textbf {\bibinfo {volume} {409}},\ \bibinfo {pages} {1172}
  (\bibinfo {year} {2010})}\BibitemShut {NoStop}%
\bibitem [{\citenamefont {Singal}\ \emph {et~al.}(2015)\citenamefont {Singal},
  \citenamefont {Kogut}, \citenamefont {Jones},\ and\ \citenamefont
  {Dunlap}}]{Singal:2015tta}%
  \BibitemOpen
  \bibfield  {author} {\bibinfo {author} {\bibfnamefont {J.}~\bibnamefont
  {Singal}}, \bibinfo {author} {\bibfnamefont {A.}~\bibnamefont {Kogut}},
  \bibinfo {author} {\bibfnamefont {E.}~\bibnamefont {Jones}},\ and\ \bibinfo
  {author} {\bibfnamefont {H.}~\bibnamefont {Dunlap}},\ }\bibfield  {title}
  {\bibinfo {title} {{Axial Ratio of Edge-on Spiral Galaxies as a Test for
  Bright Radio Halos}},\ }\href {https://doi.org/10.1088/2041-8205/799/1/L10}
  {\bibfield  {journal} {\bibinfo  {journal} {Astrophys. J. Lett.}\ }\textbf
  {\bibinfo {volume} {799}},\ \bibinfo {pages} {L10} (\bibinfo {year}
  {2015})},\ \Eprint {https://arxiv.org/abs/1501.00499} {arXiv:1501.00499
  [astro-ph.GA]} \BibitemShut {NoStop}%
\bibitem [{\citenamefont {Feng}\ and\ \citenamefont
  {Holder}(2018)}]{Feng:2018rje}%
  \BibitemOpen
  \bibfield  {author} {\bibinfo {author} {\bibfnamefont {C.}~\bibnamefont
  {Feng}}\ and\ \bibinfo {author} {\bibfnamefont {G.}~\bibnamefont {Holder}},\
  }\bibfield  {title} {\bibinfo {title} {{Enhanced global signal of neutral
  hydrogen due to excess radiation at cosmic dawn}},\ }\href
  {https://doi.org/10.3847/2041-8213/aac0fe} {\bibfield  {journal} {\bibinfo
  {journal} {Astrophys. J.}\ }\textbf {\bibinfo {volume} {858}},\ \bibinfo
  {pages} {L17} (\bibinfo {year} {2018})},\ \Eprint
  {https://arxiv.org/abs/1802.07432} {arXiv:1802.07432 [astro-ph.CO]}
  \BibitemShut {NoStop}%
%%CITATION = ARXIV:1802.07432;%%
\bibitem [{\citenamefont {Biermann}\ \emph {et~al.}(2014)\citenamefont
  {Biermann}, \citenamefont {Nath}, \citenamefont {Caramete}, \citenamefont
  {Harms}, \citenamefont {Stanev},\ and\ \citenamefont
  {Becker~Tjus}}]{Biermann:2014lna}%
  \BibitemOpen
  \bibfield  {author} {\bibinfo {author} {\bibfnamefont {P.~L.}\ \bibnamefont
  {Biermann}}, \bibinfo {author} {\bibfnamefont {B.~B.}\ \bibnamefont {Nath}},
  \bibinfo {author} {\bibfnamefont {L.~I.}\ \bibnamefont {Caramete}}, \bibinfo
  {author} {\bibfnamefont {B.~C.}\ \bibnamefont {Harms}}, \bibinfo {author}
  {\bibfnamefont {T.}~\bibnamefont {Stanev}},\ and\ \bibinfo {author}
  {\bibfnamefont {J.}~\bibnamefont {Becker~Tjus}},\ }\bibfield  {title}
  {\bibinfo {title} {{Cosmic backgrounds due to the formation of the first
  generation of supermassive black holes}},\ }\href
  {https://doi.org/10.1093/mnras/stu541} {\bibfield  {journal} {\bibinfo
  {journal} {Mon. Not. Roy. Astron. Soc.}\ }\textbf {\bibinfo {volume} {441}},\
  \bibinfo {pages} {1147} (\bibinfo {year} {2014})},\ \Eprint
  {https://arxiv.org/abs/1403.3804} {arXiv:1403.3804 [astro-ph.CO]}
  \BibitemShut {NoStop}%
\bibitem [{\citenamefont {Ewall-Wice}\ \emph {et~al.}(2018)\citenamefont
  {Ewall-Wice}, \citenamefont {Chang}, \citenamefont {Lazio}, \citenamefont
  {Dore}, \citenamefont {Seiffert},\ and\ \citenamefont
  {Monsalve}}]{Ewall-Wice:2018bzf}%
  \BibitemOpen
  \bibfield  {author} {\bibinfo {author} {\bibfnamefont {A.}~\bibnamefont
  {Ewall-Wice}}, \bibinfo {author} {\bibfnamefont {T.~C.}\ \bibnamefont
  {Chang}}, \bibinfo {author} {\bibfnamefont {J.}~\bibnamefont {Lazio}},
  \bibinfo {author} {\bibfnamefont {O.}~\bibnamefont {Dore}}, \bibinfo {author}
  {\bibfnamefont {M.}~\bibnamefont {Seiffert}},\ and\ \bibinfo {author}
  {\bibfnamefont {R.~A.}\ \bibnamefont {Monsalve}},\ }\bibfield  {title}
  {\bibinfo {title} {{Modeling the Radio Background from the First Black Holes
  at Cosmic Dawn: Implications for the 21 cm Absorption Amplitude}},\ }\href
  {https://doi.org/10.3847/1538-4357/aae51d} {\bibfield  {journal} {\bibinfo
  {journal} {Astrophys. J.}\ }\textbf {\bibinfo {volume} {868}},\ \bibinfo
  {pages} {63} (\bibinfo {year} {2018})},\ \Eprint
  {https://arxiv.org/abs/1803.01815} {arXiv:1803.01815 [astro-ph.CO]}
  \BibitemShut {NoStop}%
\bibitem [{\citenamefont {Ewall-Wice}\ \emph {et~al.}(2020)\citenamefont
  {Ewall-Wice}, \citenamefont {Chang},\ and\ \citenamefont
  {Lazio}}]{Ewall-Wice:2019may}%
  \BibitemOpen
  \bibfield  {author} {\bibinfo {author} {\bibfnamefont {A.}~\bibnamefont
  {Ewall-Wice}}, \bibinfo {author} {\bibfnamefont {T.-C.}\ \bibnamefont
  {Chang}},\ and\ \bibinfo {author} {\bibfnamefont {T.~J.~W.}\ \bibnamefont
  {Lazio}},\ }\bibfield  {title} {\bibinfo {title} {{The Radio Scream from
  black holes at Cosmic Dawn: a semi-analytic model for the impact of
  radio-loud black holes on the 21 cm global signal}},\ }\href
  {https://doi.org/10.1093/mnras/stz3501} {\bibfield  {journal} {\bibinfo
  {journal} {Mon. Not. Roy. Astron. Soc.}\ }\textbf {\bibinfo {volume} {492}},\
  \bibinfo {pages} {6086} (\bibinfo {year} {2020})},\ \Eprint
  {https://arxiv.org/abs/1903.06788} {arXiv:1903.06788 [astro-ph.GA]}
  \BibitemShut {NoStop}%
\bibitem [{\citenamefont {Mirocha}\ and\ \citenamefont
  {Furlanetto}(2019)}]{Mirocha:2018cih}%
  \BibitemOpen
  \bibfield  {author} {\bibinfo {author} {\bibfnamefont {J.}~\bibnamefont
  {Mirocha}}\ and\ \bibinfo {author} {\bibfnamefont {S.~R.}\ \bibnamefont
  {Furlanetto}},\ }\bibfield  {title} {\bibinfo {title} {{What does the first
  highly-redshifted 21-cm detection tell us about early galaxies?}},\ }\href
  {https://doi.org/10.1093/mnras/sty3260} {\bibfield  {journal} {\bibinfo
  {journal} {Mon. Not. Roy. Astron. Soc.}\ }\textbf {\bibinfo {volume} {483}},\
  \bibinfo {pages} {1980} (\bibinfo {year} {2019})},\ \Eprint
  {https://arxiv.org/abs/1803.03272} {arXiv:1803.03272 [astro-ph.GA]}
  \BibitemShut {NoStop}%
%%CITATION = ARXIV:1803.03272;%%
\bibitem [{\citenamefont {Sharma}(2018)}]{Sharma:2018agu}%
  \BibitemOpen
  \bibfield  {author} {\bibinfo {author} {\bibfnamefont {P.}~\bibnamefont
  {Sharma}},\ }\bibfield  {title} {\bibinfo {title} {{Astrophysical radio
  background cannot explain the EDGES 21-cm signal: constraints from cooling of
  non-thermal electrons}},\ }\href {https://doi.org/10.1093/mnrasl/sly147}
  {\bibfield  {journal} {\bibinfo  {journal} {Mon. Not. Roy. Astron. Soc.}\
  }\textbf {\bibinfo {volume} {481}},\ \bibinfo {pages} {L6} (\bibinfo {year}
  {2018})},\ \Eprint {https://arxiv.org/abs/1804.05843} {arXiv:1804.05843
  [astro-ph.HE]} \BibitemShut {NoStop}%
\bibitem [{\citenamefont {Bowman}\ \emph {et~al.}(2018)\citenamefont {Bowman},
  \citenamefont {Rogers}, \citenamefont {Monsalve}, \citenamefont {Mozdzen},\
  and\ \citenamefont {Mahesh}}]{Bowman:2018yin}%
  \BibitemOpen
  \bibfield  {author} {\bibinfo {author} {\bibfnamefont {J.~D.}\ \bibnamefont
  {Bowman}}, \bibinfo {author} {\bibfnamefont {A.~E.~E.}\ \bibnamefont
  {Rogers}}, \bibinfo {author} {\bibfnamefont {R.~A.}\ \bibnamefont
  {Monsalve}}, \bibinfo {author} {\bibfnamefont {T.~J.}\ \bibnamefont
  {Mozdzen}},\ and\ \bibinfo {author} {\bibfnamefont {N.}~\bibnamefont
  {Mahesh}},\ }\bibfield  {title} {\bibinfo {title} {{An absorption profile
  centred at 78 megahertz in the sky-averaged spectrum}},\ }\href
  {https://doi.org/10.1038/nature25792} {\bibfield  {journal} {\bibinfo
  {journal} {Nature}\ }\textbf {\bibinfo {volume} {555}},\ \bibinfo {pages}
  {67} (\bibinfo {year} {2018})},\ \Eprint {https://arxiv.org/abs/1810.05912}
  {arXiv:1810.05912 [astro-ph.CO]} \BibitemShut {NoStop}%
%%CITATION = ARXIV:1810.05912;%%
\bibitem [{\citenamefont {Bernardi}\ \emph {et~al.}(2016)\citenamefont
  {Bernardi}, \citenamefont {Zwart}, \citenamefont {Price}, \citenamefont
  {Greenhill}, \citenamefont {Mesinger}, \citenamefont {Dowell}, \citenamefont
  {Eftekhari}, \citenamefont {Ellingson}, \citenamefont {Kocz},\ and\
  \citenamefont {Schinzel}}]{Bernardi:2016pva}%
  \BibitemOpen
  \bibfield  {author} {\bibinfo {author} {\bibfnamefont {G.}~\bibnamefont
  {Bernardi}}, \bibinfo {author} {\bibfnamefont {J.~T.~L.}\ \bibnamefont
  {Zwart}}, \bibinfo {author} {\bibfnamefont {D.}~\bibnamefont {Price}},
  \bibinfo {author} {\bibfnamefont {L.~J.}\ \bibnamefont {Greenhill}}, \bibinfo
  {author} {\bibfnamefont {A.}~\bibnamefont {Mesinger}}, \bibinfo {author}
  {\bibfnamefont {J.}~\bibnamefont {Dowell}}, \bibinfo {author} {\bibfnamefont
  {T.}~\bibnamefont {Eftekhari}}, \bibinfo {author} {\bibfnamefont {S.~W.}\
  \bibnamefont {Ellingson}}, \bibinfo {author} {\bibfnamefont {J.}~\bibnamefont
  {Kocz}},\ and\ \bibinfo {author} {\bibfnamefont {F.}~\bibnamefont
  {Schinzel}},\ }\bibfield  {title} {\bibinfo {title} {{Bayesian constraints on
  the global 21-cm signal from the Cosmic Dawn}},\ }\href
  {https://doi.org/10.1093/mnras/stw1499} {\bibfield  {journal} {\bibinfo
  {journal} {Mon. Not. Roy. Astron. Soc.}\ }\textbf {\bibinfo {volume} {461}},\
  \bibinfo {pages} {2847} (\bibinfo {year} {2016})},\ \Eprint
  {https://arxiv.org/abs/1606.06006} {arXiv:1606.06006 [astro-ph.CO]}
  \BibitemShut {NoStop}%
\bibitem [{\citenamefont {Mertens}\ \emph {et~al.}(2020)\citenamefont {Mertens}
  \emph {et~al.}}]{Mertens:2020llj}%
  \BibitemOpen
  \bibfield  {author} {\bibinfo {author} {\bibfnamefont {F.~G.}\ \bibnamefont
  {Mertens}} \emph {et~al.},\ }\bibfield  {title} {\bibinfo {title} {{Improved
  upper limits on the 21-cm signal power spectrum of neutral hydrogen at
  $\boldsymbol{z \approx 9.1}$ from LOFAR}},\ }\href
  {https://doi.org/10.1093/mnras/staa327} {\bibfield  {journal} {\bibinfo
  {journal} {Mon. Not. Roy. Astron. Soc.}\ }\textbf {\bibinfo {volume} {493}},\
  \bibinfo {pages} {1662} (\bibinfo {year} {2020})},\ \Eprint
  {https://arxiv.org/abs/2002.07196} {arXiv:2002.07196 [astro-ph.CO]}
  \BibitemShut {NoStop}%
\bibitem [{\citenamefont {Fialkov}\ and\ \citenamefont
  {Barkana}(2019)}]{Fialkov:2019vnb}%
  \BibitemOpen
  \bibfield  {author} {\bibinfo {author} {\bibfnamefont {A.}~\bibnamefont
  {Fialkov}}\ and\ \bibinfo {author} {\bibfnamefont {R.}~\bibnamefont
  {Barkana}},\ }\bibfield  {title} {\bibinfo {title} {{Signature of Excess
  Radio Background in the 21-cm Global Signal and Power Spectrum}},\ }\href
  {https://doi.org/10.1093/mnras/stz873} {\bibfield  {journal} {\bibinfo
  {journal} {Mon. Not. Roy. Astron. Soc.}\ }\textbf {\bibinfo {volume} {486}},\
  \bibinfo {pages} {1763} (\bibinfo {year} {2019})},\ \Eprint
  {https://arxiv.org/abs/1902.02438} {arXiv:1902.02438 [astro-ph.CO]}
  \BibitemShut {NoStop}%
%%CITATION = ARXIV:1902.02438;%%
\bibitem [{\citenamefont {Holder}(2013)}]{Holder_2013}%
  \BibitemOpen
  \bibfield  {author} {\bibinfo {author} {\bibfnamefont {G.~P.}\ \bibnamefont
  {Holder}},\ }\bibfield  {title} {\bibinfo {title} {The unusual smoothness of
  the extragalactic unresolved radio background},\ }\href
  {https://doi.org/10.1088/0004-637x/780/1/112} {\bibfield  {journal} {\bibinfo
   {journal} {The Astrophysical Journal}\ }\textbf {\bibinfo {volume} {780}},\
  \bibinfo {pages} {112} (\bibinfo {year} {2013})}\BibitemShut {NoStop}%
\bibitem [{\citenamefont {{Fomalont}}\ \emph {et~al.}(1988)\citenamefont
  {{Fomalont}}, \citenamefont {{Kellermann}}, \citenamefont {{Anderson}},
  \citenamefont {{Weistrop}}, \citenamefont {{Wall}}, \citenamefont
  {{Windhorst}},\ and\ \citenamefont {{Kristian}}}]{1988AJ.....96.1187F}%
  \BibitemOpen
  \bibfield  {author} {\bibinfo {author} {\bibfnamefont {E.~B.}\ \bibnamefont
  {{Fomalont}}}, \bibinfo {author} {\bibfnamefont {K.~I.}\ \bibnamefont
  {{Kellermann}}}, \bibinfo {author} {\bibfnamefont {M.~C.}\ \bibnamefont
  {{Anderson}}}, \bibinfo {author} {\bibfnamefont {D.}~\bibnamefont
  {{Weistrop}}}, \bibinfo {author} {\bibfnamefont {J.~V.}\ \bibnamefont
  {{Wall}}}, \bibinfo {author} {\bibfnamefont {R.~A.}\ \bibnamefont
  {{Windhorst}}},\ and\ \bibinfo {author} {\bibfnamefont {J.~A.}\ \bibnamefont
  {{Kristian}}},\ }\bibfield  {title} {\bibinfo {title} {{New Limits to
  Fluctuations in the Cosmic Background Radiation at 4.86 GHz Between 12 and 60
  Arcsecond Resolution}},\ }\href {https://doi.org/10.1086/114872} {\bibfield
  {journal} {\bibinfo  {journal} {\aj}\ }\textbf {\bibinfo {volume} {96}},\
  \bibinfo {pages} {1187} (\bibinfo {year} {1988})}\BibitemShut {NoStop}%
\bibitem [{\citenamefont {{Partridge}}\ \emph {et~al.}(1997)\citenamefont
  {{Partridge}}, \citenamefont {{Richards}}, \citenamefont {{Fomalont}},
  \citenamefont {{Kellerman}},\ and\ \citenamefont
  {{Windhorst}}}]{1997ApJ...483...38P}%
  \BibitemOpen
  \bibfield  {author} {\bibinfo {author} {\bibfnamefont {R.~B.}\ \bibnamefont
  {{Partridge}}}, \bibinfo {author} {\bibfnamefont {E.~A.}\ \bibnamefont
  {{Richards}}}, \bibinfo {author} {\bibfnamefont {E.~B.}\ \bibnamefont
  {{Fomalont}}}, \bibinfo {author} {\bibfnamefont {K.~I.}\ \bibnamefont
  {{Kellerman}}},\ and\ \bibinfo {author} {\bibfnamefont {R.~A.}\ \bibnamefont
  {{Windhorst}}},\ }\bibfield  {title} {\bibinfo {title} {{Small-Scale Cosmic
  Microwave Background Observations at 8.4 GHz}},\ }\href
  {https://doi.org/10.1086/304226} {\bibfield  {journal} {\bibinfo  {journal}
  {\apj}\ }\textbf {\bibinfo {volume} {483}},\ \bibinfo {pages} {38} (\bibinfo
  {year} {1997})}\BibitemShut {NoStop}%
\bibitem [{\citenamefont {Subrahmanyan}\ \emph {et~al.}(2000)\citenamefont
  {Subrahmanyan}, \citenamefont {Kesteven}, \citenamefont {Ekers},
  \citenamefont {Sinclair},\ and\ \citenamefont {Silk}}]{Subrahmanyan:2000df}%
  \BibitemOpen
  \bibfield  {author} {\bibinfo {author} {\bibfnamefont {R.}~\bibnamefont
  {Subrahmanyan}}, \bibinfo {author} {\bibfnamefont {M.~J.}\ \bibnamefont
  {Kesteven}}, \bibinfo {author} {\bibfnamefont {R.~D.}\ \bibnamefont {Ekers}},
  \bibinfo {author} {\bibfnamefont {M.}~\bibnamefont {Sinclair}},\ and\
  \bibinfo {author} {\bibfnamefont {J.}~\bibnamefont {Silk}},\ }\bibfield
  {title} {\bibinfo {title} {{An Australia telescope survey for CMB
  anisotropies}},\ }\href {https://doi.org/10.1046/j.1365-8711.2000.03444.x}
  {\bibfield  {journal} {\bibinfo  {journal} {Mon. Not. Roy. Astron. Soc.}\
  }\textbf {\bibinfo {volume} {315}},\ \bibinfo {pages} {808} (\bibinfo {year}
  {2000})},\ \Eprint {https://arxiv.org/abs/astro-ph/0002467}
  {arXiv:astro-ph/0002467} \BibitemShut {NoStop}%
\bibitem [{\citenamefont {Choudhuri}\ \emph {et~al.}(2020)\citenamefont
  {Choudhuri}, \citenamefont {Ghosh}, \citenamefont {Roy}, \citenamefont
  {Bharadwaj}, \citenamefont {Intema},\ and\ \citenamefont
  {Ali}}]{Choudhuri:2020dgd}%
  \BibitemOpen
  \bibfield  {author} {\bibinfo {author} {\bibfnamefont {S.}~\bibnamefont
  {Choudhuri}}, \bibinfo {author} {\bibfnamefont {A.}~\bibnamefont {Ghosh}},
  \bibinfo {author} {\bibfnamefont {N.}~\bibnamefont {Roy}}, \bibinfo {author}
  {\bibfnamefont {S.}~\bibnamefont {Bharadwaj}}, \bibinfo {author}
  {\bibfnamefont {H.~T.}\ \bibnamefont {Intema}},\ and\ \bibinfo {author}
  {\bibfnamefont {S.~S.}\ \bibnamefont {Ali}},\ }\bibfield  {title} {\bibinfo
  {title} {{All-sky angular power spectrum \textendash{} I. Estimating
  brightness temperature fluctuations using the 150-MHz TGSS survey}},\ }\href
  {https://doi.org/10.1093/mnras/staa762} {\bibfield  {journal} {\bibinfo
  {journal} {Mon. Not. Roy. Astron. Soc.}\ }\textbf {\bibinfo {volume} {494}},\
  \bibinfo {pages} {1936} (\bibinfo {year} {2020})},\ \Eprint
  {https://arxiv.org/abs/2003.07869} {arXiv:2003.07869 [astro-ph.CO]}
  \BibitemShut {NoStop}%
\bibitem [{\citenamefont {Offringa}\ \emph {et~al.}(2021)\citenamefont
  {Offringa}, \citenamefont {Singal}, \citenamefont {Heston}, \citenamefont
  {Horiuchi},\ and\ \citenamefont {Lucero}}]{Offringa:2021rwp}%
  \BibitemOpen
  \bibfield  {author} {\bibinfo {author} {\bibfnamefont {A.~R.}\ \bibnamefont
  {Offringa}}, \bibinfo {author} {\bibfnamefont {J.}~\bibnamefont {Singal}},
  \bibinfo {author} {\bibfnamefont {S.}~\bibnamefont {Heston}}, \bibinfo
  {author} {\bibfnamefont {S.}~\bibnamefont {Horiuchi}},\ and\ \bibinfo
  {author} {\bibfnamefont {D.~M.}\ \bibnamefont {Lucero}},\ }\bibfield  {title}
  {\bibinfo {title} {{Measurement of the anisotropy power spectrum of the radio
  synchrotron background}},\ }\href {https://doi.org/10.1093/mnras/stab2865}
  {\bibfield  {journal} {\bibinfo  {journal} {Mon. Not. Roy. Astron. Soc.}\
  }\textbf {\bibinfo {volume} {509}},\ \bibinfo {pages} {114} (\bibinfo {year}
  {2021})},\ \Eprint {https://arxiv.org/abs/2110.00499} {arXiv:2110.00499
  [astro-ph.CO]} \BibitemShut {NoStop}%
\bibitem [{\citenamefont {Fornengo}\ \emph {et~al.}(2011)\citenamefont
  {Fornengo}, \citenamefont {Lineros}, \citenamefont {Regis},\ and\
  \citenamefont {Taoso}}]{Fornengo:2011cn}%
  \BibitemOpen
  \bibfield  {author} {\bibinfo {author} {\bibfnamefont {N.}~\bibnamefont
  {Fornengo}}, \bibinfo {author} {\bibfnamefont {R.}~\bibnamefont {Lineros}},
  \bibinfo {author} {\bibfnamefont {M.}~\bibnamefont {Regis}},\ and\ \bibinfo
  {author} {\bibfnamefont {M.}~\bibnamefont {Taoso}},\ }\bibfield  {title}
  {\bibinfo {title} {{Possibility of a Dark Matter Interpretation for the
  Excess in Isotropic Radio Emission Reported by ARCADE}},\ }\href
  {https://doi.org/10.1103/PhysRevLett.107.271302} {\bibfield  {journal}
  {\bibinfo  {journal} {Phys. Rev. Lett.}\ }\textbf {\bibinfo {volume} {107}},\
  \bibinfo {pages} {271302} (\bibinfo {year} {2011})},\ \Eprint
  {https://arxiv.org/abs/1108.0569} {arXiv:1108.0569 [hep-ph]} \BibitemShut
  {NoStop}%
\bibitem [{\citenamefont {Hooper}\ \emph {et~al.}(2012)\citenamefont {Hooper},
  \citenamefont {Belikov}, \citenamefont {Jeltema}, \citenamefont {Linden},
  \citenamefont {Profumo},\ and\ \citenamefont {Slatyer}}]{Hooper:2012jc}%
  \BibitemOpen
  \bibfield  {author} {\bibinfo {author} {\bibfnamefont {D.}~\bibnamefont
  {Hooper}}, \bibinfo {author} {\bibfnamefont {A.~V.}\ \bibnamefont {Belikov}},
  \bibinfo {author} {\bibfnamefont {T.~E.}\ \bibnamefont {Jeltema}}, \bibinfo
  {author} {\bibfnamefont {T.}~\bibnamefont {Linden}}, \bibinfo {author}
  {\bibfnamefont {S.}~\bibnamefont {Profumo}},\ and\ \bibinfo {author}
  {\bibfnamefont {T.~R.}\ \bibnamefont {Slatyer}},\ }\bibfield  {title}
  {\bibinfo {title} {{The Isotropic Radio Background and Annihilating Dark
  Matter}},\ }\href {https://doi.org/10.1103/PhysRevD.86.103003} {\bibfield
  {journal} {\bibinfo  {journal} {Phys. Rev. D}\ }\textbf {\bibinfo {volume}
  {86}},\ \bibinfo {pages} {103003} (\bibinfo {year} {2012})},\ \Eprint
  {https://arxiv.org/abs/1203.3547} {arXiv:1203.3547 [astro-ph.CO]}
  \BibitemShut {NoStop}%
\bibitem [{\citenamefont {Cline}\ and\ \citenamefont
  {Vincent}(2013)}]{Cline:2012hb}%
  \BibitemOpen
  \bibfield  {author} {\bibinfo {author} {\bibfnamefont {J.~M.}\ \bibnamefont
  {Cline}}\ and\ \bibinfo {author} {\bibfnamefont {A.~C.}\ \bibnamefont
  {Vincent}},\ }\bibfield  {title} {\bibinfo {title} {{Cosmological origin of
  anomalous radio background}},\ }\href
  {https://doi.org/10.1088/1475-7516/2013/02/011} {\bibfield  {journal}
  {\bibinfo  {journal} {JCAP}\ }\textbf {\bibinfo {volume} {02}},\ \bibinfo
  {pages} {011}},\ \Eprint {https://arxiv.org/abs/1210.2717} {arXiv:1210.2717
  [astro-ph.CO]} \BibitemShut {NoStop}%
\bibitem [{\citenamefont {Fang}\ and\ \citenamefont
  {Linden}(2015)}]{Fang:2014joa}%
  \BibitemOpen
  \bibfield  {author} {\bibinfo {author} {\bibfnamefont {K.}~\bibnamefont
  {Fang}}\ and\ \bibinfo {author} {\bibfnamefont {T.}~\bibnamefont {Linden}},\
  }\bibfield  {title} {\bibinfo {title} {{Anisotropy of the extragalactic radio
  background from dark matter annihilation}},\ }\href
  {https://doi.org/10.1103/PhysRevD.91.083501} {\bibfield  {journal} {\bibinfo
  {journal} {Phys. Rev. D}\ }\textbf {\bibinfo {volume} {91}},\ \bibinfo
  {pages} {083501} (\bibinfo {year} {2015})},\ \Eprint
  {https://arxiv.org/abs/1412.7545} {arXiv:1412.7545 [astro-ph.HE]}
  \BibitemShut {NoStop}%
\bibitem [{\citenamefont {Ajello}\ \emph {et~al.}(2015)\citenamefont {Ajello}
  \emph {et~al.}}]{Ajello:2015mfa}%
  \BibitemOpen
  \bibfield  {author} {\bibinfo {author} {\bibfnamefont {M.}~\bibnamefont
  {Ajello}} \emph {et~al.},\ }\bibfield  {title} {\bibinfo {title} {{The Origin
  of the Extragalactic Gamma-Ray Background and Implications for Dark-Matter
  Annihilation}},\ }\href {https://doi.org/10.1088/2041-8205/800/2/L27}
  {\bibfield  {journal} {\bibinfo  {journal} {Astrophys. J. Lett.}\ }\textbf
  {\bibinfo {volume} {800}},\ \bibinfo {pages} {L27} (\bibinfo {year}
  {2015})},\ \Eprint {https://arxiv.org/abs/1501.05301} {arXiv:1501.05301
  [astro-ph.HE]} \BibitemShut {NoStop}%
\bibitem [{\citenamefont {Lisanti}\ \emph {et~al.}(2016)\citenamefont
  {Lisanti}, \citenamefont {Mishra-Sharma}, \citenamefont {Necib},\ and\
  \citenamefont {Safdi}}]{Lisanti:2016jub}%
  \BibitemOpen
  \bibfield  {author} {\bibinfo {author} {\bibfnamefont {M.}~\bibnamefont
  {Lisanti}}, \bibinfo {author} {\bibfnamefont {S.}~\bibnamefont
  {Mishra-Sharma}}, \bibinfo {author} {\bibfnamefont {L.}~\bibnamefont
  {Necib}},\ and\ \bibinfo {author} {\bibfnamefont {B.~R.}\ \bibnamefont
  {Safdi}},\ }\bibfield  {title} {\bibinfo {title} {{Deciphering Contributions
  to the Extragalactic Gamma-Ray Background from 2 GeV to 2 TeV}},\ }\href
  {https://doi.org/10.3847/0004-637X/832/2/117} {\bibfield  {journal} {\bibinfo
   {journal} {Astrophys. J.}\ }\textbf {\bibinfo {volume} {832}},\ \bibinfo
  {pages} {117} (\bibinfo {year} {2016})},\ \Eprint
  {https://arxiv.org/abs/1606.04101} {arXiv:1606.04101 [astro-ph.HE]}
  \BibitemShut {NoStop}%
\bibitem [{\citenamefont {Vogel}\ and\ \citenamefont
  {Redondo}(2014)}]{Vogel:2013raa}%
  \BibitemOpen
  \bibfield  {author} {\bibinfo {author} {\bibfnamefont {H.}~\bibnamefont
  {Vogel}}\ and\ \bibinfo {author} {\bibfnamefont {J.}~\bibnamefont
  {Redondo}},\ }\bibfield  {title} {\bibinfo {title} {{Dark Radiation
  constraints on minicharged particles in models with a hidden photon}},\
  }\href {https://doi.org/10.1088/1475-7516/2014/02/029} {\bibfield  {journal}
  {\bibinfo  {journal} {JCAP}\ }\textbf {\bibinfo {volume} {02}},\ \bibinfo
  {pages} {029}},\ \Eprint {https://arxiv.org/abs/1311.2600} {arXiv:1311.2600
  [hep-ph]} \BibitemShut {NoStop}%
\bibitem [{\citenamefont {Caputo}\ \emph {et~al.}(2019)\citenamefont {Caputo},
  \citenamefont {Regis}, \citenamefont {Taoso},\ and\ \citenamefont
  {Witte}}]{Caputo:2018vmy}%
  \BibitemOpen
  \bibfield  {author} {\bibinfo {author} {\bibfnamefont {A.}~\bibnamefont
  {Caputo}}, \bibinfo {author} {\bibfnamefont {M.}~\bibnamefont {Regis}},
  \bibinfo {author} {\bibfnamefont {M.}~\bibnamefont {Taoso}},\ and\ \bibinfo
  {author} {\bibfnamefont {S.~J.}\ \bibnamefont {Witte}},\ }\bibfield  {title}
  {\bibinfo {title} {{Detecting the Stimulated Decay of Axions at
  RadioFrequencies}},\ }\href {https://doi.org/10.1088/1475-7516/2019/03/027}
  {\bibfield  {journal} {\bibinfo  {journal} {JCAP}\ }\textbf {\bibinfo
  {volume} {1903}}\bibfield  {number} {\bibinfo  {number} { (03)},\ \bibinfo
  {pages} {027}},\ }\Eprint {https://arxiv.org/abs/1811.08436}
  {arXiv:1811.08436 [hep-ph]} \BibitemShut {NoStop}%
%%CITATION = ARXIV:1811.08436;%%
\bibitem [{\citenamefont {Bolliet}\ \emph {et~al.}(2021)\citenamefont
  {Bolliet}, \citenamefont {Chluba},\ and\ \citenamefont
  {Battye}}]{Bolliet:2020ofj}%
  \BibitemOpen
  \bibfield  {author} {\bibinfo {author} {\bibfnamefont {B.}~\bibnamefont
  {Bolliet}}, \bibinfo {author} {\bibfnamefont {J.}~\bibnamefont {Chluba}},\
  and\ \bibinfo {author} {\bibfnamefont {R.}~\bibnamefont {Battye}},\
  }\bibfield  {title} {\bibinfo {title} {{Spectral distortion constraints on
  photon injection from low-mass decaying particles}},\ }\href
  {https://doi.org/10.1093/mnras/stab1997} {\bibfield  {journal} {\bibinfo
  {journal} {Mon. Not. Roy. Astron. Soc.}\ }\textbf {\bibinfo {volume} {507}},\
  \bibinfo {pages} {3148} (\bibinfo {year} {2021})},\ \Eprint
  {https://arxiv.org/abs/2012.07292} {arXiv:2012.07292 [astro-ph.CO]}
  \BibitemShut {NoStop}%
\bibitem [{\citenamefont {Mirizzi}\ \emph {et~al.}(2009)\citenamefont
  {Mirizzi}, \citenamefont {Redondo},\ and\ \citenamefont
  {Sigl}}]{Mirizzi:2009iz}%
  \BibitemOpen
  \bibfield  {author} {\bibinfo {author} {\bibfnamefont {A.}~\bibnamefont
  {Mirizzi}}, \bibinfo {author} {\bibfnamefont {J.}~\bibnamefont {Redondo}},\
  and\ \bibinfo {author} {\bibfnamefont {G.}~\bibnamefont {Sigl}},\ }\bibfield
  {title} {\bibinfo {title} {{Microwave Background Constraints on Mixing of
  Photons with Hidden Photons}},\ }\href
  {https://doi.org/10.1088/1475-7516/2009/03/026} {\bibfield  {journal}
  {\bibinfo  {journal} {JCAP}\ }\textbf {\bibinfo {volume} {0903}},\ \bibinfo
  {pages} {026}},\ \Eprint {https://arxiv.org/abs/0901.0014} {arXiv:0901.0014
  [hep-ph]} \BibitemShut {NoStop}%
%%CITATION = ARXIV:0901.0014;%%
\bibitem [{\citenamefont {Caputo}\ \emph
  {et~al.}(2020{\natexlab{a}})\citenamefont {Caputo}, \citenamefont {Liu},
  \citenamefont {Mishra-Sharma},\ and\ \citenamefont
  {Ruderman}}]{Caputo:2020rnx}%
  \BibitemOpen
  \bibfield  {author} {\bibinfo {author} {\bibfnamefont {A.}~\bibnamefont
  {Caputo}}, \bibinfo {author} {\bibfnamefont {H.}~\bibnamefont {Liu}},
  \bibinfo {author} {\bibfnamefont {S.}~\bibnamefont {Mishra-Sharma}},\ and\
  \bibinfo {author} {\bibfnamefont {J.~T.}\ \bibnamefont {Ruderman}},\
  }\bibfield  {title} {\bibinfo {title} {{Modeling Dark Photon Oscillations in
  Our Inhomogeneous Universe}},\ }\href
  {https://doi.org/10.1103/PhysRevD.102.103533} {\bibfield  {journal} {\bibinfo
   {journal} {Phys. Rev. D}\ }\textbf {\bibinfo {volume} {102}},\ \bibinfo
  {pages} {103533} (\bibinfo {year} {2020}{\natexlab{a}})},\ \Eprint
  {https://arxiv.org/abs/2004.06733} {arXiv:2004.06733 [astro-ph.CO]}
  \BibitemShut {NoStop}%
\bibitem [{\citenamefont {Caputo}\ \emph
  {et~al.}(2020{\natexlab{b}})\citenamefont {Caputo}, \citenamefont {Liu},
  \citenamefont {Mishra-Sharma},\ and\ \citenamefont
  {Ruderman}}]{Caputo:2020bdy}%
  \BibitemOpen
  \bibfield  {author} {\bibinfo {author} {\bibfnamefont {A.}~\bibnamefont
  {Caputo}}, \bibinfo {author} {\bibfnamefont {H.}~\bibnamefont {Liu}},
  \bibinfo {author} {\bibfnamefont {S.}~\bibnamefont {Mishra-Sharma}},\ and\
  \bibinfo {author} {\bibfnamefont {J.~T.}\ \bibnamefont {Ruderman}},\
  }\bibfield  {title} {\bibinfo {title} {{Dark Photon Oscillations in Our
  Inhomogeneous Universe}},\ }\href
  {https://doi.org/10.1103/PhysRevLett.125.221303} {\bibfield  {journal}
  {\bibinfo  {journal} {Phys. Rev. Lett.}\ }\textbf {\bibinfo {volume} {125}},\
  \bibinfo {pages} {221303} (\bibinfo {year} {2020}{\natexlab{b}})},\ \Eprint
  {https://arxiv.org/abs/2002.05165} {arXiv:2002.05165 [astro-ph.CO]}
  \BibitemShut {NoStop}%
\bibitem [{\citenamefont {Bondarenko}\ \emph {et~al.}(2020)\citenamefont
  {Bondarenko}, \citenamefont {Pradler},\ and\ \citenamefont
  {Sokolenko}}]{Bondarenko:2020moh}%
  \BibitemOpen
  \bibfield  {author} {\bibinfo {author} {\bibfnamefont {K.}~\bibnamefont
  {Bondarenko}}, \bibinfo {author} {\bibfnamefont {J.}~\bibnamefont
  {Pradler}},\ and\ \bibinfo {author} {\bibfnamefont {A.}~\bibnamefont
  {Sokolenko}},\ }\bibfield  {title} {\bibinfo {title} {{Constraining dark
  photons and their connection to 21 cm cosmology with CMB data}},\ }\href
  {https://doi.org/10.1016/j.physletb.2020.135420} {\bibfield  {journal}
  {\bibinfo  {journal} {Phys. Lett. B}\ }\textbf {\bibinfo {volume} {805}},\
  \bibinfo {pages} {135420} (\bibinfo {year} {2020})},\ \Eprint
  {https://arxiv.org/abs/2002.08942} {arXiv:2002.08942 [astro-ph.CO]}
  \BibitemShut {NoStop}%
\bibitem [{\citenamefont {Garcia}\ \emph {et~al.}(2020)\citenamefont {Garcia},
  \citenamefont {Bondarenko}, \citenamefont {Ploeckinger}, \citenamefont
  {Pradler},\ and\ \citenamefont {Sokolenko}}]{Garcia:2020qrp}%
  \BibitemOpen
  \bibfield  {author} {\bibinfo {author} {\bibfnamefont {A.~A.}\ \bibnamefont
  {Garcia}}, \bibinfo {author} {\bibfnamefont {K.}~\bibnamefont {Bondarenko}},
  \bibinfo {author} {\bibfnamefont {S.}~\bibnamefont {Ploeckinger}}, \bibinfo
  {author} {\bibfnamefont {J.}~\bibnamefont {Pradler}},\ and\ \bibinfo {author}
  {\bibfnamefont {A.}~\bibnamefont {Sokolenko}},\ }\bibfield  {title} {\bibinfo
  {title} {{Effective photon mass and (dark) photon conversion in the
  inhomogeneous Universe}},\ }\href
  {https://doi.org/10.1088/1475-7516/2020/10/011} {\bibfield  {journal}
  {\bibinfo  {journal} {JCAP}\ }\textbf {\bibinfo {volume} {10}},\ \bibinfo
  {pages} {011}},\ \Eprint {https://arxiv.org/abs/2003.10465} {arXiv:2003.10465
  [astro-ph.CO]} \BibitemShut {NoStop}%
\bibitem [{\citenamefont {Witte}\ \emph {et~al.}(2020)\citenamefont {Witte},
  \citenamefont {Rosauro-Alcaraz}, \citenamefont {McDermott},\ and\
  \citenamefont {Poulin}}]{Witte:2020rvb}%
  \BibitemOpen
  \bibfield  {author} {\bibinfo {author} {\bibfnamefont {S.~J.}\ \bibnamefont
  {Witte}}, \bibinfo {author} {\bibfnamefont {S.}~\bibnamefont
  {Rosauro-Alcaraz}}, \bibinfo {author} {\bibfnamefont {S.~D.}\ \bibnamefont
  {McDermott}},\ and\ \bibinfo {author} {\bibfnamefont {V.}~\bibnamefont
  {Poulin}},\ }\bibfield  {title} {\bibinfo {title} {{Dark photon dark matter
  in the presence of inhomogeneous structure}},\ }\href
  {https://doi.org/10.1007/JHEP06(2020)132} {\bibfield  {journal} {\bibinfo
  {journal} {JHEP}\ }\textbf {\bibinfo {volume} {06}},\ \bibinfo {pages}
  {132}},\ \Eprint {https://arxiv.org/abs/2003.13698} {arXiv:2003.13698
  [astro-ph.CO]} \BibitemShut {NoStop}%
\bibitem [{\citenamefont {Pospelov}\ \emph {et~al.}(2018)\citenamefont
  {Pospelov}, \citenamefont {Pradler}, \citenamefont {Ruderman},\ and\
  \citenamefont {Urbano}}]{Pospelov:2018kdh}%
  \BibitemOpen
  \bibfield  {author} {\bibinfo {author} {\bibfnamefont {M.}~\bibnamefont
  {Pospelov}}, \bibinfo {author} {\bibfnamefont {J.}~\bibnamefont {Pradler}},
  \bibinfo {author} {\bibfnamefont {J.~T.}\ \bibnamefont {Ruderman}},\ and\
  \bibinfo {author} {\bibfnamefont {A.}~\bibnamefont {Urbano}},\ }\bibfield
  {title} {\bibinfo {title} {{Room for New Physics in the Rayleigh-Jeans Tail
  of the Cosmic Microwave Background}},\ }\href
  {https://doi.org/10.1103/PhysRevLett.121.031103} {\bibfield  {journal}
  {\bibinfo  {journal} {Phys. Rev. Lett.}\ }\textbf {\bibinfo {volume} {121}},\
  \bibinfo {pages} {031103} (\bibinfo {year} {2018})},\ \Eprint
  {https://arxiv.org/abs/1803.07048} {arXiv:1803.07048 [hep-ph]} \BibitemShut
  {NoStop}%
%%CITATION = ARXIV:1803.07048;%%
\bibitem [{\citenamefont {Kogut}\ \emph {et~al.}(2011)\citenamefont {Kogut}
  \emph {et~al.}}]{Kogut:2011xw}%
  \BibitemOpen
  \bibfield  {author} {\bibinfo {author} {\bibfnamefont {A.}~\bibnamefont
  {Kogut}} \emph {et~al.},\ }\bibfield  {title} {\bibinfo {title} {{The
  Primordial Inflation Explorer (PIXIE): A Nulling Polarimeter for Cosmic
  Microwave Background Observations}},\ }\href
  {https://doi.org/10.1088/1475-7516/2011/07/025} {\bibfield  {journal}
  {\bibinfo  {journal} {JCAP}\ }\textbf {\bibinfo {volume} {1107}},\ \bibinfo
  {pages} {025}},\ \Eprint {https://arxiv.org/abs/1105.2044} {arXiv:1105.2044
  [astro-ph.CO]} \BibitemShut {NoStop}%
%%CITATION = ARXIV:1105.2044;%%
\bibitem [{\citenamefont {Abazajian}\ \emph {et~al.}(2019)\citenamefont
  {Abazajian} \emph {et~al.}}]{Abazajian:2019eic}%
  \BibitemOpen
  \bibfield  {author} {\bibinfo {author} {\bibfnamefont {K.}~\bibnamefont
  {Abazajian}} \emph {et~al.},\ }\href@noop {} {\bibinfo {title} {{CMB-S4
  Science Case, Reference Design, and Project Plan}}} (\bibinfo {year}
  {2019}),\ \Eprint {https://arxiv.org/abs/1907.04473} {arXiv:1907.04473
  [astro-ph.IM]} \BibitemShut {NoStop}%
\bibitem [{\citenamefont {Poulin}\ \emph {et~al.}(2016)\citenamefont {Poulin},
  \citenamefont {Serpico},\ and\ \citenamefont {Lesgourgues}}]{Poulin:2016nat}%
  \BibitemOpen
  \bibfield  {author} {\bibinfo {author} {\bibfnamefont {V.}~\bibnamefont
  {Poulin}}, \bibinfo {author} {\bibfnamefont {P.~D.}\ \bibnamefont
  {Serpico}},\ and\ \bibinfo {author} {\bibfnamefont {J.}~\bibnamefont
  {Lesgourgues}},\ }\bibfield  {title} {\bibinfo {title} {{A fresh look at
  linear cosmological constraints on a decaying dark matter component}},\
  }\href {https://doi.org/10.1088/1475-7516/2016/08/036} {\bibfield  {journal}
  {\bibinfo  {journal} {JCAP}\ }\textbf {\bibinfo {volume} {08}},\ \bibinfo
  {pages} {036}},\ \Eprint {https://arxiv.org/abs/1606.02073} {arXiv:1606.02073
  [astro-ph.CO]} \BibitemShut {NoStop}%
\bibitem [{\citenamefont {Nygaard}\ \emph {et~al.}(2021)\citenamefont
  {Nygaard}, \citenamefont {Tram},\ and\ \citenamefont
  {Hannestad}}]{Nygaard:2020sow}%
  \BibitemOpen
  \bibfield  {author} {\bibinfo {author} {\bibfnamefont {A.}~\bibnamefont
  {Nygaard}}, \bibinfo {author} {\bibfnamefont {T.}~\bibnamefont {Tram}},\ and\
  \bibinfo {author} {\bibfnamefont {S.}~\bibnamefont {Hannestad}},\ }\bibfield
  {title} {\bibinfo {title} {{Updated constraints on decaying cold dark
  matter}},\ }\href {https://doi.org/10.1088/1475-7516/2021/05/017} {\bibfield
  {journal} {\bibinfo  {journal} {JCAP}\ }\textbf {\bibinfo {volume} {05}},\
  \bibinfo {pages} {017}},\ \Eprint {https://arxiv.org/abs/2011.01632}
  {arXiv:2011.01632 [astro-ph.CO]} \BibitemShut {NoStop}%
\bibitem [{\citenamefont {Chen}\ \emph {et~al.}(2021)\citenamefont {Chen} \emph
  {et~al.}}]{DES:2020mpv}%
  \BibitemOpen
  \bibfield  {author} {\bibinfo {author} {\bibfnamefont {A.}~\bibnamefont
  {Chen}} \emph {et~al.} (\bibinfo {collaboration} {DES}),\ }\bibfield  {title}
  {\bibinfo {title} {{Constraints on dark matter to dark radiation conversion
  in the late universe with DES-Y1 and external data}},\ }\href
  {https://doi.org/10.1103/PhysRevD.103.123528} {\bibfield  {journal} {\bibinfo
   {journal} {Phys. Rev. D}\ }\textbf {\bibinfo {volume} {103}},\ \bibinfo
  {pages} {123528} (\bibinfo {year} {2021})},\ \Eprint
  {https://arxiv.org/abs/2011.04606} {arXiv:2011.04606 [astro-ph.CO]}
  \BibitemShut {NoStop}%
\bibitem [{\citenamefont {Mau}\ \emph {et~al.}(2022)\citenamefont {Mau} \emph
  {et~al.}}]{Mau:2022sbf}%
  \BibitemOpen
  \bibfield  {author} {\bibinfo {author} {\bibfnamefont {S.}~\bibnamefont
  {Mau}} \emph {et~al.},\ }\href@noop {} {\bibinfo {title} {{Milky Way
  Satellite Census. IV. Constraints on Decaying Dark Matter from Observations
  of Milky Way Satellite Galaxies}}} (\bibinfo {year} {2022}),\ \Eprint
  {https://arxiv.org/abs/2201.11740} {arXiv:2201.11740 [astro-ph.CO]}
  \BibitemShut {NoStop}%
\bibitem [{\citenamefont {Simon}\ \emph {et~al.}(2022)\citenamefont {Simon},
  \citenamefont {Abell\'an}, \citenamefont {Du}, \citenamefont {Poulin},\ and\
  \citenamefont {Tsai}}]{Simon:2022ftd}%
  \BibitemOpen
  \bibfield  {author} {\bibinfo {author} {\bibfnamefont {T.}~\bibnamefont
  {Simon}}, \bibinfo {author} {\bibfnamefont {G.~F.}\ \bibnamefont
  {Abell\'an}}, \bibinfo {author} {\bibfnamefont {P.}~\bibnamefont {Du}},
  \bibinfo {author} {\bibfnamefont {V.}~\bibnamefont {Poulin}},\ and\ \bibinfo
  {author} {\bibfnamefont {Y.}~\bibnamefont {Tsai}},\ }\href@noop {} {\bibinfo
  {title} {{Constraining decaying dark matter with BOSS data and the effective
  field theory of large-scale structures}}} (\bibinfo {year} {2022}),\ \Eprint
  {https://arxiv.org/abs/2203.07440} {arXiv:2203.07440 [astro-ph.CO]}
  \BibitemShut {NoStop}%
\bibitem [{\citenamefont {Alvi}\ \emph {et~al.}(2022)\citenamefont {Alvi},
  \citenamefont {Brinckmann}, \citenamefont {Gerbino}, \citenamefont
  {Lattanzi},\ and\ \citenamefont {Pagano}}]{Alvi:2022aam}%
  \BibitemOpen
  \bibfield  {author} {\bibinfo {author} {\bibfnamefont {S.}~\bibnamefont
  {Alvi}}, \bibinfo {author} {\bibfnamefont {T.}~\bibnamefont {Brinckmann}},
  \bibinfo {author} {\bibfnamefont {M.}~\bibnamefont {Gerbino}}, \bibinfo
  {author} {\bibfnamefont {M.}~\bibnamefont {Lattanzi}},\ and\ \bibinfo
  {author} {\bibfnamefont {L.}~\bibnamefont {Pagano}},\ }\href@noop {}
  {\bibinfo {title} {{Do you smell something decaying? Updated linear
  constraints on decaying dark matter scenarios}}} (\bibinfo {year} {2022}),\
  \Eprint {https://arxiv.org/abs/2205.05636} {arXiv:2205.05636 [astro-ph.CO]}
  \BibitemShut {NoStop}%
\bibitem [{\citenamefont {Skilling}(2006)}]{10.1214/06-BA127}%
  \BibitemOpen
  \bibfield  {author} {\bibinfo {author} {\bibfnamefont {J.}~\bibnamefont
  {Skilling}},\ }\bibfield  {title} {\bibinfo {title} {{Nested sampling for
  general Bayesian computation}},\ }\href {https://doi.org/10.1214/06-BA127}
  {\bibfield  {journal} {\bibinfo  {journal} {Bayesian Analysis}\ }\textbf
  {\bibinfo {volume} {1}},\ \bibinfo {pages} {833 } (\bibinfo {year}
  {2006})}\BibitemShut {NoStop}%
\bibitem [{\citenamefont {Feroz}\ \emph {et~al.}(2009)\citenamefont {Feroz},
  \citenamefont {Hobson},\ and\ \citenamefont {Bridges}}]{Feroz:2008xx}%
  \BibitemOpen
  \bibfield  {author} {\bibinfo {author} {\bibfnamefont {F.}~\bibnamefont
  {Feroz}}, \bibinfo {author} {\bibfnamefont {M.~P.}\ \bibnamefont {Hobson}},\
  and\ \bibinfo {author} {\bibfnamefont {M.}~\bibnamefont {Bridges}},\
  }\bibfield  {title} {\bibinfo {title} {{MultiNest: an efficient and robust
  Bayesian inference tool for cosmology and particle physics}},\ }\href
  {https://doi.org/10.1111/j.1365-2966.2009.14548.x} {\bibfield  {journal}
  {\bibinfo  {journal} {Mon. Not. Roy. Astron. Soc.}\ }\textbf {\bibinfo
  {volume} {398}},\ \bibinfo {pages} {1601} (\bibinfo {year} {2009})},\ \Eprint
  {https://arxiv.org/abs/0809.3437} {arXiv:0809.3437 [astro-ph]} \BibitemShut
  {NoStop}%
\bibitem [{\citenamefont {{Skilling}}(2004)}]{2004AIPC..735..395S}%
  \BibitemOpen
  \bibfield  {author} {\bibinfo {author} {\bibfnamefont {J.}~\bibnamefont
  {{Skilling}}},\ }\bibfield  {title} {\bibinfo {title} {{Nested Sampling}},\
  }in\ \href {https://doi.org/10.1063/1.1835238} {\emph {\bibinfo {booktitle}
  {Bayesian Inference and Maximum Entropy Methods in Science and Engineering:
  24th International Workshop on Bayesian Inference and Maximum Entropy Methods
  in Science and Engineering}}},\ \bibinfo {series} {American Institute of
  Physics Conference Series}, Vol.\ \bibinfo {volume} {735},\ \bibinfo {editor}
  {edited by\ \bibinfo {editor} {\bibfnamefont {R.}~\bibnamefont {{Fischer}}},
  \bibinfo {editor} {\bibfnamefont {R.}~\bibnamefont {{Preuss}}},\ and\
  \bibinfo {editor} {\bibfnamefont {U.~V.}\ \bibnamefont {{Toussaint}}}}\
  (\bibinfo {year} {2004})\ pp.\ \bibinfo {pages} {395--405}\BibitemShut
  {NoStop}%
\bibitem [{\citenamefont {Speagle}(2020)}]{Speagle_2020}%
  \BibitemOpen
  \bibfield  {author} {\bibinfo {author} {\bibfnamefont {J.~S.}\ \bibnamefont
  {Speagle}},\ }\bibfield  {title} {\bibinfo {title} {dynesty: a dynamic nested
  sampling package for estimating bayesian posteriors and evidences},\ }\href
  {https://doi.org/10.1093/mnras/staa278} {\bibfield  {journal} {\bibinfo
  {journal} {Monthly Notices of the Royal Astronomical Society}\ }\textbf
  {\bibinfo {volume} {493}},\ \bibinfo {pages} {3132} (\bibinfo {year}
  {2020})}\BibitemShut {NoStop}%
\bibitem [{\citenamefont {{Massardi}}\ \emph {et~al.}(2010)\citenamefont
  {{Massardi}}, \citenamefont {{Bonaldi}}, \citenamefont {{Negrello}},
  \citenamefont {{Ricciardi}}, \citenamefont {{Raccanelli}},\ and\
  \citenamefont {{de Zotti}}}]{2010MNRAS.404..532M}%
  \BibitemOpen
  \bibfield  {author} {\bibinfo {author} {\bibfnamefont {M.}~\bibnamefont
  {{Massardi}}}, \bibinfo {author} {\bibfnamefont {A.}~\bibnamefont
  {{Bonaldi}}}, \bibinfo {author} {\bibfnamefont {M.}~\bibnamefont
  {{Negrello}}}, \bibinfo {author} {\bibfnamefont {S.}~\bibnamefont
  {{Ricciardi}}}, \bibinfo {author} {\bibfnamefont {A.}~\bibnamefont
  {{Raccanelli}}},\ and\ \bibinfo {author} {\bibfnamefont {G.}~\bibnamefont
  {{de Zotti}}},\ }\bibfield  {title} {\bibinfo {title} {{A model for the
  cosmological evolution of low-frequency radio sources}},\ }\href
  {https://doi.org/10.1111/j.1365-2966.2010.16305.x} {\bibfield  {journal}
  {\bibinfo  {journal} {\mnras}\ }\textbf {\bibinfo {volume} {404}},\ \bibinfo
  {pages} {532} (\bibinfo {year} {2010})},\ \Eprint
  {https://arxiv.org/abs/1001.1069} {arXiv:1001.1069 [astro-ph.CO]}
  \BibitemShut {NoStop}%
\bibitem [{\citenamefont {{Limber}}(1953)}]{Limber}%
  \BibitemOpen
  \bibfield  {author} {\bibinfo {author} {\bibfnamefont {D.~N.}\ \bibnamefont
  {{Limber}}},\ }\bibfield  {title} {\bibinfo {title} {{The Analysis of Counts
  of the Extragalactic Nebulae in Terms of a Fluctuating Density Field.}},\
  }\href {https://doi.org/10.1086/145672} {\bibfield  {journal} {\bibinfo
  {journal} {\apj}\ }\textbf {\bibinfo {volume} {117}},\ \bibinfo {pages} {134}
  (\bibinfo {year} {1953})}\BibitemShut {NoStop}%
\bibitem [{\citenamefont {Kaiser}(1992)}]{Kaiser:1991qi}%
  \BibitemOpen
  \bibfield  {author} {\bibinfo {author} {\bibfnamefont {N.}~\bibnamefont
  {Kaiser}},\ }\bibfield  {title} {\bibinfo {title} {{Weak gravitational
  lensing of distant galaxies}},\ }\href {https://doi.org/10.1086/171151}
  {\bibfield  {journal} {\bibinfo  {journal} {Astrophys. J.}\ }\textbf
  {\bibinfo {volume} {388}},\ \bibinfo {pages} {272} (\bibinfo {year}
  {1992})}\BibitemShut {NoStop}%
\bibitem [{\citenamefont {Kaiser}(1998)}]{Kaiser:1996tp}%
  \BibitemOpen
  \bibfield  {author} {\bibinfo {author} {\bibfnamefont {N.}~\bibnamefont
  {Kaiser}},\ }\bibfield  {title} {\bibinfo {title} {{Weak lensing and
  cosmology}},\ }\href {https://doi.org/10.1086/305515} {\bibfield  {journal}
  {\bibinfo  {journal} {Astrophys. J.}\ }\textbf {\bibinfo {volume} {498}},\
  \bibinfo {pages} {26} (\bibinfo {year} {1998})},\ \Eprint
  {https://arxiv.org/abs/astro-ph/9610120} {arXiv:astro-ph/9610120}
  \BibitemShut {NoStop}%
\bibitem [{\citenamefont {Fornengo}\ and\ \citenamefont
  {Regis}(2014)}]{Fornengo:2013rga}%
  \BibitemOpen
  \bibfield  {author} {\bibinfo {author} {\bibfnamefont {N.}~\bibnamefont
  {Fornengo}}\ and\ \bibinfo {author} {\bibfnamefont {M.}~\bibnamefont
  {Regis}},\ }\bibfield  {title} {\bibinfo {title} {{Particle dark matter
  searches in the anisotropic sky}},\ }\href
  {https://doi.org/10.3389/fphy.2014.00006} {\bibfield  {journal} {\bibinfo
  {journal} {Front. Physics}\ }\textbf {\bibinfo {volume} {2}},\ \bibinfo
  {pages} {6} (\bibinfo {year} {2014})},\ \Eprint
  {https://arxiv.org/abs/1312.4835} {arXiv:1312.4835 [astro-ph.CO]}
  \BibitemShut {NoStop}%
\bibitem [{\citenamefont {Kumar}\ \emph {et~al.}(2019)\citenamefont {Kumar},
  \citenamefont {Carroll}, \citenamefont {Hartikainen},\ and\ \citenamefont
  {Martin}}]{arviz_2019}%
  \BibitemOpen
  \bibfield  {author} {\bibinfo {author} {\bibfnamefont {R.}~\bibnamefont
  {Kumar}}, \bibinfo {author} {\bibfnamefont {C.}~\bibnamefont {Carroll}},
  \bibinfo {author} {\bibfnamefont {A.}~\bibnamefont {Hartikainen}},\ and\
  \bibinfo {author} {\bibfnamefont {O.}~\bibnamefont {Martin}},\ }\bibfield
  {title} {\bibinfo {title} {Arviz a unified library for exploratory analysis
  of bayesian models in python},\ }\href {https://doi.org/10.21105/joss.01143}
  {\bibfield  {journal} {\bibinfo  {journal} {Journal of Open Source Software}\
  }\textbf {\bibinfo {volume} {4}},\ \bibinfo {pages} {1143} (\bibinfo {year}
  {2019})}\BibitemShut {NoStop}%
\bibitem [{\citenamefont {Price-Whelan}\ \emph {et~al.}(2018)\citenamefont
  {Price-Whelan} \emph {et~al.}}]{Price-Whelan:2018hus}%
  \BibitemOpen
  \bibfield  {author} {\bibinfo {author} {\bibfnamefont {A.~M.}\ \bibnamefont
  {Price-Whelan}} \emph {et~al.},\ }\bibfield  {title} {\bibinfo {title} {{The
  Astropy Project: Building an Open-science Project and Status of the v2.0 Core
  Package}},\ }\href {https://doi.org/10.3847/1538-3881/aabc4f} {\bibfield
  {journal} {\bibinfo  {journal} {Astron. J.}\ }\textbf {\bibinfo {volume}
  {156}},\ \bibinfo {pages} {123} (\bibinfo {year} {2018})},\ \Eprint
  {https://arxiv.org/abs/1801.02634} {arXiv:1801.02634} \BibitemShut {NoStop}%
%%CITATION = ARXIV:1801.02634;%%
\bibitem [{\citenamefont {Robitaille}\ \emph {et~al.}(2013)\citenamefont
  {Robitaille} \emph {et~al.}}]{Robitaille:2013mpa}%
  \BibitemOpen
  \bibfield  {author} {\bibinfo {author} {\bibfnamefont {T.~P.}\ \bibnamefont
  {Robitaille}} \emph {et~al.} (\bibinfo {collaboration} {Astropy}),\
  }\bibfield  {title} {\bibinfo {title} {{Astropy: A Community Python Package
  for Astronomy}},\ }\href {https://doi.org/10.1051/0004-6361/201322068}
  {\bibfield  {journal} {\bibinfo  {journal} {Astron. Astrophys.}\ }\textbf
  {\bibinfo {volume} {558}},\ \bibinfo {pages} {A33} (\bibinfo {year}
  {2013})},\ \Eprint {https://arxiv.org/abs/1307.6212} {arXiv:1307.6212
  [astro-ph.IM]} \BibitemShut {NoStop}%
%%CITATION = ARXIV:1307.6212;%%
\bibitem [{\citenamefont {Blas}\ \emph {et~al.}(2011)\citenamefont {Blas},
  \citenamefont {Lesgourgues},\ and\ \citenamefont {Tram}}]{Blas:2011rf}%
  \BibitemOpen
  \bibfield  {author} {\bibinfo {author} {\bibfnamefont {D.}~\bibnamefont
  {Blas}}, \bibinfo {author} {\bibfnamefont {J.}~\bibnamefont {Lesgourgues}},\
  and\ \bibinfo {author} {\bibfnamefont {T.}~\bibnamefont {Tram}},\ }\bibfield
  {title} {\bibinfo {title} {{The Cosmic Linear Anisotropy Solving System
  (CLASS) II: Approximation schemes}},\ }\href
  {https://doi.org/10.1088/1475-7516/2011/07/034} {\bibfield  {journal}
  {\bibinfo  {journal} {JCAP}\ }\textbf {\bibinfo {volume} {1107}},\ \bibinfo
  {pages} {034}},\ \Eprint {https://arxiv.org/abs/1104.2933} {arXiv:1104.2933
  [astro-ph.CO]} \BibitemShut {NoStop}%
%%CITATION = ARXIV:1104.2933;%%
\bibitem [{\citenamefont {Ali-Haimoud}\ and\ \citenamefont
  {Hirata}(2011)}]{AliHaimoud:2010dx}%
  \BibitemOpen
  \bibfield  {author} {\bibinfo {author} {\bibfnamefont {Y.}~\bibnamefont
  {Ali-Haimoud}}\ and\ \bibinfo {author} {\bibfnamefont {C.~M.}\ \bibnamefont
  {Hirata}},\ }\bibfield  {title} {\bibinfo {title} {{HyRec: A fast and highly
  accurate primordial hydrogen and helium recombination code}},\ }\href
  {https://doi.org/10.1103/PhysRevD.83.043513} {\bibfield  {journal} {\bibinfo
  {journal} {Phys. Rev.}\ }\textbf {\bibinfo {volume} {D83}},\ \bibinfo {pages}
  {043513} (\bibinfo {year} {2011})},\ \Eprint
  {https://arxiv.org/abs/1011.3758} {arXiv:1011.3758 [astro-ph.CO]}
  \BibitemShut {NoStop}%
%%CITATION = ARXIV:1011.3758;%%
\bibitem [{\citenamefont {{Perez}}\ and\ \citenamefont
  {{Granger}}(2007)}]{PER-GRA:2007}%
  \BibitemOpen
  \bibfield  {author} {\bibinfo {author} {\bibfnamefont {F.}~\bibnamefont
  {{Perez}}}\ and\ \bibinfo {author} {\bibfnamefont {B.~E.}\ \bibnamefont
  {{Granger}}},\ }\bibfield  {title} {\bibinfo {title} {{IPython: A System for
  Interactive Scientific Computing}},\ }\href
  {https://doi.org/10.1109/MCSE.2007.53} {\bibfield  {journal} {\bibinfo
  {journal} {Computing in Science and Engineering}\ }\textbf {\bibinfo {volume}
  {9}},\ \bibinfo {pages} {21} (\bibinfo {year} {2007})}\BibitemShut {NoStop}%
\bibitem [{\citenamefont {Kluyver}\ \emph {et~al.}(2016)\citenamefont {Kluyver}
  \emph {et~al.}}]{Kluyver2016JupyterN}%
  \BibitemOpen
  \bibfield  {author} {\bibinfo {author} {\bibfnamefont {T.}~\bibnamefont
  {Kluyver}} \emph {et~al.},\ }\bibfield  {title} {\bibinfo {title} {Jupyter
  notebooks - a publishing format for reproducible computational workflows},\
  }in\ \href@noop {} {\emph {\bibinfo {booktitle} {ELPUB}}}\ (\bibinfo {year}
  {2016})\BibitemShut {NoStop}%
\bibitem [{\citenamefont {Hunter}(2007)}]{Hunter:2007}%
  \BibitemOpen
  \bibfield  {author} {\bibinfo {author} {\bibfnamefont {J.~D.}\ \bibnamefont
  {Hunter}},\ }\bibfield  {title} {\bibinfo {title} {Matplotlib: A 2d graphics
  environment},\ }\href@noop {} {\bibfield  {journal} {\bibinfo  {journal}
  {Computing In Science \& Engineering}\ }\textbf {\bibinfo {volume} {9}},\
  \bibinfo {pages} {90} (\bibinfo {year} {2007})}\BibitemShut {NoStop}%
\bibitem [{\citenamefont {Harris}\ \emph {et~al.}(2020)\citenamefont {Harris},
  \citenamefont {Millman}, \citenamefont {van~der Walt}, \citenamefont
  {Gommers}, \citenamefont {Virtanen}, \citenamefont {Cournapeau},
  \citenamefont {Wieser}, \citenamefont {Taylor}, \citenamefont {Berg},
  \citenamefont {Smith}, \citenamefont {Kern}, \citenamefont {Picus},
  \citenamefont {Hoyer}, \citenamefont {van Kerkwijk}, \citenamefont {Brett},
  \citenamefont {Haldane}, \citenamefont {del R{\'{i}}o}, \citenamefont
  {Wiebe}, \citenamefont {Peterson}, \citenamefont {G{\'{e}}rard-Marchant},
  \citenamefont {Sheppard}, \citenamefont {Reddy}, \citenamefont {Weckesser},
  \citenamefont {Abbasi}, \citenamefont {Gohlke},\ and\ \citenamefont
  {Oliphant}}]{harris2020array}%
  \BibitemOpen
  \bibfield  {author} {\bibinfo {author} {\bibfnamefont {C.~R.}\ \bibnamefont
  {Harris}}, \bibinfo {author} {\bibfnamefont {K.~J.}\ \bibnamefont {Millman}},
  \bibinfo {author} {\bibfnamefont {S.~J.}\ \bibnamefont {van~der Walt}},
  \bibinfo {author} {\bibfnamefont {R.}~\bibnamefont {Gommers}}, \bibinfo
  {author} {\bibfnamefont {P.}~\bibnamefont {Virtanen}}, \bibinfo {author}
  {\bibfnamefont {D.}~\bibnamefont {Cournapeau}}, \bibinfo {author}
  {\bibfnamefont {E.}~\bibnamefont {Wieser}}, \bibinfo {author} {\bibfnamefont
  {J.}~\bibnamefont {Taylor}}, \bibinfo {author} {\bibfnamefont
  {S.}~\bibnamefont {Berg}}, \bibinfo {author} {\bibfnamefont {N.~J.}\
  \bibnamefont {Smith}}, \bibinfo {author} {\bibfnamefont {R.}~\bibnamefont
  {Kern}}, \bibinfo {author} {\bibfnamefont {M.}~\bibnamefont {Picus}},
  \bibinfo {author} {\bibfnamefont {S.}~\bibnamefont {Hoyer}}, \bibinfo
  {author} {\bibfnamefont {M.~H.}\ \bibnamefont {van Kerkwijk}}, \bibinfo
  {author} {\bibfnamefont {M.}~\bibnamefont {Brett}}, \bibinfo {author}
  {\bibfnamefont {A.}~\bibnamefont {Haldane}}, \bibinfo {author} {\bibfnamefont
  {J.~F.}\ \bibnamefont {del R{\'{i}}o}}, \bibinfo {author} {\bibfnamefont
  {M.}~\bibnamefont {Wiebe}}, \bibinfo {author} {\bibfnamefont
  {P.}~\bibnamefont {Peterson}}, \bibinfo {author} {\bibfnamefont
  {P.}~\bibnamefont {G{\'{e}}rard-Marchant}}, \bibinfo {author} {\bibfnamefont
  {K.}~\bibnamefont {Sheppard}}, \bibinfo {author} {\bibfnamefont
  {T.}~\bibnamefont {Reddy}}, \bibinfo {author} {\bibfnamefont
  {W.}~\bibnamefont {Weckesser}}, \bibinfo {author} {\bibfnamefont
  {H.}~\bibnamefont {Abbasi}}, \bibinfo {author} {\bibfnamefont
  {C.}~\bibnamefont {Gohlke}},\ and\ \bibinfo {author} {\bibfnamefont {T.~E.}\
  \bibnamefont {Oliphant}},\ }\bibfield  {title} {\bibinfo {title} {Array
  programming with {NumPy}},\ }\href
  {https://doi.org/10.1038/s41586-020-2649-2} {\bibfield  {journal} {\bibinfo
  {journal} {Nature}\ }\textbf {\bibinfo {volume} {585}},\ \bibinfo {pages}
  {357} (\bibinfo {year} {2020})}\BibitemShut {NoStop}%
\bibitem [{\citenamefont {Talman}(1978)}]{TALMAN197835}%
  \BibitemOpen
  \bibfield  {author} {\bibinfo {author} {\bibfnamefont {J.~D.}\ \bibnamefont
  {Talman}},\ }\bibfield  {title} {\bibinfo {title} {Numerical fourier and
  bessel transforms in logarithmic variables},\ }\href
  {https://doi.org/https://doi.org/10.1016/0021-9991(78)90107-9} {\bibfield
  {journal} {\bibinfo  {journal} {Journal of Computational Physics}\ }\textbf
  {\bibinfo {volume} {29}},\ \bibinfo {pages} {35} (\bibinfo {year}
  {1978})}\BibitemShut {NoStop}%
\bibitem [{\citenamefont {Hamilton}(2000)}]{Hamilton:1999uv}%
  \BibitemOpen
  \bibfield  {author} {\bibinfo {author} {\bibfnamefont {A.~J.~S.}\
  \bibnamefont {Hamilton}},\ }\bibfield  {title} {\bibinfo {title}
  {{Uncorrelated modes of the nonlinear power spectrum}},\ }\href
  {https://doi.org/10.1046/j.1365-8711.2000.03071.x} {\bibfield  {journal}
  {\bibinfo  {journal} {Mon. Not. Roy. Astron. Soc.}\ }\textbf {\bibinfo
  {volume} {312}},\ \bibinfo {pages} {257} (\bibinfo {year} {2000})},\ \Eprint
  {https://arxiv.org/abs/astro-ph/9905191} {arXiv:astro-ph/9905191}
  \BibitemShut {NoStop}%
\bibitem [{\citenamefont
  {Werthm{\"u}ller}(2020)}]{dieter_werthmuller_2020_3830476}%
  \BibitemOpen
  \bibfield  {author} {\bibinfo {author} {\bibfnamefont {D.}~\bibnamefont
  {Werthm{\"u}ller}},\ }\href {https://doi.org/10.5281/zenodo.3830476}
  {\bibinfo {title} {prisae/pyfftlog: First packaged release}} (\bibinfo {year}
  {2020})\BibitemShut {NoStop}%
\bibitem [{\citenamefont {Waskom}\ \emph {et~al.}(2017)\citenamefont {Waskom}
  \emph {et~al.}}]{seaborn}%
  \BibitemOpen
  \bibfield  {author} {\bibinfo {author} {\bibfnamefont {M.}~\bibnamefont
  {Waskom}} \emph {et~al.},\ }\href {https://doi.org/10.5281/zenodo.883859}
  {\bibinfo {title} {mwaskom/seaborn: v0.8.1 (september 2017)}} (\bibinfo
  {year} {2017})\BibitemShut {NoStop}%
\bibitem [{\citenamefont {McKinney}(2010)}]{pandas:2010}%
  \BibitemOpen
  \bibfield  {author} {\bibinfo {author} {\bibfnamefont {W.}~\bibnamefont
  {McKinney}},\ }\bibfield  {title} {\bibinfo {title} {Data structures for
  statistical computing in python},\ }in\ \href@noop {} {\emph {\bibinfo
  {booktitle} {Proceedings of the 9th Python in Science Conference}}},\
  \bibinfo {editor} {edited by\ \bibinfo {editor} {\bibfnamefont
  {S.}~\bibnamefont {van~der Walt}}\ and\ \bibinfo {editor} {\bibfnamefont
  {J.}~\bibnamefont {Millman}}}\ (\bibinfo {year} {2010})\ pp.\ \bibinfo
  {pages} {51 -- 56}\BibitemShut {NoStop}%
\bibitem [{\citenamefont {{Virtanen}}\ \emph {et~al.}(2020)\citenamefont
  {{Virtanen}} \emph {et~al.}}]{2020SciPy-NMeth}%
  \BibitemOpen
  \bibfield  {author} {\bibinfo {author} {\bibfnamefont {P.}~\bibnamefont
  {{Virtanen}}} \emph {et~al.},\ }\bibfield  {title} {\bibinfo {title} {{SciPy
  1.0: Fundamental Algorithms for Scientific Computing in Python}},\ }\bibfield
   {journal} {\bibinfo  {journal} {Nature Methods}\ }\href
  {https://doi.org/https://doi.org/10.1038/s41592-019-0686-2}
  {https://doi.org/10.1038/s41592-019-0686-2} (\bibinfo {year}
  {2020})\BibitemShut {NoStop}%
\bibitem [{\citenamefont {da~Costa-Luis}(2019)}]{da2019tqdm}%
  \BibitemOpen
  \bibfield  {author} {\bibinfo {author} {\bibfnamefont {C.~O.}\ \bibnamefont
  {da~Costa-Luis}},\ }\bibfield  {title} {\bibinfo {title} {tqdm: A fast,
  extensible progress meter for python and cli},\ }\href@noop {} {\bibfield
  {journal} {\bibinfo  {journal} {JOSS}\ }\textbf {\bibinfo {volume} {4}},\
  \bibinfo {pages} {1277} (\bibinfo {year} {2019})}\BibitemShut {NoStop}%
\bibitem [{\citenamefont {Grzadkowski}\ \emph {et~al.}(2021)\citenamefont
  {Grzadkowski}, \citenamefont {Iglicki},\ and\ \citenamefont
  {Mr\'owczy\'nski}}]{Grzadkowski:2021kgi}%
  \BibitemOpen
  \bibfield  {author} {\bibinfo {author} {\bibfnamefont {B.}~\bibnamefont
  {Grzadkowski}}, \bibinfo {author} {\bibfnamefont {M.}~\bibnamefont
  {Iglicki}},\ and\ \bibinfo {author} {\bibfnamefont {S.}~\bibnamefont
  {Mr\'owczy\'nski}},\ }\href@noop {} {\bibinfo {title} {{$t$-channel
  singularities in cosmology and particle physics}}} (\bibinfo {year} {2021}),\
  \Eprint {https://arxiv.org/abs/2108.01757} {arXiv:2108.01757 [hep-ph]}
  \BibitemShut {NoStop}%
\bibitem [{\citenamefont {Redondo}\ and\ \citenamefont
  {Postma}(2009)}]{Redondo:2008ec}%
  \BibitemOpen
  \bibfield  {author} {\bibinfo {author} {\bibfnamefont {J.}~\bibnamefont
  {Redondo}}\ and\ \bibinfo {author} {\bibfnamefont {M.}~\bibnamefont
  {Postma}},\ }\bibfield  {title} {\bibinfo {title} {{Massive hidden photons as
  lukewarm dark matter}},\ }\href
  {https://doi.org/10.1088/1475-7516/2009/02/005} {\bibfield  {journal}
  {\bibinfo  {journal} {JCAP}\ }\textbf {\bibinfo {volume} {0902}},\ \bibinfo
  {pages} {005}},\ \Eprint {https://arxiv.org/abs/0811.0326} {arXiv:0811.0326
  [hep-ph]} \BibitemShut {NoStop}%
%%CITATION = ARXIV:0811.0326;%%
\bibitem [{\citenamefont {LoVerde}\ and\ \citenamefont
  {Afshordi}(2008)}]{LoVerde:2008re}%
  \BibitemOpen
  \bibfield  {author} {\bibinfo {author} {\bibfnamefont {M.}~\bibnamefont
  {LoVerde}}\ and\ \bibinfo {author} {\bibfnamefont {N.}~\bibnamefont
  {Afshordi}},\ }\bibfield  {title} {\bibinfo {title} {{Extended Limber
  Approximation}},\ }\href {https://doi.org/10.1103/PhysRevD.78.123506}
  {\bibfield  {journal} {\bibinfo  {journal} {Phys. Rev. D}\ }\textbf {\bibinfo
  {volume} {78}},\ \bibinfo {pages} {123506} (\bibinfo {year} {2008})},\
  \Eprint {https://arxiv.org/abs/0809.5112} {arXiv:0809.5112 [astro-ph]}
  \BibitemShut {NoStop}%
\bibitem [{\citenamefont {Kayo}\ \emph {et~al.}(2001)\citenamefont {Kayo},
  \citenamefont {Taruya},\ and\ \citenamefont {Suto}}]{Kayo:2001gu}%
  \BibitemOpen
  \bibfield  {author} {\bibinfo {author} {\bibfnamefont {I.}~\bibnamefont
  {Kayo}}, \bibinfo {author} {\bibfnamefont {A.}~\bibnamefont {Taruya}},\ and\
  \bibinfo {author} {\bibfnamefont {Y.}~\bibnamefont {Suto}},\ }\bibfield
  {title} {\bibinfo {title} {{Probability distribution function of cosmological
  density fluctuations from Gaussian initial condition: comparison of one- and
  two-point log-normal model predictions with n-body simulations}},\ }\href
  {https://doi.org/10.1086/323227} {\bibfield  {journal} {\bibinfo  {journal}
  {Astrophys. J.}\ }\textbf {\bibinfo {volume} {561}},\ \bibinfo {pages} {22}
  (\bibinfo {year} {2001})},\ \Eprint {https://arxiv.org/abs/astro-ph/0105218}
  {arXiv:astro-ph/0105218 [astro-ph]} \BibitemShut {NoStop}%
%%CITATION = ASTRO-PH/0105218;%%
\bibitem [{\citenamefont {{Alsing}}\ and\ \citenamefont
  {{Handley}}(2021)}]{2021arXiv210212478A}%
  \BibitemOpen
  \bibfield  {author} {\bibinfo {author} {\bibfnamefont {J.}~\bibnamefont
  {{Alsing}}}\ and\ \bibinfo {author} {\bibfnamefont {W.}~\bibnamefont
  {{Handley}}},\ }\bibfield  {title} {\bibinfo {title} {{Nested sampling with
  any prior you like}},\ }\href@noop {} {\bibfield  {journal} {\bibinfo
  {journal} {arXiv e-prints}\ ,\ \bibinfo {eid} {arXiv:2102.12478}} (\bibinfo
  {year} {2021})},\ \Eprint {https://arxiv.org/abs/2102.12478}
  {arXiv:2102.12478 [astro-ph.IM]} \BibitemShut {NoStop}%
\bibitem [{\citenamefont {Rezende}\ and\ \citenamefont
  {Mohamed}(2015)}]{pmlr-v37-rezende15}%
  \BibitemOpen
  \bibfield  {author} {\bibinfo {author} {\bibfnamefont {D.}~\bibnamefont
  {Rezende}}\ and\ \bibinfo {author} {\bibfnamefont {S.}~\bibnamefont
  {Mohamed}},\ }\bibfield  {title} {\bibinfo {title} {Variational inference
  with normalizing flows},\ }in\ \href
  {http://proceedings.mlr.press/v37/rezende15.html} {\emph {\bibinfo
  {booktitle} {Proceedings of the 32nd International Conference on Machine
  Learning}}},\ \bibinfo {series} {Proceedings of Machine Learning Research},
  Vol.~\bibinfo {volume} {37},\ \bibinfo {editor} {edited by\ \bibinfo {editor}
  {\bibfnamefont {F.}~\bibnamefont {Bach}}\ and\ \bibinfo {editor}
  {\bibfnamefont {D.}~\bibnamefont {Blei}}}\ (\bibinfo  {publisher} {PMLR},\
  \bibinfo {address} {Lille, France},\ \bibinfo {year} {2015})\ pp.\ \bibinfo
  {pages} {1530--1538}\BibitemShut {NoStop}%
\bibitem [{\citenamefont {Papamakarios}\ \emph {et~al.}(2019)\citenamefont
  {Papamakarios}, \citenamefont {Nalisnick}, \citenamefont {Rezende},
  \citenamefont {Mohamed},\ and\ \citenamefont
  {Lakshminarayanan}}]{papamakarios2019normalizing}%
  \BibitemOpen
  \bibfield  {author} {\bibinfo {author} {\bibfnamefont {G.}~\bibnamefont
  {Papamakarios}}, \bibinfo {author} {\bibfnamefont {E.}~\bibnamefont
  {Nalisnick}}, \bibinfo {author} {\bibfnamefont {D.~J.}\ \bibnamefont
  {Rezende}}, \bibinfo {author} {\bibfnamefont {S.}~\bibnamefont {Mohamed}},\
  and\ \bibinfo {author} {\bibfnamefont {B.}~\bibnamefont {Lakshminarayanan}},\
  }\bibfield  {title} {\bibinfo {title} {Normalizing flows for probabilistic
  modeling and inference},\ }\href@noop {} {\bibfield  {journal} {\bibinfo
  {journal} {arXiv preprint arXiv:1912.02762}\ } (\bibinfo {year}
  {2019})}\BibitemShut {NoStop}%
\bibitem [{\citenamefont {Durkan}\ \emph {et~al.}(2019)\citenamefont {Durkan},
  \citenamefont {Bekasov}, \citenamefont {Murray},\ and\ \citenamefont
  {Papamakarios}}]{NEURIPS2019_7ac71d43}%
  \BibitemOpen
  \bibfield  {author} {\bibinfo {author} {\bibfnamefont {C.}~\bibnamefont
  {Durkan}}, \bibinfo {author} {\bibfnamefont {A.}~\bibnamefont {Bekasov}},
  \bibinfo {author} {\bibfnamefont {I.}~\bibnamefont {Murray}},\ and\ \bibinfo
  {author} {\bibfnamefont {G.}~\bibnamefont {Papamakarios}},\ }\bibfield
  {title} {\bibinfo {title} {Neural spline flows},\ }in\ \href
  {https://proceedings.neurips.cc/paper/2019/file/7ac71d433f282034e088473244df8c02-Paper.pdf}
  {\emph {\bibinfo {booktitle} {Advances in Neural Information Processing
  Systems}}},\ Vol.~\bibinfo {volume} {32},\ \bibinfo {editor} {edited by\
  \bibinfo {editor} {\bibfnamefont {H.}~\bibnamefont {Wallach}}, \bibinfo
  {editor} {\bibfnamefont {H.}~\bibnamefont {Larochelle}}, \bibinfo {editor}
  {\bibfnamefont {A.}~\bibnamefont {Beygelzimer}}, \bibinfo {editor}
  {\bibfnamefont {F.}~\bibnamefont {d'Alch\'{e} Buc}}, \bibinfo {editor}
  {\bibfnamefont {E.}~\bibnamefont {Fox}},\ and\ \bibinfo {editor}
  {\bibfnamefont {R.}~\bibnamefont {Garnett}}}\ (\bibinfo  {publisher} {Curran
  Associates, Inc.},\ \bibinfo {year} {2019})\BibitemShut {NoStop}%
\bibitem [{\citenamefont {Isi}\ \emph {et~al.}(2022)\citenamefont {Isi},
  \citenamefont {Farr},\ and\ \citenamefont {Chatziioannou}}]{Isi:2022cii}%
  \BibitemOpen
  \bibfield  {author} {\bibinfo {author} {\bibfnamefont {M.}~\bibnamefont
  {Isi}}, \bibinfo {author} {\bibfnamefont {W.~M.}\ \bibnamefont {Farr}},\ and\
  \bibinfo {author} {\bibfnamefont {K.}~\bibnamefont {Chatziioannou}},\
  }\bibfield  {title} {\bibinfo {title} {{Comparing Bayes factors and
  hierarchical inference for testing general relativity with gravitational
  waves}},\ }\href@noop {} {\bibfield  {journal} {\bibinfo  {journal} {arXiv
  e-prints}\ } (\bibinfo {year} {2022})},\ \Eprint
  {https://arxiv.org/abs/2204.10742} {arXiv:2204.10742 [gr-qc]} \BibitemShut
  {NoStop}%
\bibitem [{\citenamefont {{Rybicki}}\ and\ \citenamefont
  {{Lightman}}(1979)}]{1979rpa..book.....R}%
  \BibitemOpen
  \bibfield  {author} {\bibinfo {author} {\bibfnamefont {G.~B.}\ \bibnamefont
  {{Rybicki}}}\ and\ \bibinfo {author} {\bibfnamefont {A.~P.}\ \bibnamefont
  {{Lightman}}},\ }\href@noop {} {\emph {\bibinfo {title} {{Radiative processes
  in astrophysics}}}}\ (\bibinfo {year} {1979})\BibitemShut {NoStop}%
\bibitem [{\citenamefont {Bell}(1978)}]{Bell:1978fj}%
  \BibitemOpen
  \bibfield  {author} {\bibinfo {author} {\bibfnamefont {A.~R.}\ \bibnamefont
  {Bell}},\ }\bibfield  {title} {\bibinfo {title} {{The acceleration of cosmic
  rays in shock fronts. II.}},\ }\href@noop {} {\bibfield  {journal} {\bibinfo
  {journal} {Mon. Not. Roy. Astron. Soc.}\ }\textbf {\bibinfo {volume} {182}},\
  \bibinfo {pages} {443} (\bibinfo {year} {1978})}\BibitemShut {NoStop}%
\bibitem [{\citenamefont {{Bell}}(1978)}]{BellI}%
  \BibitemOpen
  \bibfield  {author} {\bibinfo {author} {\bibfnamefont {A.~R.}\ \bibnamefont
  {{Bell}}},\ }\bibfield  {title} {\bibinfo {title} {{The acceleration of
  cosmic rays in shock fronts - I.}},\ }\href
  {https://doi.org/10.1093/mnras/182.2.147} {\bibfield  {journal} {\bibinfo
  {journal} {\mnras}\ }\textbf {\bibinfo {volume} {182}},\ \bibinfo {pages}
  {147} (\bibinfo {year} {1978})}\BibitemShut {NoStop}%
\bibitem [{\citenamefont {Fermi}(1949)}]{Fermi49}%
  \BibitemOpen
  \bibfield  {author} {\bibinfo {author} {\bibfnamefont {E.}~\bibnamefont
  {Fermi}},\ }\bibfield  {title} {\bibinfo {title} {On the origin of the cosmic
  radiation},\ }\href {https://doi.org/10.1103/PhysRev.75.1169} {\bibfield
  {journal} {\bibinfo  {journal} {Phys. Rev.}\ }\textbf {\bibinfo {volume}
  {75}},\ \bibinfo {pages} {1169} (\bibinfo {year} {1949})}\BibitemShut
  {NoStop}%
\bibitem [{\citenamefont {Blandford}\ and\ \citenamefont
  {Ostriker}(1980)}]{Blandford:1980hv}%
  \BibitemOpen
  \bibfield  {author} {\bibinfo {author} {\bibfnamefont {R.~D.}\ \bibnamefont
  {Blandford}}\ and\ \bibinfo {author} {\bibfnamefont {J.~P.}\ \bibnamefont
  {Ostriker}},\ }\bibfield  {title} {\bibinfo {title} {{Supernova Shock
  Acceleration of Cosmic Rays in the Galaxy}},\ }\href
  {https://doi.org/10.1086/157926} {\bibfield  {journal} {\bibinfo  {journal}
  {Astrophys. J.}\ }\textbf {\bibinfo {volume} {237}},\ \bibinfo {pages} {793}
  (\bibinfo {year} {1980})}\BibitemShut {NoStop}%
\bibitem [{\citenamefont {Blandford}\ and\ \citenamefont
  {Ostriker}(1978)}]{Blandford:1978ky}%
  \BibitemOpen
  \bibfield  {author} {\bibinfo {author} {\bibfnamefont {R.~D.}\ \bibnamefont
  {Blandford}}\ and\ \bibinfo {author} {\bibfnamefont {J.~P.}\ \bibnamefont
  {Ostriker}},\ }\bibfield  {title} {\bibinfo {title} {{Particle Acceleration
  by Astrophysical Shocks}},\ }\href {https://doi.org/10.1086/182658}
  {\bibfield  {journal} {\bibinfo  {journal} {Astrophys. J. Lett.}\ }\textbf
  {\bibinfo {volume} {221}},\ \bibinfo {pages} {L29} (\bibinfo {year}
  {1978})}\BibitemShut {NoStop}%
\bibitem [{\citenamefont {Malkov}\ and\ \citenamefont
  {Drury}(2001)}]{Malkov:2001kya}%
  \BibitemOpen
  \bibfield  {author} {\bibinfo {author} {\bibfnamefont {M.}~\bibnamefont
  {Malkov}}\ and\ \bibinfo {author} {\bibfnamefont {L.~O.}\ \bibnamefont
  {Drury}},\ }\bibfield  {title} {\bibinfo {title} {{Nonlinear theory of
  diffusive acceleration of particles by shock waves}},\ }\href
  {https://doi.org/10.1088/0034-4885/64/4/201} {\bibfield  {journal} {\bibinfo
  {journal} {Rept. Prog. Phys.}\ }\textbf {\bibinfo {volume} {64}},\ \bibinfo
  {pages} {429} (\bibinfo {year} {2001})}\BibitemShut {NoStop}%
\bibitem [{\citenamefont {Dom\v{c}ek}\ \emph {et~al.}(2021)\citenamefont
  {Dom\v{c}ek}, \citenamefont {Vink}, \citenamefont {Hern\'andez~Santisteban},
  \citenamefont {Delaney},\ and\ \citenamefont {Zhou}}]{Domcek:2020pfz}%
  \BibitemOpen
  \bibfield  {author} {\bibinfo {author} {\bibfnamefont {V.}~\bibnamefont
  {Dom\v{c}ek}}, \bibinfo {author} {\bibfnamefont {J.}~\bibnamefont {Vink}},
  \bibinfo {author} {\bibfnamefont {J.~V.}\ \bibnamefont
  {Hern\'andez~Santisteban}}, \bibinfo {author} {\bibfnamefont
  {T.}~\bibnamefont {Delaney}},\ and\ \bibinfo {author} {\bibfnamefont
  {P.}~\bibnamefont {Zhou}},\ }\bibfield  {title} {\bibinfo {title} {{Mapping
  the spectral index of Cassiopeia A: evidence for flattening from radio to
  infrared}},\ }\href {https://doi.org/10.1093/mnras/staa3896} {\bibfield
  {journal} {\bibinfo  {journal} {Mon. Not. Roy. Astron. Soc.}\ }\textbf
  {\bibinfo {volume} {502}},\ \bibinfo {pages} {1026} (\bibinfo {year}
  {2021})},\ \Eprint {https://arxiv.org/abs/2005.12677} {arXiv:2005.12677
  [astro-ph.HE]} \BibitemShut {NoStop}%
\end{thebibliography}%
